\documentclass[graybox, secnum]{svmult}

\usepackage{mathptmx}       
\usepackage{helvet}         
\usepackage{courier}        
\usepackage{type1cm}        
\usepackage{makeidx}         
\usepackage{graphicx}        

\usepackage[usenames,dvipsnames]{xcolor}
\usepackage[margin=10pt,font=small,labelfont=bf,labelsep=endash]{caption}
\usepackage{subcaption}
\usepackage{amsmath}
\usepackage{amssymb}

\graphicspath{{Figures_gok/}}
\usepackage{multicol}        
\usepackage[bottom]{footmisc}
\usepackage{hyperref}        
\usepackage{soul}            
\hypersetup{colorlinks=true,urlcolor=blue,citecolor = blue, linkcolor = blue}
\usepackage[square,numbers]{natbib}

\makeindex             

\begin{document}
\title*{Compton Telescopes for Gamma-ray Astrophysics}
\author{Carolyn Kierans\thanks{corresponding author}, Tadayuki Takahashi, and Gottfried Kanbach}
\institute{
Carolyn Kierans
 \at NASA Goddard Space Flight Center, 8800 Greenbelt Way, Maryland 20771, USA \\ \email{carolyn.a.kierans@nasa.gov}
\and Tadayuki Takahashi \at Kavli Institute for the Physics and Mathematics of the Universe (WPI), The University of Tokyo,
5-1-5 Kashiwa-no-Ha, Kashiwa, 
Chiba, 277-8583, Japan
\and Gottfried Kanbach \at Max Planck Institute for Extraterrestrial Physics, Giessenbachstrasse 1, 85748 Garching, Bayern, Germany
}
%
%
\maketitle
\abstract{
Compton telescopes rely on the dominant interaction mechanism in the MeV gamma-ray energy range: Compton scattering. 
By precisely recording the position and energy of multiple Compton scatter interactions in a detector volume, a photon's original direction and energy can be recovered. 
These powerful survey instruments can have wide fields of view, good spectroscopy, and polarization capabilities, and can address many of the open science questions in the MeV range, and in particular, from multimessenger astrophysics. 
The first space-based Compton telescope was launched in 1991 and progress in the field continues with advancements in detector technology.
This chapter will give an overview of the physics of Compton scattering and the basic principles of operation of Compton telescopes; electron tracking and polarization capabilities will be discussed. A brief introduction to Compton event reconstruction and imaging reconstruction is given. The point spread function for Compton telescopes and standard performance parameters are described, and notable instrument designs are introduced. 
}

\vspace{1cm}

\section{Introduction}
\label{sec:intro}

Megaelectronvolt (MeV) gamma rays are an excellent tool to explore the cosmos: they travel long distances without deviation or absorption, they probe deeper into dense regions than radiation at other wavelengths, and they are emitted by the most extreme and energetic processes in the universe.
In particular, the soft and medium energy regime, from $\sim$100~keV to 100~MeV, is ripe with scientific opportunities. 
With recent advances in multimessenger astrophysics, the community's demand for sensitive medium energy gamma-ray telescopes has grown as the sources which generate gravitational waves, neutrinos, and cosmic rays shine bright in the low-end of the gamma-ray spectrum. 

The MeV range is rich with scientific potential.
Neutron stars, black holes, magnetars with extreme magnetic fields, and active galactic nuclei with relativistic jets are natural accelerators that emit gamma rays through bremsstrahlung, inverse Compton, and synchrotron radiation.
Emissions in the MeV range constrain the thermal and non-thermal energy budget for high-energy Galactic and extra-galactic sources.
There are also compelling reasons to believe that indirect signatures of dark matter will be found at these energies. 
And finally, the MeV range is the natural scale to probe nuclear processes in the Galaxy and beyond. 
Many radioactive isotopes decay by emitting gamma rays up to the nuclear binding energies $\lesssim 8$~MeV per nucleon. 
Radioactive isotopes generated in stellar and explosive nucleosynthesis are proton rich and often undergo $\beta ^{+}$ decay, emitting a positron which later annihilates and generates signature 511~keV photons. 

\begin{figure}[tb]
    \centering
    \includegraphics[width=0.85\textwidth]{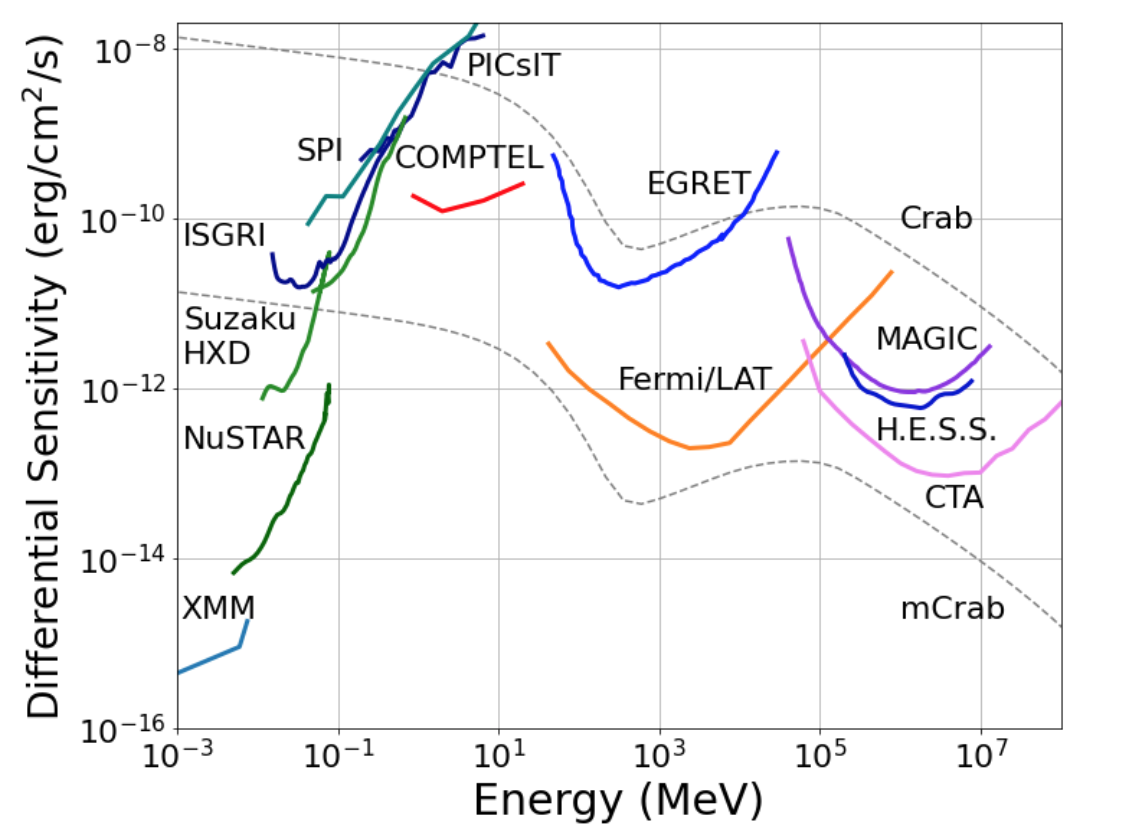}
    \caption{
    The differential sensitivity for current and past X-ray and gamma-ray missions shows the limited performance achieved in the MeV regime. 
    The reduced sensitivity in the range from 100~keV to 100~MeV is referred to as the \textit{MeV Gap}.
    The XMM sensitivity is for a 1.6 Ms observation 
    at 4$\sigma$ detection significance~\cite{hasinger2001xmm,XMM_Web}.
    The NuSTAR~\cite{koglin2005nustar}, Suzaku HXD~\cite{FukazawaHXD}, INTEGRAL ISGRI~\cite{LebrunISGRI}, PICsIT~\cite{CoccoPICsIT}, and SPI~\cite{roques2003spi} sensitivities are shown for 3$\sigma$ detections with 100~ks exposure and $\Delta E/E= 0.5$, assuming statistical error only.
    The sensitivity of COMPTEL corresponds to the 9~yr all-sky survey of CGRO~\cite{Schonfelder2000}. 
    CGRO/EGRET sensitivity (3$\sigma$) is taken from \cite{2005AIPC..745..210K} for a 300~hr exposure, which is consistent with the numbers given in \cite{ThompsonEGRET}. 
    The sensitivity of Fermi/LAT is for the 10 yr survey~\cite{LAT10year,Fermi}. 
    The 5$\sigma$ sensitivities for MAGIC, H.E.S.S, and CTA (simulated) correspond to 50 hr observations \cite{Zanin:2021tx,CTA}.
    The dashed grey line representing the Crab flux is calculated from Naima~\cite{naima}. 
     }
    \label{fig:MeVGap}
\end{figure}

However, the soft to medium energy gamma-ray regime remains one of the least explored energy ranges in multi-wavelength astrophysics. 
This gap in sensitivity, often referred to as the \textit{MeV Gap}, has greatly hindered progress of science due to limited observations in this thermal to non-thermal transition regime. 
Figure~\ref{fig:MeVGap} shows the continuum sensitivity of current and past telescopes in the X-ray and high-energy gamma-ray regime, and it is clear that the sensitivity in the \textit{MeV Gap} is orders of magnitude less than neighboring bands.

The lack of sensitive telescopes in the MeV range is not due to a lack of scientific interest, as demonstrated throughout Volume 3 and 4 of this handbook, but instead due to physical, technological, and environmental challenges (i.e., background radiation). 
The interaction cross-section between photons and matter is lowest in this range, and the dominant mechanism is Compton scattering, which produces long-range secondaries. 
Large detector volumes are needed to stop and contain photons with interaction depths on the order of $\sim$10~g/cm$^{2}$. 
And with no focusing element, these large detector volumes result in high backgrounds, not only from atmospheric albedo emission, but also from activation within and around the instrument.
While this high background does not significantly impact the detection sensitivity to high-energy transients due to the short bright signals, studies of steady-state sources in the MeV range require sophisticated background rejection techniques.

To build a sensitive telescope to explore the MeV Gap, it must make use of the dominant interaction mechanism in this range: Compton scattering. 
In Compton scattering, gamma rays will partially transfer their energy to bound electrons, and by doing so, the electron recoils and the gamma ray is scattered at an angle relative to its initial direction with lower energy.
The Compton-scattered photon will then interact a second time, and potentially a third and fourth, before finally absorbing all its energy in the detector.
Only the complete measurement of the secondary products, i.e. the energized target electron and the de-energized scattered photon, allow for the determination of the initial energy and direction of the incident photon. 

The pioneering telescope that opened the MeV range as a new window to astronomy was the Imaging Compton Telescope (COMPTEL \cite{schonfelder1993}) aboard the Compton Gamma Ray Observatory (CGRO). 
COMPTEL covered the energy range 0.8--30~MeV from its launch in 1991 until its termination in 2000. 
It produced the first all-sky MeV survey, mapped the diffuse Galactic emission in the continuum and in the light of radioactive gamma-ray lines (e.g. $^{26}$Al, $^{44}$Ti), and detected more than 30 gamma-ray sources (pulsars, pulsar wind nebulae, black holes, and active galaxies).
COMPTEL revolutionized gamma-ray astrophysics, and many of the techniques and tools developed for the mission are still being used in modern Compton telescopes.
As the last instrument to observe much of the MeV Gap,  there are continuing efforts to build upon COMPTEL's legacy.


The power of Compton telescopes come from the single photon detection and imaging capabilities. 
With a large field of view and modest angular resolutions ($\sim$1$^{\circ}$), Compton telescopes make powerful survey instruments.
The single photon detections and necessary event reconstruction allow for effective background rejection capabilities. 
Additionally, Compton scattering is inherently sensitive to the linear polarization of incoming gamma rays, and there is a class of Compton telescopes designed specifically for polarization measurements of transients, such as gamma-ray bursts.
Recent advances in technology have shown the breadth and power of Compton telescopes, and looking 
forward the future is bright with the Compton Spectrometer and Imager (COSI) mission~\cite{Tomsick2021} selected by NASA to launch in 2026.


This chapter aims to provide a concise, comprehensive, and up-to date overview of Compton telescopes for astrophysics. 
The physics of Compton scattering, which is the basis of telescope operation, will be introduced in Section~\ref{sec:physics}. 
An overview of the operating principles of Compton telescopes including the COMPTEL-like double-scattering approach, modern compact Compton telescopes, and electron tracking capabilities, will be given in Section~\ref{sec:operating}. 
Section~\ref{sec:eventrecon} provides a discussion of event reconstruction techniques, and Section~\ref{sec:performance} covers the standard performance parameters for a Compton telescope. 
And finally, notable Compton telescope instruments will be briefly introduced in Section~\ref{sec:instruments}. 

\section{Physics of Compton Scattering}
\label{sec:physics}

A major debate was conducted after the discovery of X-rays (also called R\"ontgen-rays in the early days) in the 1900's on the nature of this new, penetrating radiation. 
Arthur H. Compton, who received the Nobel prize in Physics in 1927 for the discovery of Compton scattering, summarized the profound findings in his Nobel lecture~\cite{ComptonNobel}:  
\begin{quote}
    All phenomena in the realm of light are found in parallel in the realm of X-rays. 
    Reflection, refraction, diffuse scattering, polarization, diffraction, emission and absorption spectra, photoelectric effect, all of the essential characteristics of light have been found also to be characteristic of X-rays. 
    At the same time it has been found that some of these phenomena undergo a gradual change as we proceed to the extreme frequencies of X-rays.
\end{quote}  
For high-energy X-rays, a modified component of scattered radiation was found at large angles ($\varphi$) offset from the primary beam ($\lambda _0$) and shifted to longer wavelengths ($\lambda _{scat}$), i.e. lower energies. 
This experimental result  could not be explained by coherent scattering of electromagnetic waves on free or bound electrons (Thomson or Rayleigh scattering). 
Arthur Compton introduced the new concept that quanta of light, photons, scattering off electrons can be described by the law of mechanics~\cite{Compton1923}.

\begin{figure}[tb]
    \centering
    \includegraphics[width=7cm]{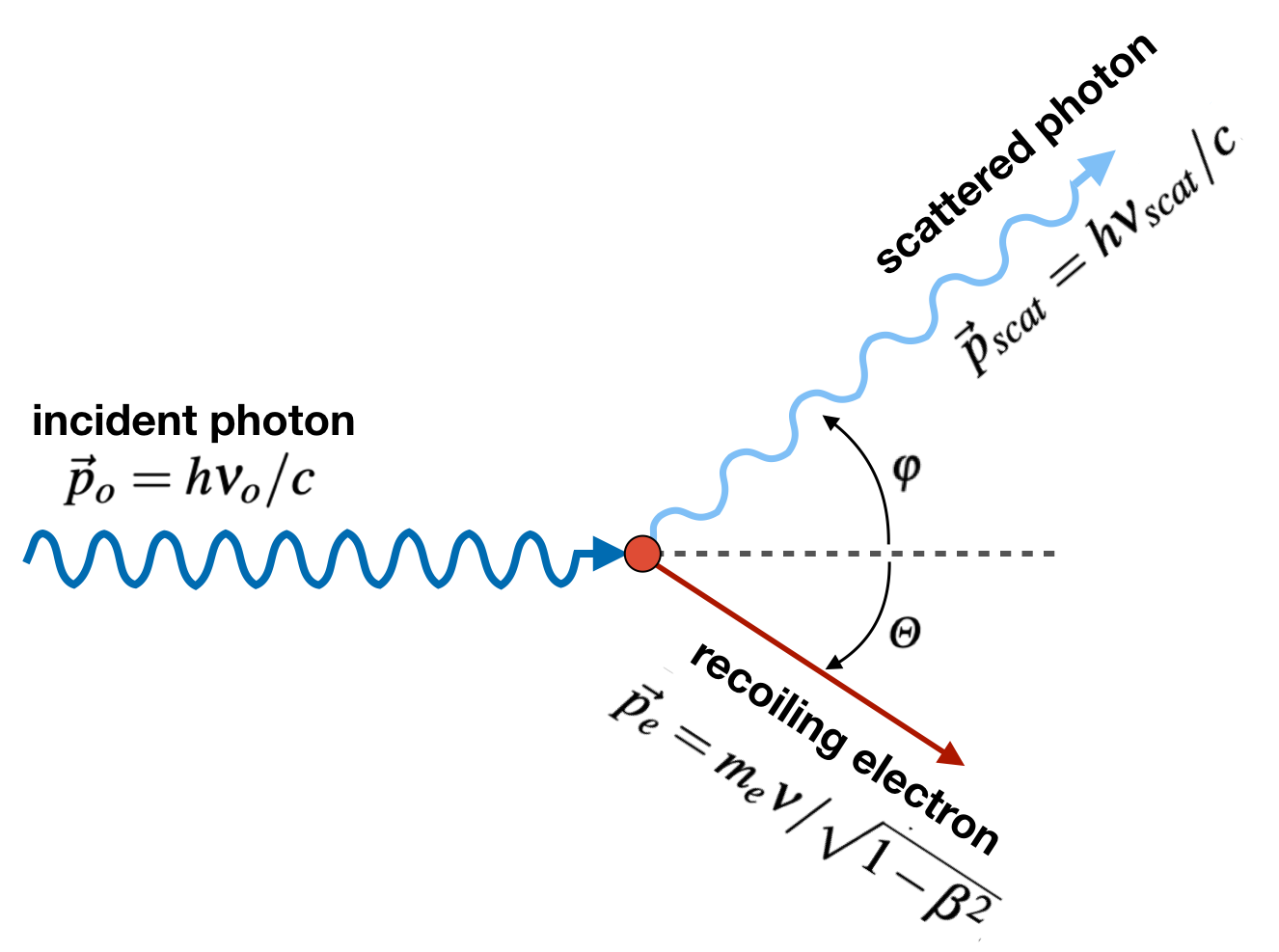}
    \caption{A photon with initial momentum $\vec{p}_0 = h \nu_0/c$ scatters off an electron assumed to be unbound and at rest. The photon scatters at an angle $\varphi$ relative to the initial direction, and has a momentum $\vec{p}_{scat} = h \nu_{scat} /c$. The electron recoils from the scatter at an angle $\Theta$ and with momentum $\vec{p_e} = m_e \nu/\sqrt{1-\beta^2}$. Through the conservation of energy and momentum, the relation between the scattering angle and the final photon energy can be determined.}
    \label{fig:Compton_scattering}
\end{figure}

Compton showed that scattering photons conserve energy and momentum, as shown schematically in Figure~\ref{fig:Compton_scattering}. For a photon with initial energy $E_0$, energy conservation gives
\begin{equation}
    E_0 = E_{scat} + E_e,
\end{equation}
where $E_{scat}$ is the scattered photon energy, and the electron, with final energy $E_e$, is assumed to initially be free and at rest (this limitation is discussed in Section~\ref{sec:doppler}).
Conservation of momentum gives
\begin{equation}
    \vec{p_0} = \vec {p}_{scat} + \vec{p_e},
\end{equation}
where the initial momentum of the photon is $ \left| \vec{p_0} \right| = h \nu_0 / c$, and the final reduced photon momentum is $\left| \vec{p}_{scat} \right| = {h \nu_{scat} / c} $. The recoil electron is accelerated and the final momentum is given by
$ \left| \vec{p_e} \right| = m_e v \gamma$, with 
 $\ \gamma = {1 / {\sqrt{1-\beta ^2}} }$ and $\beta = v/c$. 
This leads to the well-known Compton equation:
\begin{equation}
\label{eq:Compton_wavelength}
\lambda _{scat} - \lambda _0 = \frac{h}{m_e c} (1-\cos \varphi),
\end{equation}
where $h$, $m_e$, and $c$ are Planck's constant, the electron rest mass, and the speed of light, respectively. The fraction $ {h/ {m_e c}} = 2.426\ 10^{-12}$~m is often called the Compton wavelength, which is the wavelength shift for a $90^\circ$ scattering. It is important to note that a photon can never lose all its energy in a Compton scattering process, even if it is completely backscattered. 

Equation~\ref{eq:Compton_wavelength} can be solved for $\varphi$ and with wavelength converted to energy:
\begin{equation}\label{eq:Compton_scatter_angle}
\cos \varphi =  1 - m_e c^2 \left( \frac{1}{E _{scat}} - \frac{1}{E _0} \right),
\end{equation}
where the energy of the scattered photon is
\begin{equation} \label{eq:Photon_scattered}
 E_{scat} = h \nu _{scat} = \frac{E_0}{1+ \frac{E_0}{m_e c^2} {(1- \cos \varphi)}}.
\end{equation}
As one can see in Equation~\ref{eq:Compton_scatter_angle} and \ref{eq:Photon_scattered}, the final energy of the Compton-scattered photon can be uniquely determined with the initial energy and Compton scatter angle information.

\begin{figure}[tb]
    \centering
    \includegraphics[width=0.8\textwidth]{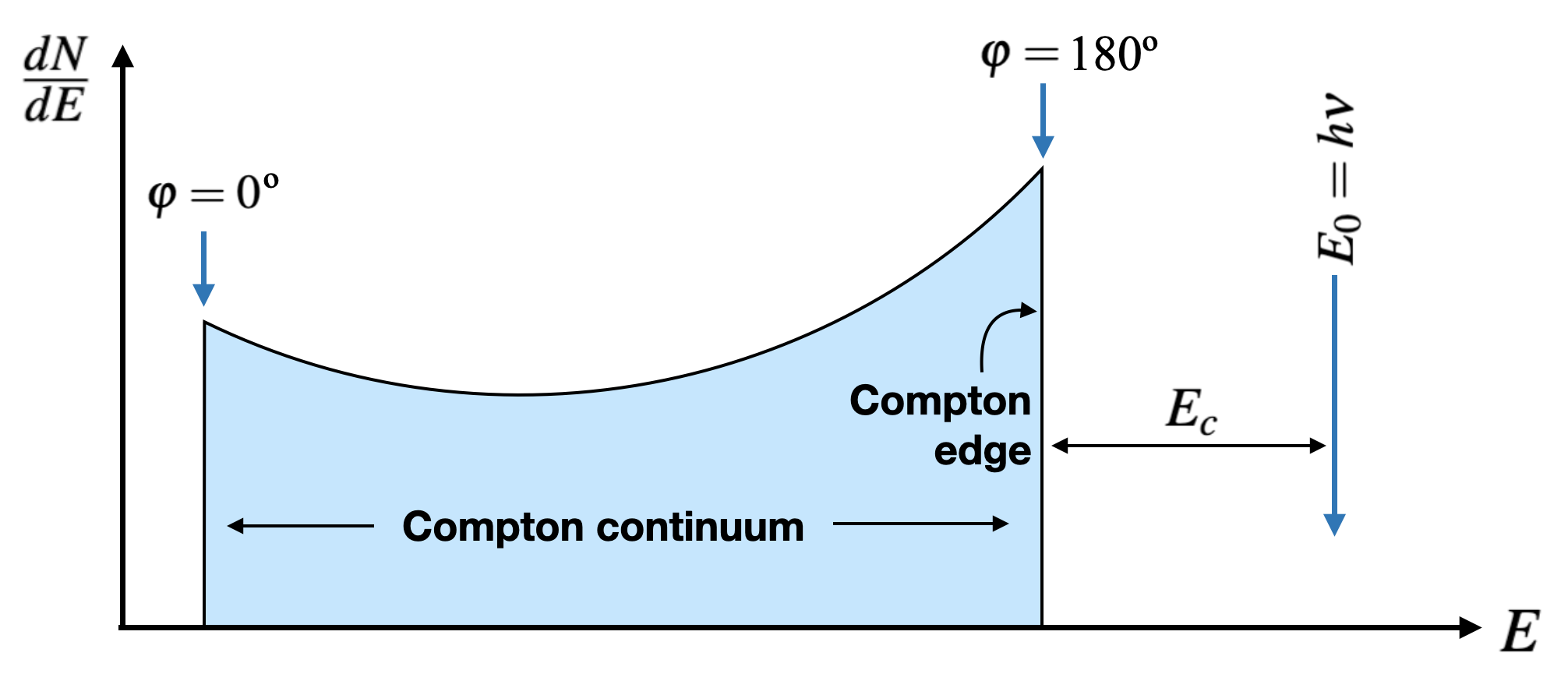}
    \caption{The spectral shape of the recoil electron energy for a photon with initial energy $E_0 = h\nu$. There is a continuum of electron energies observed due to the range of allowable Compton scatter angles. The maximum recoil electron energy is achieved for backscatter interactions when $\varphi = 180^{\circ}$; this defines what is referred to as the Compton edge in a measured spectrum. Modified from \citet{knoll}.}
    \label{fig:compton_spectrum}
\end{figure}

As detectors are sensitive to the ionization from the recoil electron, it is important to consider the kinetic energy of the Compton-scattered electron:
\begin{equation} \label{eq:Electron_scattered_1}  
E_e = E_0 - E_{scat} = E_0 \left( \frac{E_0 (1-\cos \varphi )}{{m_e c^2} + E_0 (1-\cos \varphi)} \right).
\end{equation}
The spectral shape of the electron energy is shown in Figure~\ref{fig:compton_spectrum}. 
To better understand this distribution, we consider the extreme cases of grazing-angle scattering and backscattering. For $\varphi \approx 0$, Equation~\ref{eq:Electron_scattered_1} gives the recoil electron energy $E_e \approx 0$, and Equation~\ref{eq:Photon_scattered} gives $E_{scat} \approx E_0$.
For a head-on collision where the photon backscatters at $\varphi = 180^{\circ}$, the maximum amount of energy is transferred to the electron and Equation~\ref{eq:Electron_scattered_1} reduces to
\begin{equation}
\label{eq:comptonedge}
    \left. E_e \right\rvert_{\varphi = 180^{\circ}} = E_0\left( \frac{2E_0}{m_ec^2 + 2E_0} \right) = E_0\left( 1 - \frac{1}{1 + \frac{2E_0}{m_e c^2}} \right).
\end{equation}
The difference between the initial photon energy and the maximum recoil electron energy is given by
\begin{equation}
    E_c = E_0 - \left. E_e \right\rvert_{\varphi = 180^{\circ}} = \frac{E_0}{1 + \frac{2E_0}{m_ec^2}},
\end{equation}
which is often called the Compton edge and is depicted in Figure~\ref{fig:compton_spectrum}.

The Compton-scattered electron energy $E_e$ can also be expressed in terms of the angle $\Theta$ between the incident photon and the direction of the recoil electron (see Figure~\ref{fig:Compton_scattering}), making use of the relation 
$$ \cot (\Theta) = (1+\alpha)  \tan (\varphi /2),$$
where $\alpha = {E_0 / {m_e c^2}}$
\begin{equation} \label{eq:Electron_scattered_2}
 E_e = {{2 E_0 \alpha \cos^2 \Theta} \over {(1+\alpha)^2 - \alpha ^2 \cos^2 \Theta}}.
\end{equation}

\begin{figure}[tb]
    \begin{minipage}{0.55\textwidth}
    \centering
    \includegraphics[width=0.95\textwidth]{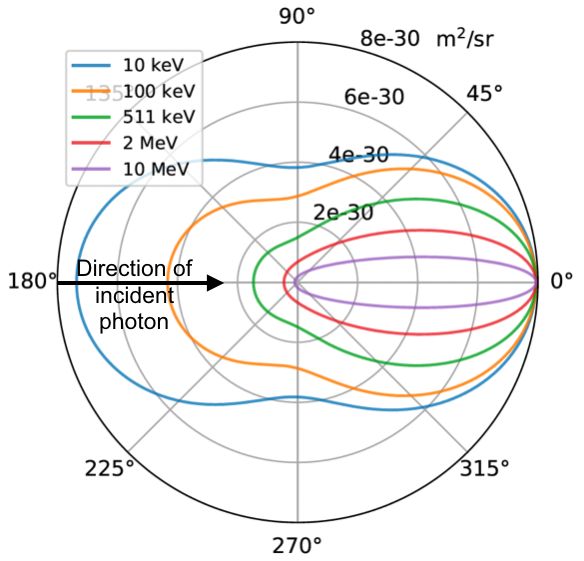}
    \end{minipage}
    \hfill
    \begin{minipage}{0.45\textwidth}
    \caption{The Klein-Nishina differential cross-section (m$^2$/sr) as a function of Compton scatter angle shows the energy dependence of the scattering direction. For lower energy photons, such as 10~keV shown with the blue curve, the magnitude of the differential cross-section only changes by a factor of two for all Compton scatter angles. For higher energy photons, such as 2~MeV shown in red, the cross-section is maximized for small scattering angles.}
    \label{fig:klein-nishina-diff}
    \end{minipage}
\end{figure}

The full quantum-mechanical treatment of the scattering of unpolarized energetic photons on single, free target electrons $\gamma + e^- \rightarrow \gamma + e^-$ was derived by Oskar Klein and Yoshio Nishina in 1928. 
The differential cross-section for Compton scattering 
is given by the Klein-Nishina formula~\cite{klein1929}:
\begin{equation} \label{eq:klein-nishina}
	\frac{d\sigma}{d\Omega} = \frac{{r_0}^2}{2} \left( \frac{E_{scat}}{E_0} \right)^2 \left( \frac{E_{scat}}{E_0} + \frac{E_0}{E_{scat}} - \sin^2 \varphi  \right) ,  \\ 
\end{equation}
where $r_0 = 2.818\times10^{-15} {\rm\ m}$ is the classical electron radius, $E_0$ is the initial energy of the incident photon, $E_{scat}$ is the energy of the scattered photon, $\varphi$ is the scattering polar angle, as shown in Figure~\ref{fig:Compton_scattering}. 
From this equation it can be seen that higher energy photons will, in general, have smaller Compton scatter angles, and lower energy photons will result in larger scatter angles.
The Klein-Nishina differential cross-section is shown graphically in Figure~\ref{fig:klein-nishina-diff} for a number of different energies.

Integration of Equation~\ref{eq:klein-nishina} over $d\Omega$ gives the total cross-section for Compton scattering. In terms of Thomson units ($ \sigma _T = {8\pi e^2} / {3 m_e c^2}$ = $0.665 \times 10^{-28}$~m$^2$), and using dimensionless units for the photon energy ($ \epsilon = E_{0} / {m_e c^2}$), the total cross-section is:
\begin{equation} \label{eq:KN_total cross section}
\sigma _C = {3 \over 4} \sigma _T \left[ {{1+\epsilon} \over {\epsilon ^3}} \left( {{2\epsilon (1+\epsilon)} \over {1+2\epsilon}} - \ln (1+2\epsilon)  \right) + {\ln (1+2\epsilon) \over {2\epsilon}} - {{1+3\epsilon} \over {(1+2\epsilon)^2}} \right]
\end{equation}
If the gamma-ray energy is much smaller than electron mass, the total cross-section is reduced to the Thomson cross-section, $\sigma_{T}$.

\begin{figure}[tb]
\begin{minipage}{0.55\textwidth}
    \centering
    \includegraphics[width = 0.8\textwidth]{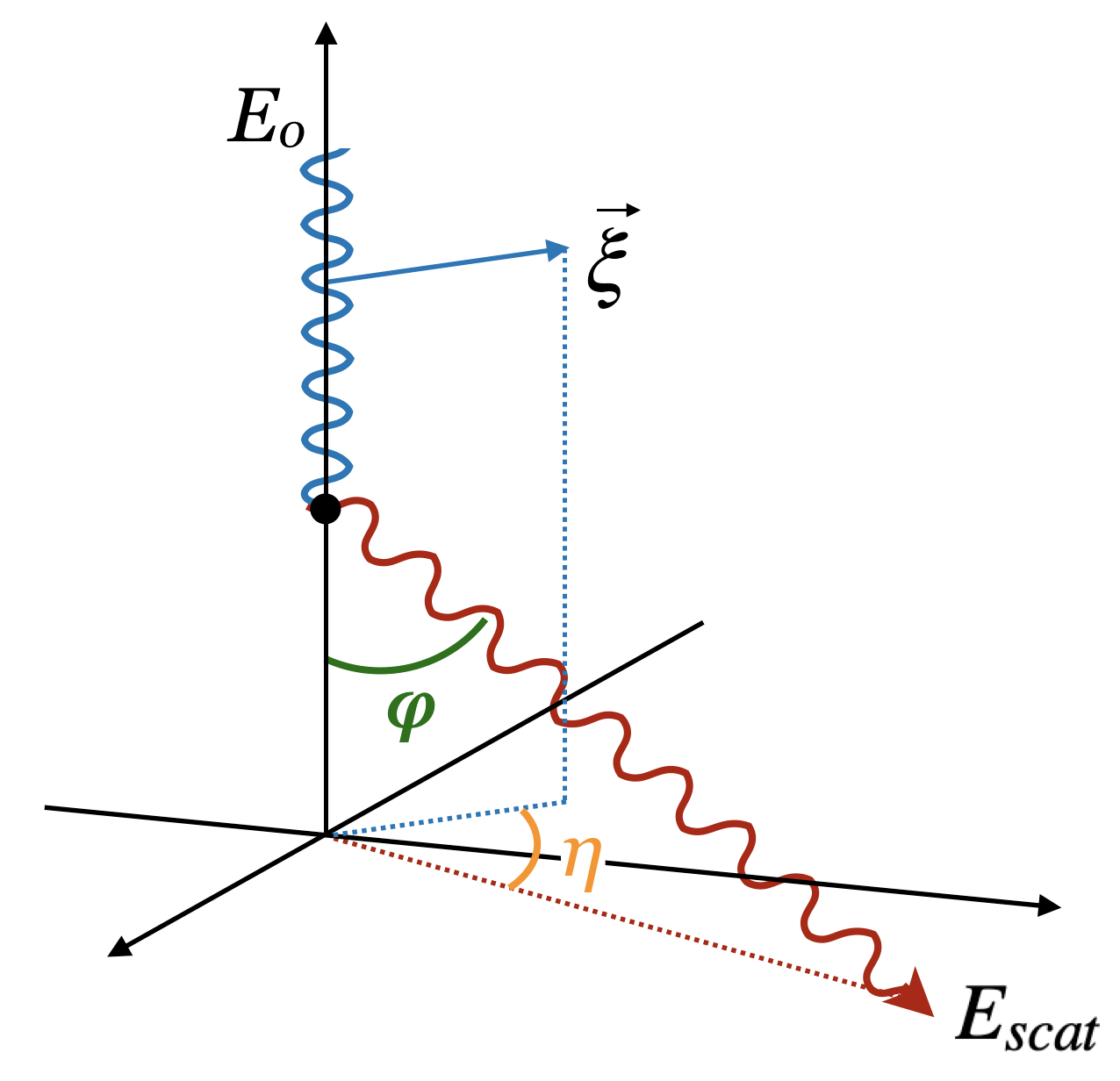}
\end{minipage}
\hfill
\begin{minipage}[b!]{0.45\textwidth}
\captionof{figure}{The Klein-Nishina differential cross-section for polarized photons (Equation~\ref{eq:klein-nishina-pol}) is maximized when the azimuthal scattering angle $\eta$ is perpendicular to the photon's initial electric field vector $\vec{\xi}$. 
By measuring the azimuthal scattering angle $\eta$, an instrument can be sensitive to the linear polarization of incoming photons.}
\label{fig:polarized_angles}
\end{minipage}
\end{figure}

The differential cross-section for linearly polarized photons was derived by \citet{Heitler1954} in 1954:
\begin{equation} \label{eq:klein-nishina-pol}
	\frac{d\sigma}{d\Omega} = \frac{{r_0}^2}{2} \left( \frac{E_{scat}}{E_0} \right)^2 \left( \frac{E_{scat}}{E_0} + \frac{E_0}{E_{scat}} - 2 \sin^2 \varphi \cos^2 \eta \right) ,  \\ 
\end{equation}
where the polar Compton scatter angle $\varphi$ and the azimuthal scattering angle $\eta$ are explicitly written~\cite{Evans1955}. 
There is a similar dependence on the Compton scatter angle $\varphi$ in Equation~\ref{eq:klein-nishina}; however, the polarized differential cross-section is maximized when the $\cos^2\eta$ term equals zero. 
In other words, photons will predominantly scatter at $90^{\circ}$ relative to the the initial photon’s electric field vector, defined as $\eta = 0$. 
This results in an azimuthal-dependence of the cross-section, and therefore, Compton telescopes can inherently detect polarization if a measurement of $\eta$ is made~\cite{lei1997}. 
The definition of the scattering angles relative to the initial photon electric field vector $\vec{\xi}$ is shown in Figure~\ref{fig:polarized_angles}.

Equation \ref{eq:klein-nishina} not only describes the energy dependence of the Compton scattering cross-section on single electrons, but also implicitly the total cross-section for scattering in material of atomic number ($Z$) and density $\rho$. 
Since each single scattering event is independent, the total cross-section is the sum of all events and depends directly on the electron density in the target material. Detailed compilations of gamma-ray cross-sections for various materials can be accessed with the internet based application $XCOM$~\cite{xcom}.

It is convenient to describe the total cross-section in terms of the mass attenuation coefficient $\mu$, particularly when comparing Compton scattering to the dominant photon interactions at lower and higher energies, i.e., photoelectric absorption and pair production. 
These coefficients, given in units of cm$^2$/g, are most useful to estimate what fraction of photons from an incoming beam with intensity $I_0$  are lost by interactions, either by absorption or scattering. The intensity $I$ of a beam  after traversing  matter with thickness $x$~(g/cm$^2$) is 
\begin{equation} \label{eq:mass attenuation}
 I = I_0 e^{-\mu x},
\end{equation}
where $x$ is the product of the geometrical path length $l$ and density $\rho$ of the transversed material.

As an illustrative example, we can calculate the amount of silicon needed to attenuate 50\% of incident gamma rays with initial energy of 511~keV. From $XCOM$, the $\mu$ for silicon at 511~keV is $8.67\times 10^{-2}$~cm$^2$/g, and the density $\rho$ is 2.33~g/cm$^3$.
Rearranging Equation~\ref{eq:mass attenuation} for the path length gives
\begin{equation}
    l = \frac{1}{\rho \mu} \ln\left(\frac{I}{I_0} \right) = \frac{\ln(0.5)}{(2.33~\text{g/cm}^3) (8.67 \times 10^{-2}~\text{cm$^2$/g})} = 3.4~\text{cm}.
\end{equation}
Therefore, over 3~cm of silicon are needed to scatter or absorb only half of incident 511~keV photons. This exemplifies the challenges of  low interaction cross-sections for Compton telescopes where large detector volumes are required.


\begin{figure}[tb]
    \centering
    \begin{subfigure}[b]{0.32\textwidth}
    \includegraphics[width=\textwidth]{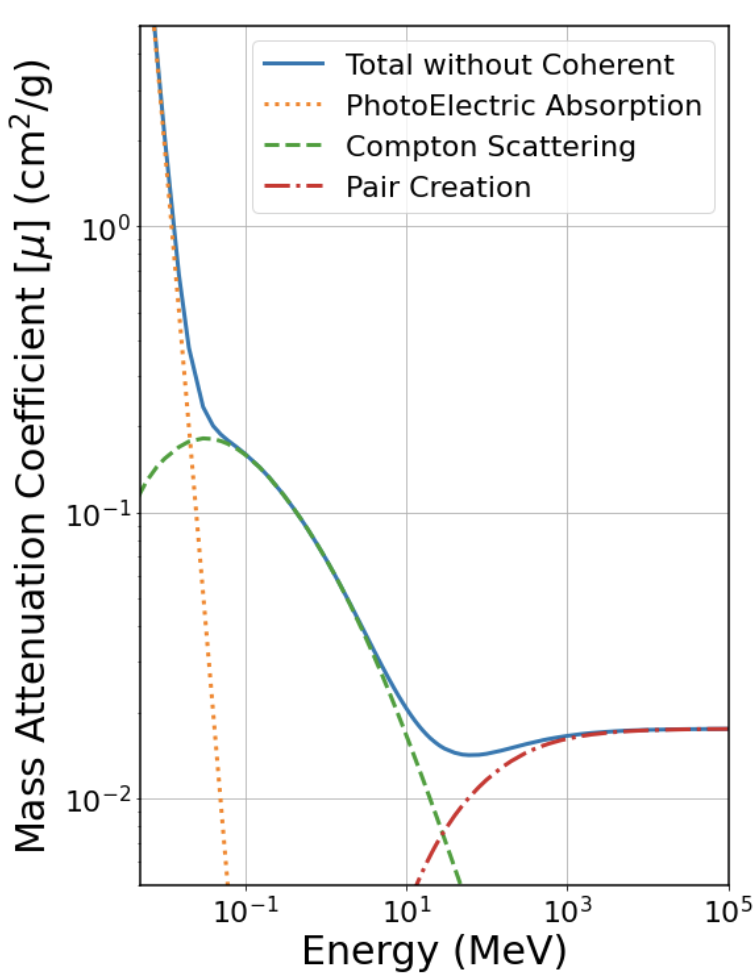}
    \caption{Plastic Scintillator}
    \label{fig:massattenuation_plastic}
    \end{subfigure}
    \begin{subfigure}[b]{0.32\textwidth}
    \includegraphics[width=\textwidth]{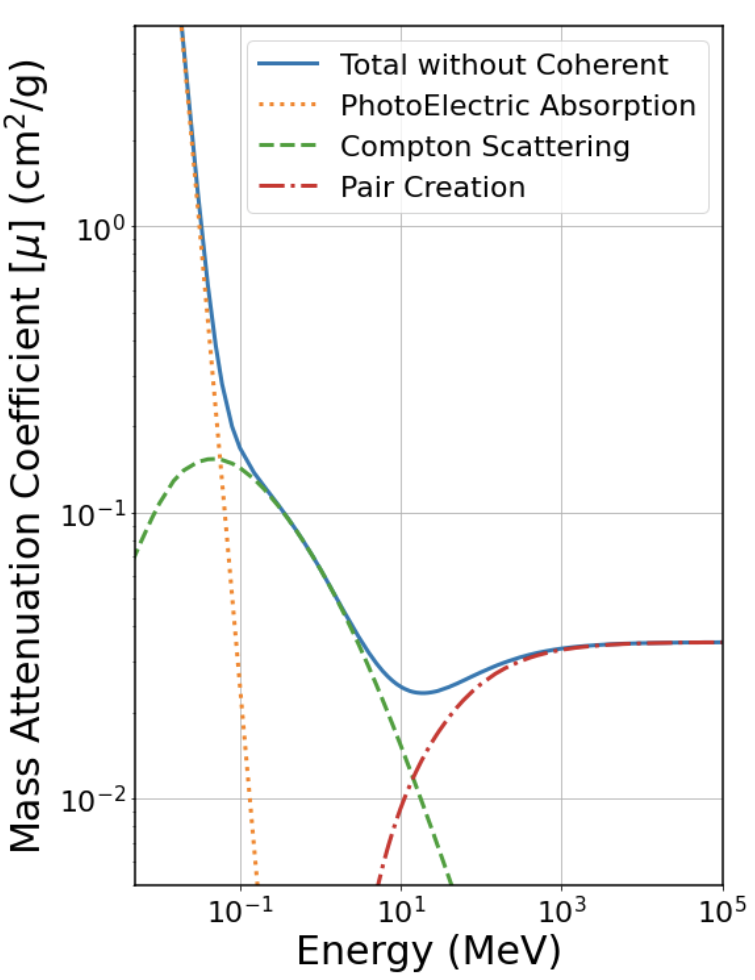}
    \caption{Silicon}
    \label{fig:massattenuation_si}
    \end{subfigure}
    \begin{subfigure}[b]{0.32\textwidth}
    \includegraphics[width=\textwidth]{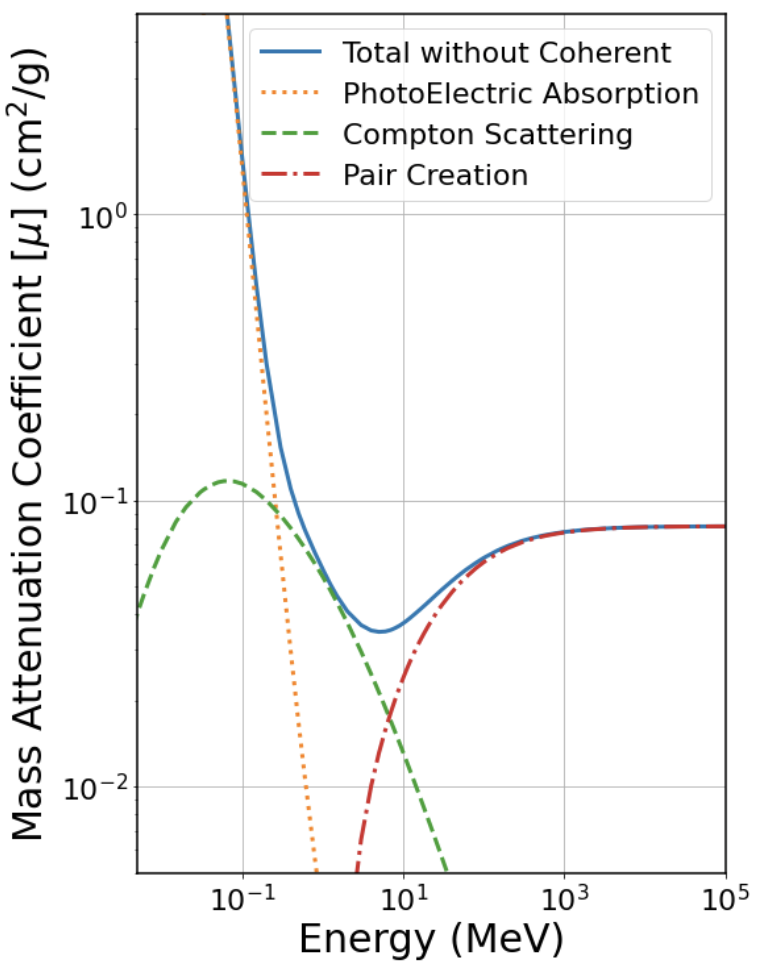}
    \caption{Sodium Iodide (thallium doped)}
    \label{fig:massattenuation_nai}
    \end{subfigure}
    \caption{The mass attenuation coefficients for three common gamma-ray detector materials: (\textbf{a}) plastic scintillator (hydrocarbon-based polymers), (\textbf{b}) silicon, and \textbf{(c)} sodium iodide NaI(Tl). Compton scattering dominates over a much larger energy range from $\sim$30~keV to 30~MeV for low-Z materials, as shown here for plastic scintillator. For high-Z materials, such as NaI, Compton scattering only is dominant from $\sim$300~keV to 7~MeV. Calculations are from $XCOM$~\cite{xcom}.}
    \label{fig:mass_attenuation_coefficients}
\end{figure}

\begin{figure}[tp]
    \centering
    \includegraphics[width=8cm]{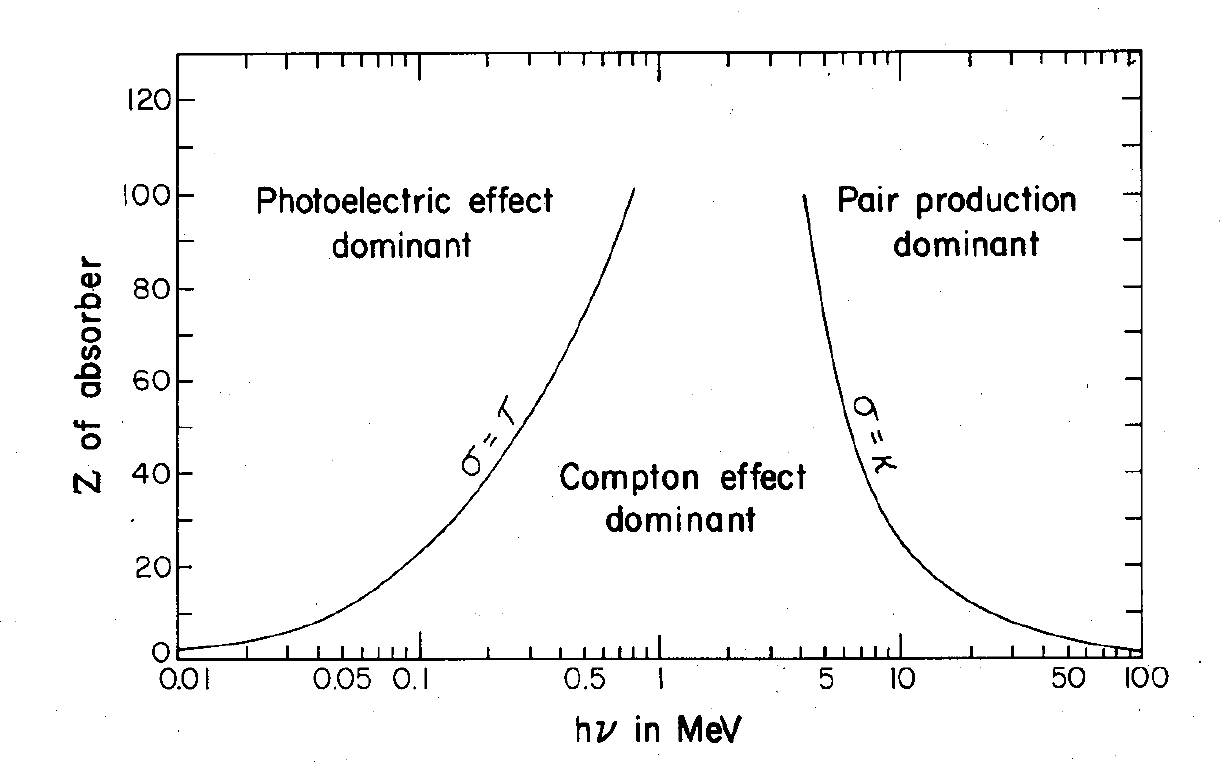}
    \caption{The regions of dominant interaction cross-sections for the photoelectric effect, Compton scattering, and pair production are shown as a function of photon energy and for detector materials of atomic number Z. 
    Photoelectric effect and Compton interactions have the same cross-section along the boundary $\sigma = \tau $, and pair creation dominates beyond $\sigma = \kappa $. Figure From \citet{Evans1955}.}
    \label{fig:Photo_Compton_Pair_boundaries}
\end{figure}

Figure~\ref{fig:mass_attenuation_coefficients} depicts the mass attenuation coefficient $\mu$ as a function of energy for three common detector materials: a low-Z plastic scintillator (hydrocarbon-based material of density $\rho \approx 1$~g/cm$^3$), the ubiquitous semiconductor silicon  ($\rho = 2.33$~g/cm$^3$), and the common high-Z sodium iodide doped with thallium NaI(Tl) ($\rho = 3.67$~g/cm$^3$). Compton scattering dominates in the range 30 keV to 30 MeV for the plastic scintillator, while for Na(Tl) this range is reduced to 300 keV to 8 MeV. These differences are the basis for many Compton telescope designs that separate scattering and absorption processes in multiple interactions, and
Figure~\ref{fig:Photo_Compton_Pair_boundaries} 
shows the interplay between the three major interaction processes as a function of the target material atomic number and initial photon energy. 
It is evident that materials with low to medium Z, e.g. hydrocarbon-based scintillators or solid state detectors like silicon ($Z = 14$), are the most efficient Compton scatterers over a wide energy range.  
Photoabsorption of the Compton-scattered photon is most efficient in a high-Z and high density detector, such as NaI(Tl) or cesium iodide CsI(Tl) scintillators, and solid state detectors made of cadmium  telluride (CdTe) or cadmium zinc telluride (CZT).

\section{Basic Operating Principles of Compton Telescopes}
\label{sec:operating}

In the 1960's and early 1970's, the imaging and spectroscopy of celestial  sources in the energy range from  100~keV to 10's of  MeV was considered a very difficult task.  
Instruments relying primarily on photoabsorption and pair production were developed, but were too inefficient at these energies. 
The only physical process with reasonable interaction probability in this energy range is Compton scattering (see Figure~\ref{fig:Photo_Compton_Pair_boundaries}).
However, it is impossible to derive the initial gamma ray arrival direction and energy (i.e. for imaging or spectroscopy) from only the first Compton-scattering interaction; the photon must be fully absorbed and the energy and position of each interaction must be recorded to reconstruct a Compton-scattering event.

Various work-arounds were attempted to obtain a coarse angular resolution through the use of massive collimators and anti-coincidence shields surrounding detectors large enough to contain the full energy of a Compton event and its secondaries. 
For sources characterized by a clear temporal signal (e.g. pulsars, solar flares, gamma-ray bursts), these instruments could provide important scientific results; however, a more general astronomical imaging instrument at MeV energies was elusive. 
Progress towards a sensitive MeV telescope was only made with a dedicated effort to measure the locations of energy deposits from Compton interactions.

\begin{figure}[tb]
    \centering
    \includegraphics[width=0.4\textwidth]{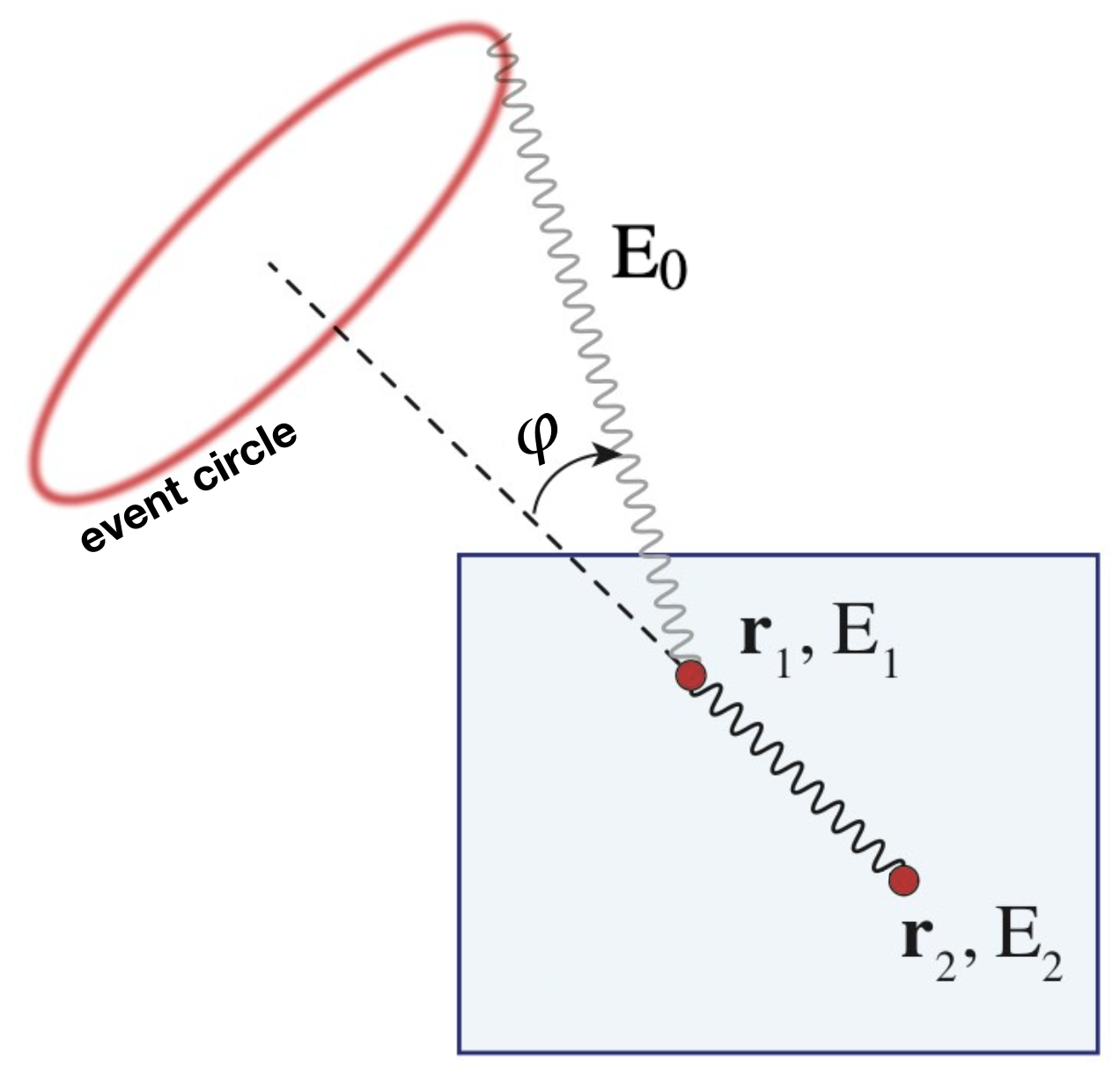}
    \hspace{1cm}
    \includegraphics[width=0.4\textwidth]{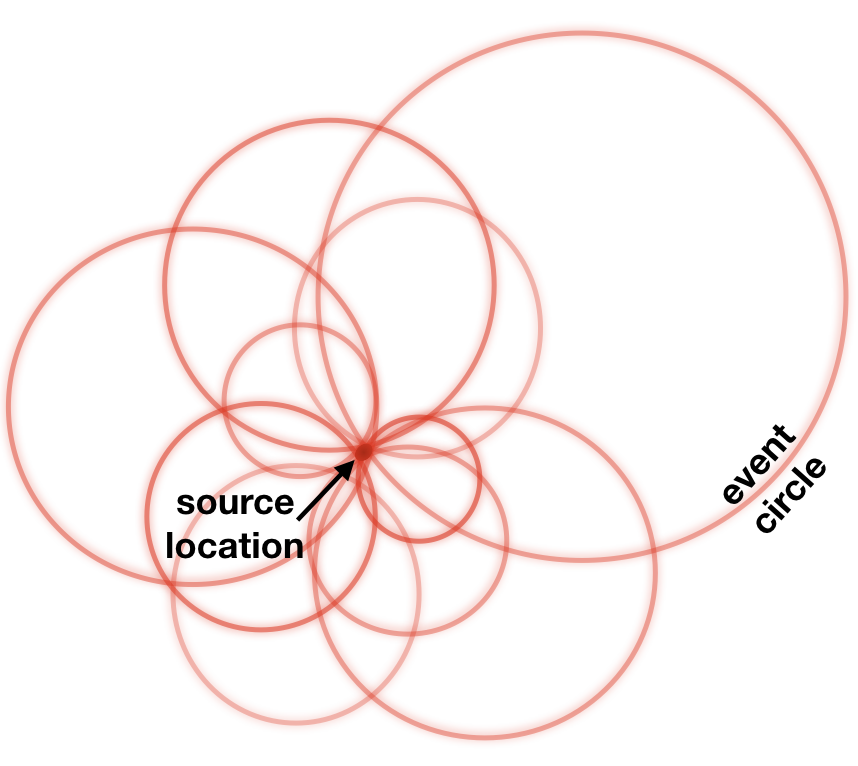}
    \caption{A Compton telescope works by measuring the energy and position of one or more Compton scattering interactions. For a two-site event, the initial energy of the photon is determined from the total absorbed energy $E_0 = E_1 + E_2$, where as the direction of the incoming photon is constrained to a circle on the sky defined by the Compton scatter angle $\varphi$ of the first interaction. When multiple photons are detected originating from the same source, the location of the source is determined by the overlap of each back-projected event circle, as is visualized for 10 events on the right. Modified from \citet{kieransthesis}.}
    \label{fig:compton_basics}
\end{figure}

Compton telescopes rely on the precise measurement of the energy and location of Compton scattering interactions which allow for the arrival direction of the primary gamma ray to be constrained. 
For example, let us consider a two-site event, that is, a photon event that consists of a single Compton scatter followed by photoelectric absorption.
The basic operating principle is shown in Figure~\ref{fig:compton_basics}, where a detector volume measures the two interactions, fully containing the event.
The Compton scatter angle $\varphi$ of the first interaction can be derived from the energy measurements. 
For a two-site event, 
Equation~\ref{eq:Compton_scatter_angle} gives
\begin{equation}
    \cos \varphi = 1 - \frac{m_ec^2}{E_2} + \frac{m_ec^2}{E_1 + E_2},
    \label{eq:twosite}
\end{equation}
where $E_1 + E_2$ is the total deposited energy, which is equal to the initial photon energy $E_0$ from Equation~\ref{eq:Compton_scatter_angle}, $E_{scat} = E_2$, and $E_e = E_1$.
The direction of the scattered gamma ray between $\vec{r}_1$ and $\vec{r}_2$ defines a cone on the sky with opening angle given by $\varphi$. 
The back-projection of this cone on the celestial sphere is referred to as the event circle.
If the recoil electron direction is not measured (see electron-tracking Compton telescopes in Section~\ref{sec:electron_tracking}), one cannot determine where on the event circle the single gamma-ray photon originates from.
Multiple photons from the same source will have overlapping event circles in image space, revealing the location of the source (see Section~\ref{sec:imaging} for an overview of imaging reconstruction techniques).

\subsection{The Classic Double Scattering Compton Telescope}

\begin{figure}[tb]
    \begin{minipage}{0.5\linewidth}
    \centering
    \includegraphics[height=7cm]{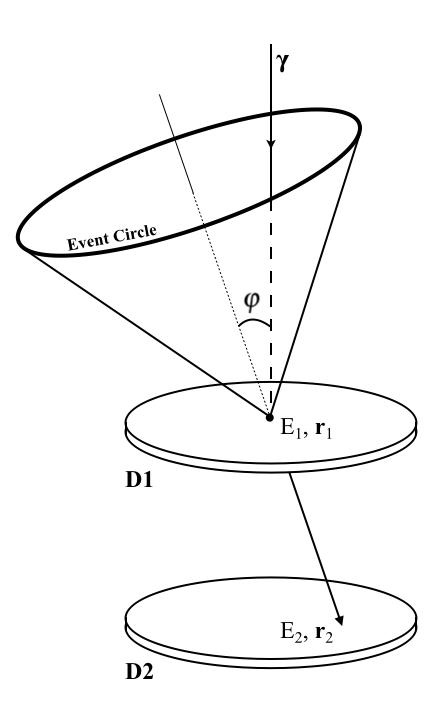}
    \subcaption{}\label{fig:ClassicCompton}
    \end{minipage}
    \hfill
    \begin{minipage}{0.5\linewidth}
    \centering
    \includegraphics[height=8cm]{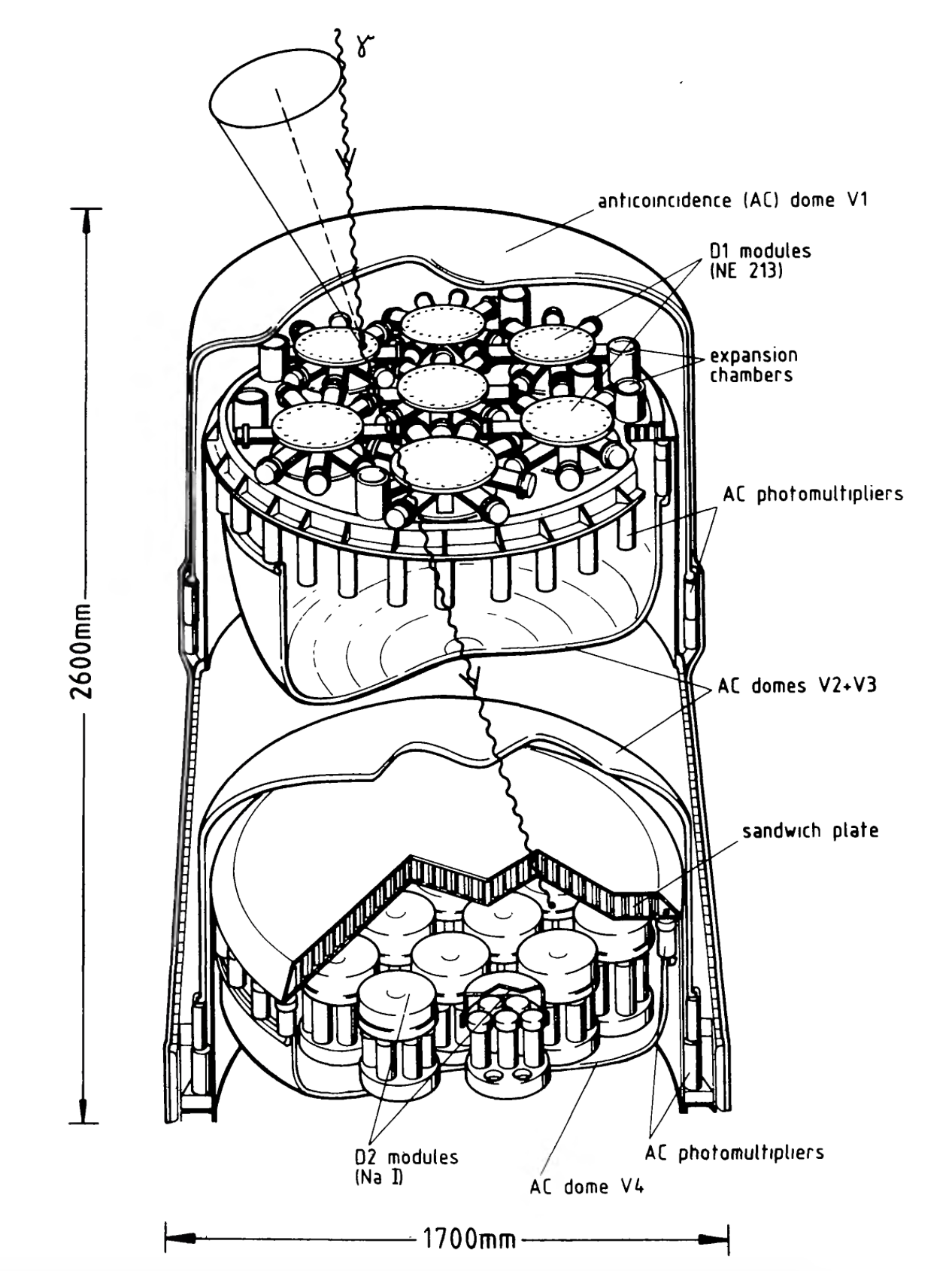}
    \subcaption{}\label{fig:COMPTELSch}
    \end{minipage}
    \caption{\textbf{(a)} A classic Compton telescope uses a scattering plane (D1) and an absorption plane (D2) to measure the position and energy of two interactions in a Compton event. From these measurements, the original photon direction can be constrained to a circle on the sky called the ``event circle.'' Adapted from \citet{Schonfelder2013}.
    \textbf{(b)} The first Compton telescope flown was COMPTEL onboard CGRO. COMPTEL used the classic Compton telescope configuration with two planes to map the MeV sky in the 1990's. Figure from \citet{schonfelder1993}.}
    \label{fig:COMPTEL}
\end{figure}

The first Compton telescopes for astrophysics were developed in the 1970's \cite{HERZO1975583, SCHONFELDER1973385} and were described as ``double scatter'' telescopes.
These classic Compton telescopes consisted of two detectors, referred to as D1 and D2, to measure a Compton scattering interaction and subsequent photoabsorption in two separated planes, as shown in Figure \ref{fig:ClassicCompton}. The trajectory of the scattered photon is known by recording the energy and location  ($E_1,\ 
\vec{r}_1, \ E_2,\ \vec{r}_2$) of the two interaction sites. 
Equation \ref{eq:twosite} then affords an estimate for the scattering angle of the primary photon.

The first successful Compton telescope operating in space was COMPTEL~\cite{schonfelder1993} on NASA's Compton Gamma Ray Observatory ({\it CGRO}, 1991-2000). 
COMPTEL consisted of two detector arrays: D1 contained a low-Z liquid scintillator  (NE 213A) to enhance the Compton scattering cross-section, and D2 contains a high-Z NaI(Tl) scintillator to enhance photoabsorption (see Figure~\ref{fig:mass_attenuation_coefficients}). COMPTEL was a very large instrument, with D1 and D2 being $\sim$1.5~m in diameter with a separation of 1.5~m between the two planes, and together weighing almost 600~kg. The instrument is shown in Figure~\ref{fig:COMPTELSch}.

COMPTEL opened the MeV range with 9 years of groundbreaking observations and provided the first all-sky survey in the  0.75 to 30 MeV energy range. The COMPTEL source catalog~\cite{Schonfelder2000} includes 63 sources, 32 of which are steady state, from spin-down pulsars and stellar black-hole candidates to supernova remnants and active galactic nuclei.
Potentially the most notable science achievement from COMPTEL observations was the all-sky map of the 1.8 MeV line from $^{26}$Al decay~\cite{1993A&AS...97..181D, 2001ESASP.459...55P}, which showed structure and hot spots consistent with star forming regions. While significant in its contributions, COMPTEL's achieved sensitivity was still modest~\cite{schonfelder2003}.

While the instrument was massive, COMPTEL's efficiency was low. The effective area, which is a measure of the equivalent area of an ideal detector, ranged from only 10 to 50 cm$^2$ depending on energy and event selections, i.e. the efficiency of detecting incident photons was less than 1\%. 
COMPTEL's angular resolution of a few degrees was limited at low energies by the energy resolution of the scintillators, and at higher energies by the position resolution of D1 and D2 (see Section~\ref{sec:psf_errors}). COMPTEL's field of view (FOV; $\sim$1~sr) was limited by the lower-energy thresholds of the two detector planes. 

COMPTEL employed two essential background reduction features. 
Due to the large separation between D1 and D2, a time-of-flight measurement was possible to determine the direction of the scattered photons. 
A selection of downward scattered photons was performed with a time-delay window of 4.5~ns. 
Secondly, the D1 liquid scintillator allowed for a discrimination of photon and neutron interactions based on differing pulse shapes. 
This further reduced the background observed in flight. 
These powerful background-rejection techniques allowed COMPTEL to achieve reasonable sensitivity despite its low effective area, and ultimately led to COMPTEL's unprecedented mapping of the MeV sky.

\subsection{Modern Compton Telescopes}
\label{sec:modern_telescopes}

COMPTEL's double-scattering concept with large separated planes was a limiting factor for the FOV and measurable Compton scatter angles, and thus led to the low efficiency of the instrument.
Additionally, COMPTEL was only able to measure two-site events, i.e. a single Compton scatter in D1 and a photoabsorption in D2. However, depending on the atomic number of the detector material, a photon can scatter multiple times before finally stopping in a photoabsorption event. 
Ideally, one would want to track the photon trajectory by measuring all secondary and tertiary (and beyond) interactions after the initial Compton scatter.
While COMPTEL effectively measured interactions in two planes, most modern Compton telescopes are based around the design philosophy of building a 3D position-sensitive detector volume that acts both as the scatterer and absorber.
With modern advancements in detector technology achieving (sub)millimetre spatial resolution, the efficiency and performance of Compton telescopes have significantly improved in the past few decades since COMPTEL's development.

\begin{figure}[tb]
    \centering
    \begin{minipage}[b]{0.48\textwidth}
       \centering
       \includegraphics[width=0.95\textwidth]{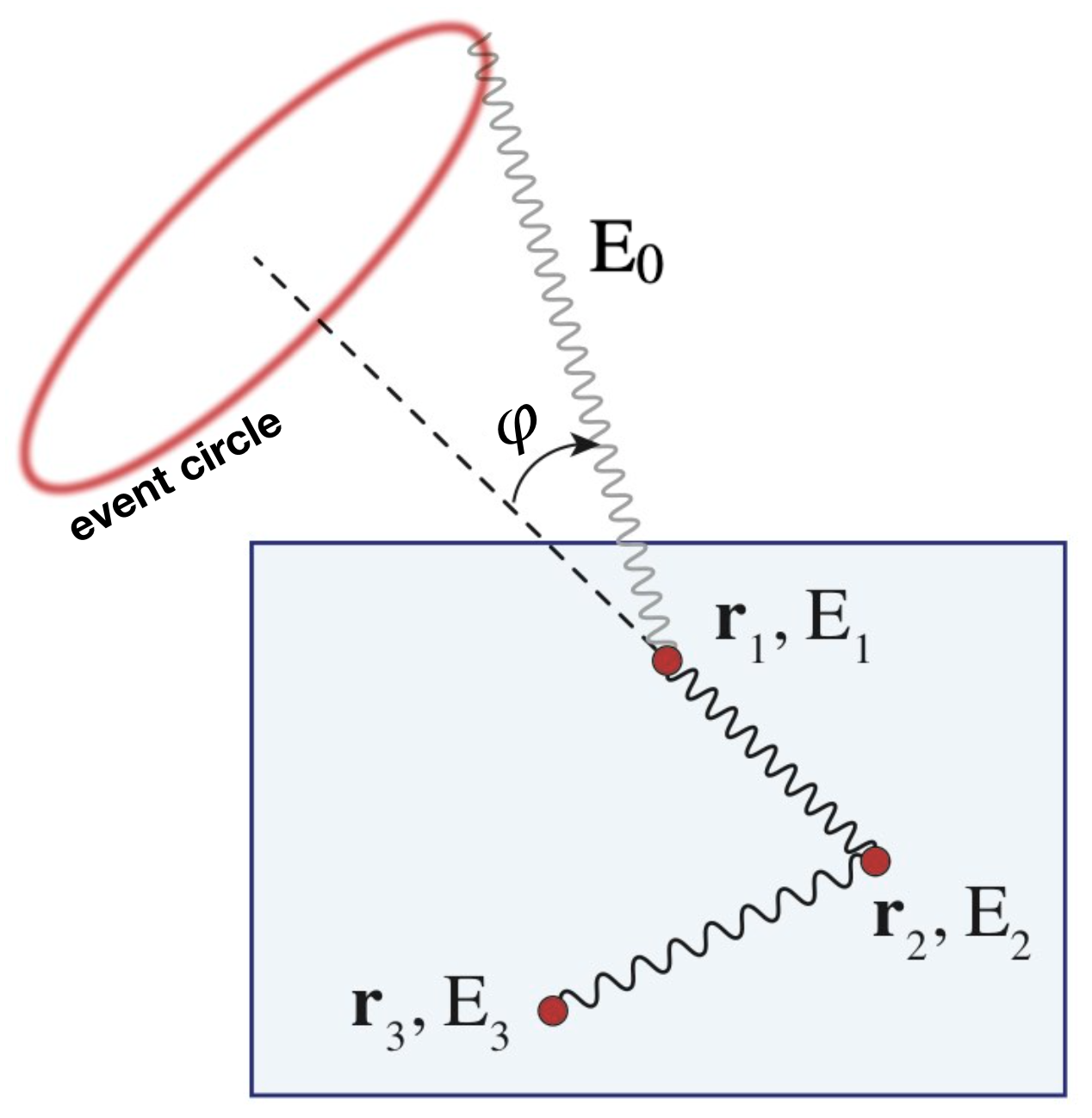}
    \end{minipage}
    \hfill
    \begin{minipage}[b!]{0.5\textwidth}
        \strut\vspace*{-3.5cm}\newline\captionof{figure}{Most modern Compton telescopes use high position-resolution detectors with good energy resolution to measure the position and energy of multiple interactions, with no geometrical limit on the Compton scatter angle. As in Figure~\ref{fig:compton_basics}, the direction of the photon is determined by the Compton scattering angle of the first interaction. Modified from \citet{kieransthesis}.}
        \label{fig:modern_compton3}
    \end{minipage}
\end{figure}

Combining the scattering and absorbing detector capabilities without large separation allows for the detection of multiple Compton scatters with no geometric limitation on the scattering angle, as long as the detector volume contains the interactions. 
Both of these features increase the efficiency of the telescope and improve upon the polarization capabilities (discussed in Section~\ref{sec:polarimeter_basics} and \ref{sec:polarization_caps}) compared to the classic Compton telescope design. An example of this is shown in Figure~\ref{fig:modern_compton3}, where a photon Compton scatters twice before fully depositing its energy with a photoabsorption event. All three interactions are measured within a large 3D position-sensitive detector volume, indicated by the blue area. 

For a three-site event, as in Figure~\ref{fig:modern_compton3}, the Compton scattering angle of the first interaction can be determined in the same way as for two interactions; however, the total absorbed energy is now $E_0 = E_1 + E_2 + E_3$, and the energy deposited after the second interaction must be accounted for in $E_{scat}$ in Equation~\ref{eq:Compton_scatter_angle}:
\begin{equation}
    \cos \varphi = 1 - \frac{m_ec^2}{E_2 + E_3} + \frac{m_ec^2}{E_1 + E_2 + E_3}.
    \label{eq:threesite}
\end{equation}
This formula can be scaled to a large number of interactions as long as the initial energy includes all energy deposits from the fully-absorbed photon $E_0 = E_1 + E_2 + ... + E_n$, and the scattered photon energy is equal to the total deposited energy excluding $E_1$: $E_{scat} = E_2 + ... + E_n$. 
The additional event pattern recognition achievable with these multi-scatter Compton telescopes enables sophisticated background rejection techniques, as discussed further in Section~\ref{sec:eventrecon}.

After COMPTEL, there were multiple competing technologies that were trying to achieve the next-generation MeV telescope design with position-sensitive detectors~\cite{kurfess2000}. 
One such instrument developed in the 1990's was the Liquid Xenon Gamma-Ray Imaging Telescope (LXeGRIT)~\cite{2000AIPC..510..799A,2000SPIE.4140..333A}. 
LXeGRIT used a time projection chamber (TPC) filled with liquid xenon to achieve the event-by-event Compton imaging within one large detector volume. 
Gamma rays interact with xenon via ionization and a release of ultraviolet scintillation photons. 
When a high voltage is applied, the charges drift to the anode and cathode, and allow for the total energy deposit to be recorded. 
The X-Y interaction location is measured by orthogonal sensing wires embedded in the liquid xenon, and the Z coordinate is determined via the drift time. 
The homogeneous detector volume allowed for a much more efficient design than the classic double-scattering Compton telescope.
Similar techniques are used in other modern Compton telescopes SMILE~\cite{Tanimori_2015, Takada_2022} and GRAMS~\cite{Aramaki2020}. 

The large instrumented detector volume realized by liquid or gaseous TPCs have their advantages, but are generally limited in stopping power and energy resolution.
Other instrument designs in the 1990's and early 2000's started employing semiconductor detectors, which have the advantage of superior energy resolution. 
However, it is not possible to fabricate or instrument arbitrarily large active volumes of semiconductors; the detection volume must be segmented into smaller detectors each with energy and internal position sensitivity.
This was achieved in the early days through the use of the ubiquitous semiconductor silicon~\cite{Kamae_etal_1987}. 
The telescope volume was broken up into thin planes of silicon with 2-D resolution, where the Z-dimension of the interaction was defined as the location of the hit detector plane with no internal depth resolution. 
A multilayered design was developed by both the Medium Energy Gamma-ray Astronomy (MEGA) telescope~\cite{2002NewAR..46..611B, Kanbach_etal_2005} and the Tracking and Imaging Gamma-Ray Telescope (TIGRE)~\cite{10.1117/12.187266} and proved to be a viable detection technique for Compton telescopes. The added advantage of the thin silicon detector planes was the potential to measure the direction of the Compton-scattered electron in more than one of the silicon planes, which can then be used to further constrain the kinematics of the scatter, as will be discussed further in Section~\ref{sec:electron_tracking}.

Other methods of building a 3D position-sensitive detector volume with good energy resolution have been explored since COMPTEL showed the power of Compton imaging.
Scintillators, where the internal interaction position within a crystal can be inferred from the light intensity measured at the ends of long scintillating bars, have been used when energy resolution is not a priority, but a large effective area or stopping power is a driver. 
Semiconductor detectors, with their high spectral resolution, are ideal for Compton telescopes, since the better energy resolution provides a more accurate measure of the Compton scattering angle $\varphi$. 
Other than the thin multi-layer silicon detector approach, both germanium~\cite{amman2000, coburn2003} and CZT~\cite{XuPhD,DU2001203, 7407479} with a large form factor have been demonstrated. 
High-purity double-sided strip germanium detectors will be used in COSI~\cite{Tomsick2021}. 
With an individual crystal size of $8 \times 8 \times 1.5$~cm$^3$, the total telescope consists of 16 tightly packed germanium crystals with independent readout.
The COSI detectors achieve an internal position resolution of $\lesssim 1.5$~mm$^{3}$: the X-Y position is determined by the charge collection on orthogonal strips, and the Z dimension is determined from the charge collection timing~\cite{amman2000}.
Other semiconductor-based Compton telescopes are discussed in Section~\ref{sec:instruments}. 

\begin{figure}[bt]
    \centering
    \includegraphics[width=0.95\columnwidth]{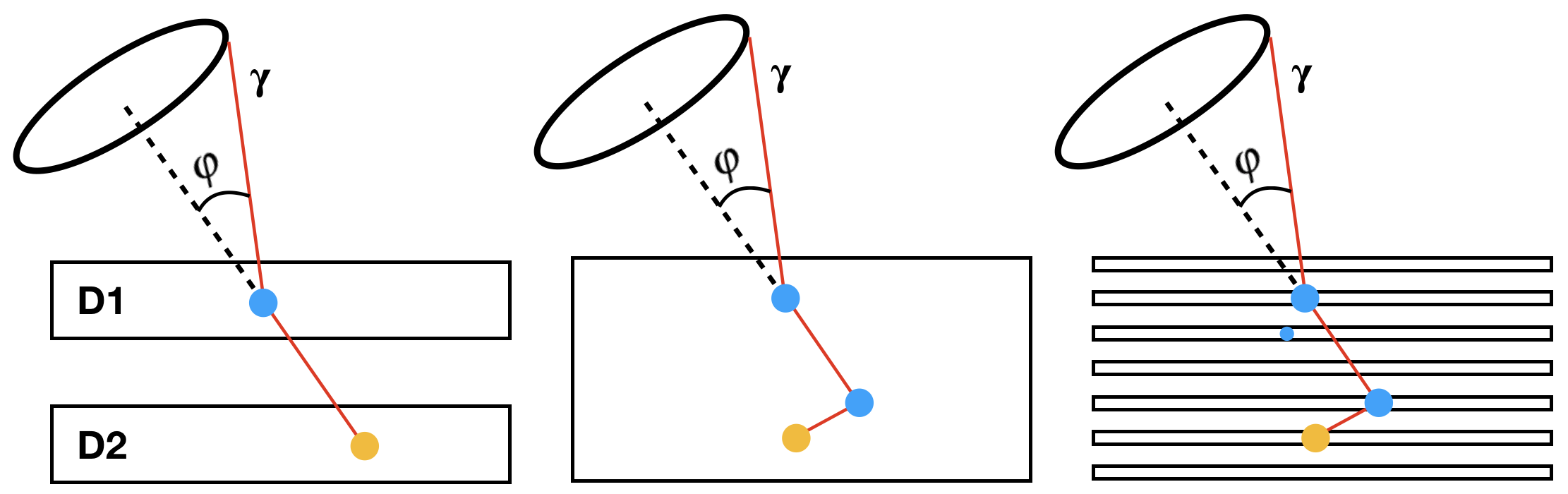}
    \caption{There are three common configurations for Compton telescopes. \textbf{(a)} The classic Compton telescope uses two detector planes as a scatterer and absorber to measure the interaction of double-scattering events. 
    \textbf{(b)}  The compact Compton telescope uses a single 3-D position-sensitive volume to measure multiple Compton scatter interactions. 
    \textbf{(c)} The multi-layer design uses many thin layers of detectors, with the added advantage of electron tracking. Modified from \citet{YabuPhD}.}
    \label{fig:modern_compton}
\end{figure}

As long as a precise measurement of the energy and 3D location of each interaction is achieved, the actual detector implementation for a Compton telescope can be quite diverse.
A summary of three common configurations for modern Compton telescopes are shown Figure~\ref{fig:modern_compton}.
The classic Compton telescope uses two detector planes, as demonstrated with COMPTEL. Alternatively, a large 3D position-sensitive volume can measure the position and energy of multiple Compton scatter interactions, increasing the detection efficiency, as demonstrated with LXeGRIT. 
The third configuration achieves more precise energy and/or position information from each scatter using a detection volume segmented into multiple smaller position-sensitive detectors.
One standard design uses thin detectors with 2D resolution, as demonstrated with MEGA.

Due to the increased efficiency and smaller resolution elements, most modern Compton telescope are compact in design.
One important point to note is that compact designs no longer have time-of-flight capabilities to determine the temporal order of interactions. 
While there are still efforts to modernize time-of-flight Compton capabilities~\cite{2018SPIE10699E..5XB}, most modern Compton telescopes require an extra step for event reconstruction; the relative order of interactions can be determined by the Klein-Nishina cross section and by redundant information in the energy deposits and relative positions of the interactions. 
The process for determining the most probable sequence of interactions is referred to as Compton sequencing or event reconstruction and is discussed further in Section~\ref{sec:eventrecon}.

\subsection{Electron Tracking}
\label{sec:electron_tracking}

In previous discussions, the direction of the incident gamma ray has been restricted to the Compton event circle 
where the uncertainty in the azimuthal direction is due to a lack of knowledge of the Compton-scattered electron's direction.
Now, if the direction of the recoil electron is known, momentum conservation can further restrict the direction of each incident gamma ray. 
With a more accurate determination of the origin of the gamma ray, background rejection can be improved~\cite{AKYUZ2004127}.

\begin{figure}[tb]
\begin{minipage}{0.5\textwidth}
    \centering
    \includegraphics[width=\textwidth]{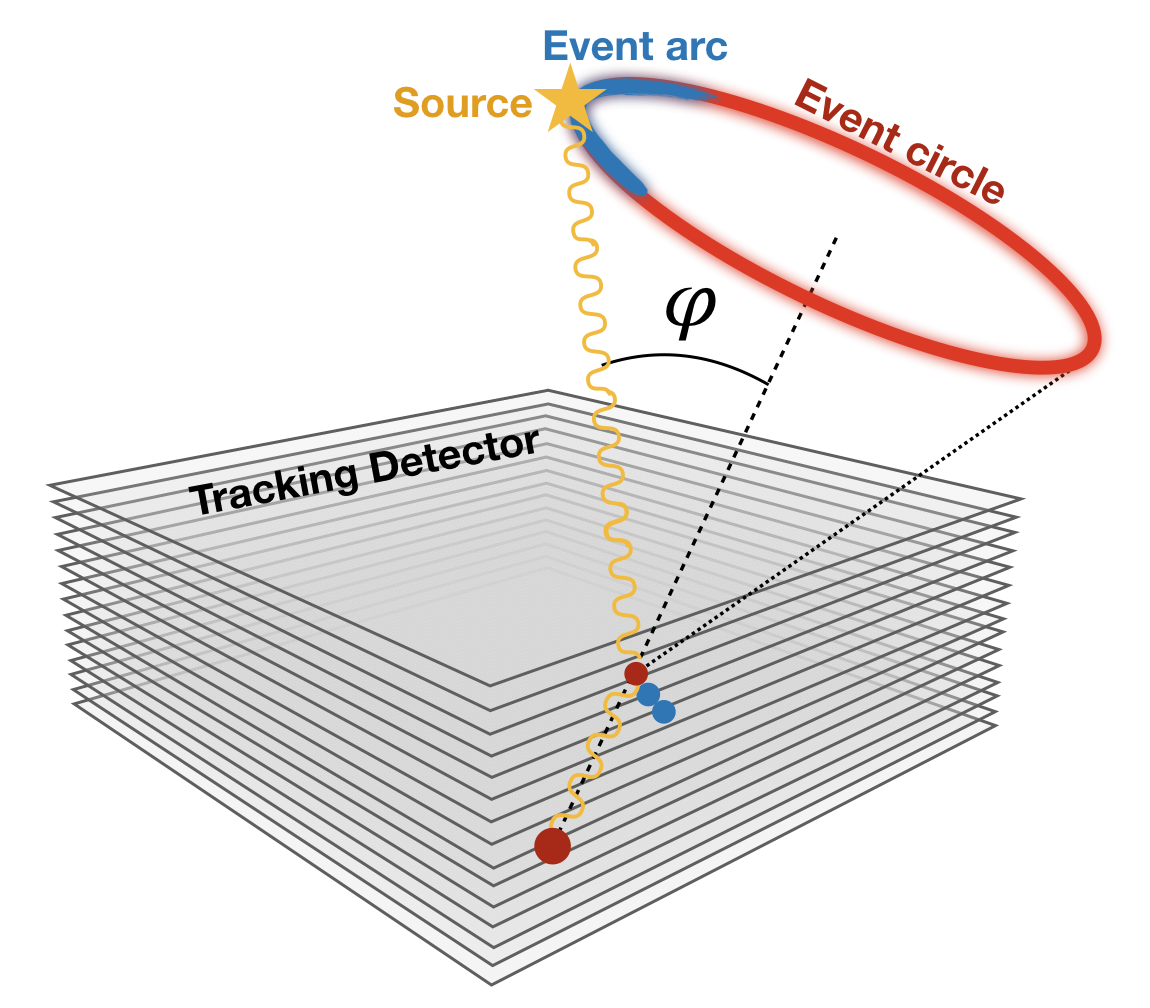}
    \end{minipage}
    \hfill
    \begin{minipage}{0.5\textwidth}
    \centering
    \captionof{figure}{By measuring the direction of the Compton-scattered electron of the first interaction, the incident photon direction can be constrained within a reduced arc segment of the original Compton event circle. The measured recoil electron direction is depicted with blue dots representing energy deposits in multiple detector layers, and the length of the event arc is dictated by the precision of the electron trajectory measurement. Modified from \citet{2021NIMPA101965795B}.}
    \label{fig:electrontrackingbasics}
    \end{minipage}
\end{figure}

\begin{figure}[tb]
    \centering
    \begin{subfigure}[b]{0.48\textwidth}
    \centering
    \includegraphics[width=0.95\textwidth]{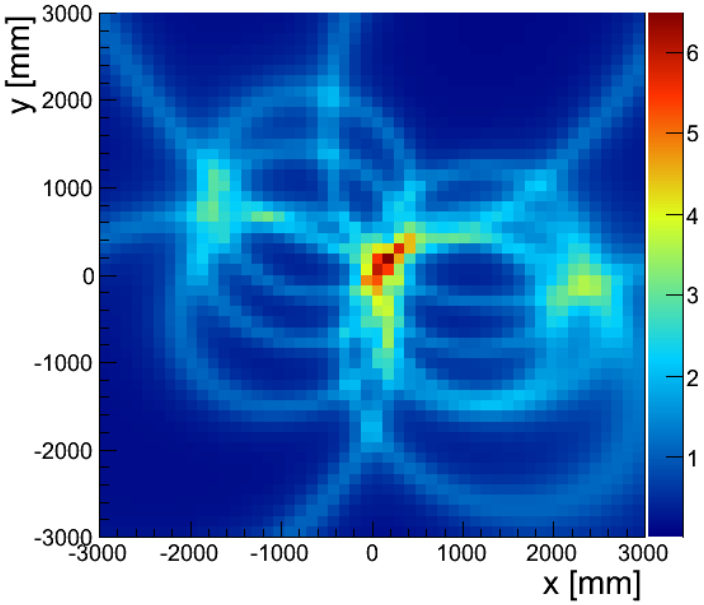}
    \caption{10 untracked Compton events.}
    \label{fig:electrontrack_before}
    \end{subfigure}
    \hfill
    \begin{subfigure}[b]{0.48\textwidth}
    \centering
    \includegraphics[width=0.95\textwidth]{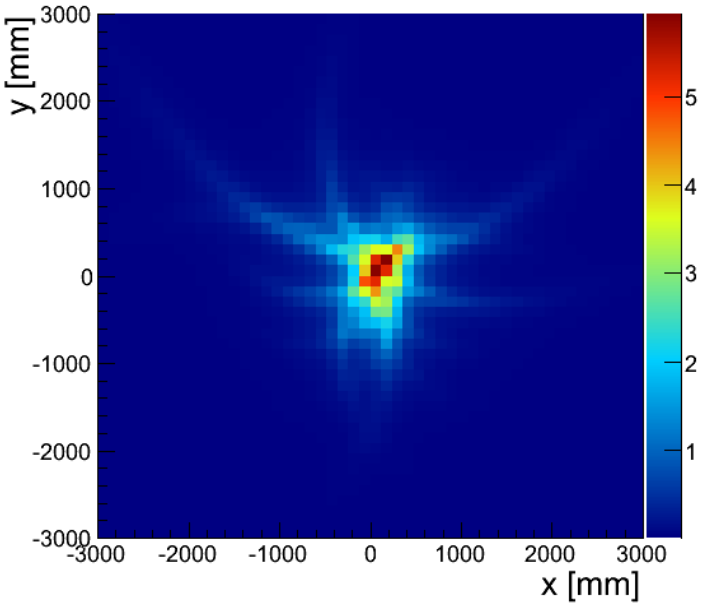}
    \caption{10 Compton events with electron tracking.}
    \label{fig:electrontrack_after}
    \end{subfigure}
    \caption{The back-projection of  event circles from untracked Compton events compared with tracked events highlights the benefit of the additional kinematic information. While the angular resolution does not change, the point source is more easily discernible, and background contamination can be significantly reduced. Therefore, electron tracking can enhance the sensitivity of a telescope. Figure from \citet{Mizumura_2014}.} 
    \label{fig:electron_tracking_images}
\end{figure}

The kinematics of the recoil electron from a Compton interaction can be calculated in an analogous way to the scattered photon as derived in Section~\ref{sec:physics}. 
However, since the direction of the Compton-scattered electron $\Theta$ is less precise than the direction of the Compton-scattered photon due to a short penetration depth and Moli\`{e}ere scattering~\cite{molieres}, it is preferable to define the additional kinematic constraint relative to the original photon event circle. 
With the recoil electron direction, the initial event circle can be reduced to a small segment, defined as the event arc. Figure~\ref{fig:electrontrackingbasics} shows this concept, where a multi-layer Compton telescope is used to measure energy deposited from the Compton-scattered electron through multiple detector layers. 
The length of the event arc is related to the uncertainty in the electron's recoil direction. 
Figure~\ref{fig:electron_tracking_images} shows the back-projected Compton event circles in image space with and without electron tracking; by using the recoil electron trajectory, the source position is easily discernible with fewer photons.
It is important to note that electron-tracking is a powerful background reduction technique and allows for a more clearly defined point spread function in image space, but does not in general improve the angular resolution of a telescope due to the large uncertainties in the recoil electron direction. 
The details on how electron tracking affects the point spread function of the instrument will be discussed in detail in Section~\ref{sec:performance_electrontracking}. 

The TIGRE and MEGA groups were the first to demonstrate electron tracking capabilities for Compton imaging; with  multiple layers of thin silicon detectors, they were able to measure the momentum direction of the Compton-scattered electrons based on the ionization tracks through multiple silicon layers. Furthermore, electron-tracking Compton imaging has also been demonstrated in large gaseous time-projection chambers~\cite{tanimori2004mev,KABUKI20071031,1596289}. 


Any telescope that is capable of detecting the track of the Compton-scattered electron also has the enhanced capability of being sensitive to  pair-production events.
As the gamma-ray energy increases and pair-production become dominant, the direction and energy of an incident gamma rays can be constrained by measuring the electron and positron trajectories to reconstruct the vertex and opening angle of the pair conversion. A combination electron-tracking Compton and pair telescope is an enticing technology that can span the majority of the MeV Gap with one telescope. This has been explored by a number of different telescope concepts, as discussed further in Section~\ref{sec:instruments}.

\subsection{Dedicated Polarimeter}
\label{sec:polarimeter_basics}

There also exists a class of Compton telescopes that is not optimized for imaging, but for polarimetery. 
As Equation~\ref{eq:klein-nishina-pol} shows, there is an asymmetry in the cross-section for photons that scatter parallel and perpendicular to the initial photon's electric field vector. 
This asymmetry, or the amplitude of the polarization response, is maximized when the $\sin^2\varphi$ term is maximized, or at polar Compton scattering angles $\varphi = 90^{\circ}$.
A polarized source will result in a sinusoidal distribution in the azimuthal scattering angle ($\eta$ in Figure~\ref{fig:polarized_angles}).
Therefore, instruments that measure the azimuthal scattering angle, especially at large Compton scatter angles, can ultimately detect the polarization of a gamma-ray source. 

\begin{figure}[t]
\begin{minipage}{0.35\textwidth}
    \centering
    \includegraphics[width=\textwidth]{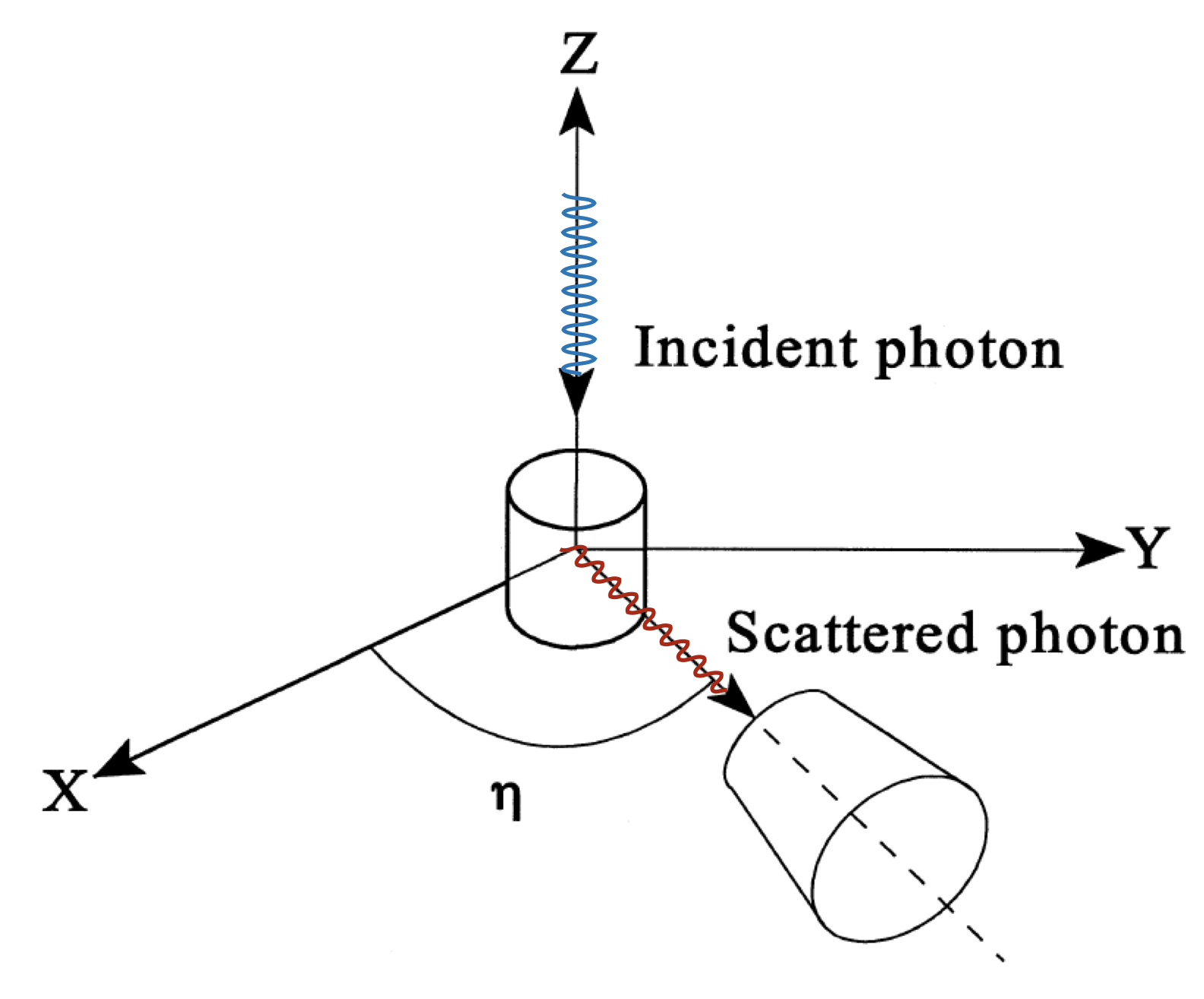}
    \subcaption{}\label{fig:lei_pol}
\end{minipage}
\hfill
\begin{minipage}{0.64\textwidth}
    \centering
    \includegraphics[width=\textwidth]{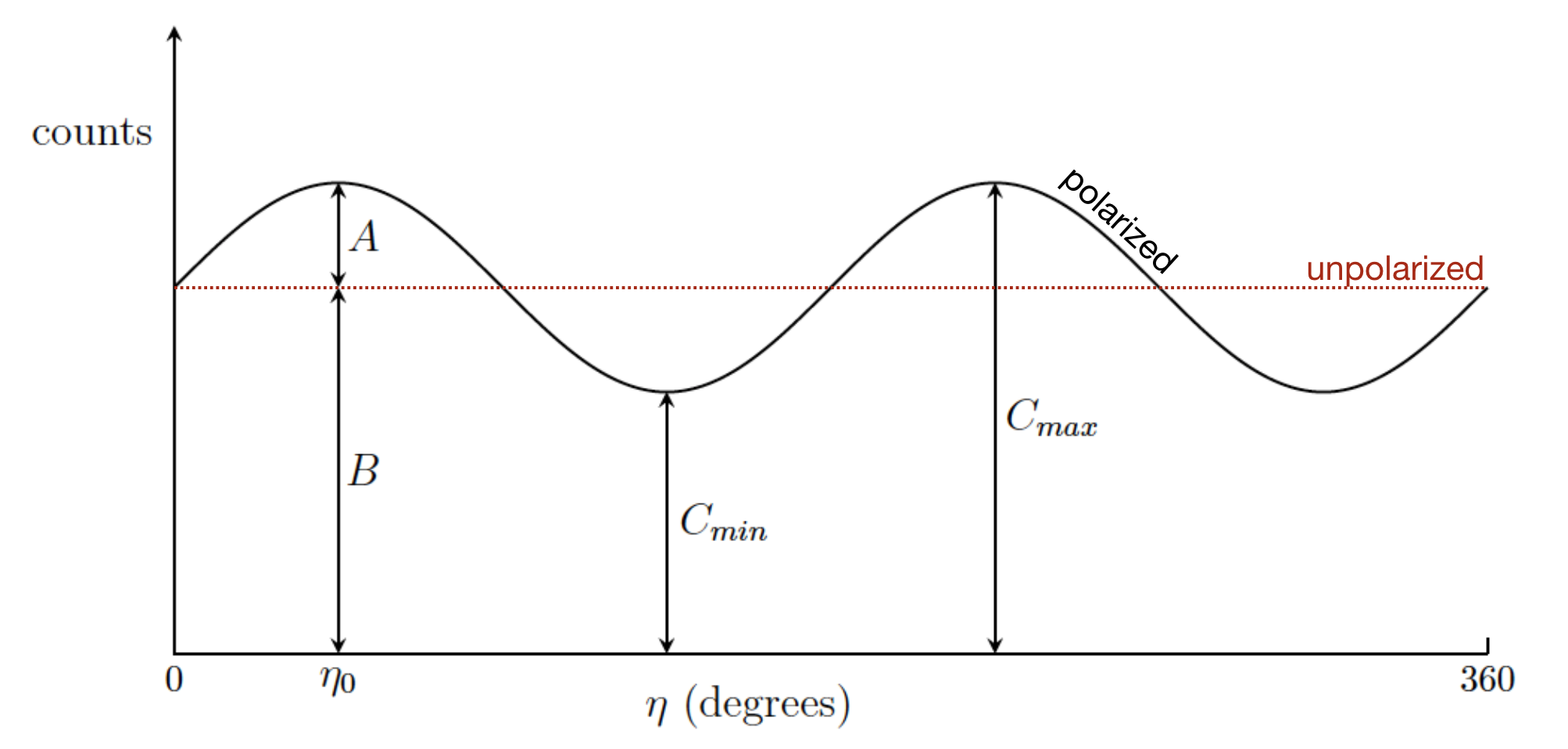}
    \subcaption{}\label{fig:ASAD}
\end{minipage}
\caption{\textbf{(a)} Photons will predominately scatter at 90$^{\circ}$ relative to their initial electric field vector, according to Equation~\ref{eq:klein-nishina-pol}. Therefore, a gamma-ray polarimeter can determine the degree of polarization of a source by measuring the azimuthal scattering direction $\eta$ of photons. Modified from \citet{lei1997}. 
\textbf{(b)} The azimuthal scattering direction for a polarized source gives a sinusoidal response, referred to as the ASAD (azimuthal scattering angle distribution), which is maximized at 90$^{\circ}$ relative to the direction of the initial electric field vector. Modified from \citet{2011PhDT.......139B}.}
\label{fig:pol_basics}
\end{figure}

The simplest version of a Compton polarimeter is shown in Figure~\ref{fig:lei_pol}, where the initial photon is scattered in a central detector element and absorbed in a detector placed at $90^{\circ}$. If this secondary detector is rotated to sample the counts in the azimuthal (Figure~\ref{fig:ASAD}), the direction and degree of polarization can be measured, as described in detail in Section~\ref{sec:polarization_caps}.

Most dedicated polarimeters are compact in design to be sensitive to 90$^{\circ}$ Compton scatter angles. As the azimuthal scatter distribution is sinusoidal, the distribution can be coarsely sampled without a significant loss of sensitivity. In addition, the measured energy resolution for dedicated polarimeters is not as crucial as for imaging Compton telescopes as the energy information is not used in the reconstruction (but can provide spectral information for the incoming photons). Therefore, most dedicated polarimeters take advantage of scintillator detectors, which are simpler than segmented semiconductor detectors used in most compact and multi-layer Compton telescopes.

\begin{figure}[tb]
    \centering
    \begin{minipage}{0.5\textwidth}
    \includegraphics[width = 0.9\textwidth]{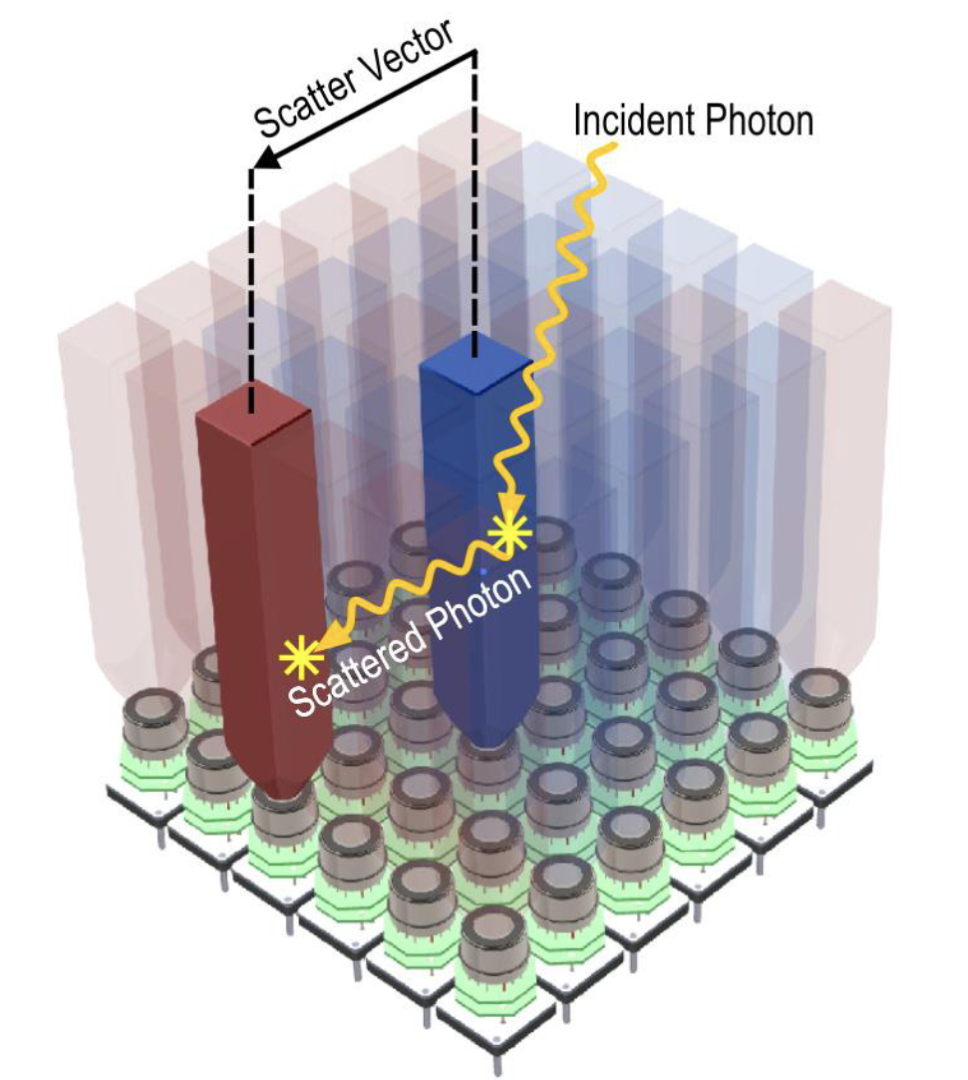}
    \end{minipage}
    \hfill
    \begin{minipage}{0.48\textwidth}
    \captionof{figure}{The LargE Area busrt Polarimeter (LEAP) telescope is a dedicated Compton polarimeter that measures the azimuthal scattering direction with scintillating bars. LEAP is compact by design and even with fairly coarse position knowledge of each interaction, the scatter vector can be used to precisely measure the polarization of incoming photons.
    Figure from \citet{leap2021}.}
    \label{fig:leap}
    \end{minipage}
\end{figure}

An example of a dedicated polarimeter telescope is the LargE Area burst Polarimeter (LEAP) shown in Figure~\ref{fig:leap}. LEAP is a NASA Mission of Opportunity concept considered in the 2019 selection round, and is similar to other Compton polarimeter designs.
LEAP uses two different types of scintillators to measure the azimuthal scattering direction: a low-Z plastic scintillator is optimized for scattering and a high-Z scintillator optimized for absorption.
The use of scintillators allows for a simple, high-efficiency, and large field-of-view design. 
However, without the high-resolution 3D position and energy information for each interaction, Compton imaging is not possible in dedicated polarimeter. 
These instruments are therefore optimized for transient detections when a large signal-to-background is available.

\section{Event Reconstruction}
\label{sec:eventrecon}

Event reconstruction for Compton telescopes is the process of translating the measured positions and energies of detected photon interactions to the initial gamma-ray direction defined by the event circle (or the event arc for electron-tracking Compton telescopes). 
For classic Compton telescopes with a limit of two interactions and time-of-flight information to determine the temporal sequence of hits, event reconstruction is a fairly straightforward process. 
With the measured energy and Equation~\ref{eq:Compton_scatter_angle}, the Compton scatter angle of the first interaction can be determined, and the axis of the event circle can be defined from the measured positions (see Figure~\ref{fig:ClassicCompton}). 

As modern Compton telescopes become more compact and sophisticated, so too have the event reconstruction techniques. 
The timing resolution is not sufficient to determine the temporal sequence of interactions in compact Compton telescopes. 
Therefore, in order to determine the Compton event circle, one must first determine which energy deposits correspond to the first and second interactions for each gamma ray. 
This is done by testing all $N!$ combinations of the interactions for an event with $N$ energy deposits and determining the most probable sequence.

The process of event reconstruction for modern telescopes is computationally intensive; however, there are also significant benefits. 
The first is that with the added number of allowable interactions compared to a classic double-scatter Compton telescope, the efficiency of the instrument increases significantly. 
The average number of interactions depends on the photon energy and the detector material, but higher energy photons can often scatter $>$5 times in a detector before being photoabsorbed; sophisticated reconstruction techniques are needed to properly trace a high-energy gamma-ray path through a large detector volume. 
Second, and more importantly, the main advantage of event reconstruction is background rejection~\cite{APRILE1993216, boggs2000}. 
Events that have any missing energy, from incomplete absorption or interactions in passive material, or events associated with $\beta^+$ decays within the instrument are effectively background events, but with a filtering step in the reconstruction, these unusable events are rejected and the signal-to-background ratio increases.

The first step of reconstruction is event identification, outlined in Section~\ref{sec:trackidentification}, where it must be determined whether or not the measured energy deposits are consistent with a Compton scattering event (or a charged particle track or pair event), and, when applicable, whether there are measured electron tracks.
Then the Compton sequence determination can be achieved by taking advantage of redundant information in the kinematics and geometrical information measured for each event. 
The standard approach is referred to as Compton Kinematic Discrimination~\cite{boggs2000}, or mean squared difference method. This straightforward and fast sequence reconstruction technique is employed by various research groups and will be detailed in Section~\ref{sec:sequencing}. 
It is difficult to unambiguously determine the correct order for events which only have two interactions and no discernible temporal separation, and thus two-site event reconstruction will be considered separately in Section~\ref{sec:twositerecon}.

\subsection{Event Identification and Track Recognition}
\label{sec:trackidentification}

Before determining the correct sequence of interactions for an event, one must characterize the measured energy deposits and determine if they are consistent with a Compton-scattering event, a pair event, a charged-particle track, or any other potential interaction. 
Telescopes that have tracking capabilities have more possible event types to identify and will be the basis of the presented approach, which follows~\citet{zoglauerthesis}. Detectors without electron tracking  may use a simplified version of the steps listed below.

Photons which interact with multiple Compton scatters within a detector volume leave separated, isolated energy deposits as shown in Figure~\ref{fig:modern_compton3}, i.e., there is no easily identifiable straight path or pattern. Therefore, it is easiest to first confirm if the measured interactions are consistent with event types that do have easily recognizable patterns (i.e. pair events or charged-particle tracks); if no clear pattern is found, then compatibility with Compton interactions can be checked.
The general steps for event identification for a tracking Compton telescope are as follows:
\begin{enumerate}
    \item Search for pair event vertex
    \item Search for high-energy charged particle straight tracks
    \item Search for recoil electron tracks from Compton interactions
    \item Search for Compton interaction sequence
\end{enumerate}

Pair events are recognizable by the characteristic inverted ``V'' from the electron and positron (i.e., the pair) ionization tracks in the detector volume.
In a multi-layer Compton telescope, as described in Section~\ref{sec:modern_telescopes}, the interactions from the electron and positron are measured after conversion in subsequent layers with increasing separation. 
If a vertex is found with a pattern recognition algorithm, the event can be flagged as a pair event and the initial photon direction can be determined by the trajectory of each particle, weighted by their relative energies. Pair reconstruction is described in detail in \cite{doi:10.1142/9789811203817_0005} and has been matured for the Fermi Large Area Telescope~\cite{Atwood_2009}. 

The second easily identifiable event type is the straight tracks left by high-energy charged particles, such as cosmic rays and muons. With linear ionization tracks, the reconstruction of high-energy charged particles is straightforward and can be accomplished by fitting a straight line to the measured energy deposits.

\subsubsection{Recoil Electron Track Reconstruction}

The third step in event identification is to search for tracks left by Compton recoil electrons. 
In Compton scattering, when a photon interacts with an electron, it will transfer momentum in the scatter, causing the electron to be ejected from its atom and recoil. 
The recoil electron will then leave an ionization track in the detector material, and the length of the track is based on the initial electron momentum and the density of the material. 
The track lengths in high-Z materials, such as germanium, CdTe, and CZT, are too short to measure the direction of motion and the full electron energy is deposited within one position resolution element of the detector. For lower Z material, such as silicon, or the low densities found in gaseous time projection chambers, the track length can be longer than a few millimeters and span multiple resolution elements. For example, in multi-layer Compton telescopes consisting of thin 2D sensitive silicon detectors, the recoil electron can deposit energy in 3-4 layers with $\sim$1~cm spacing.

\begin{figure}[tb]
    \centering
    \begin{minipage}{0.5\textwidth}
    \includegraphics[width = \textwidth]{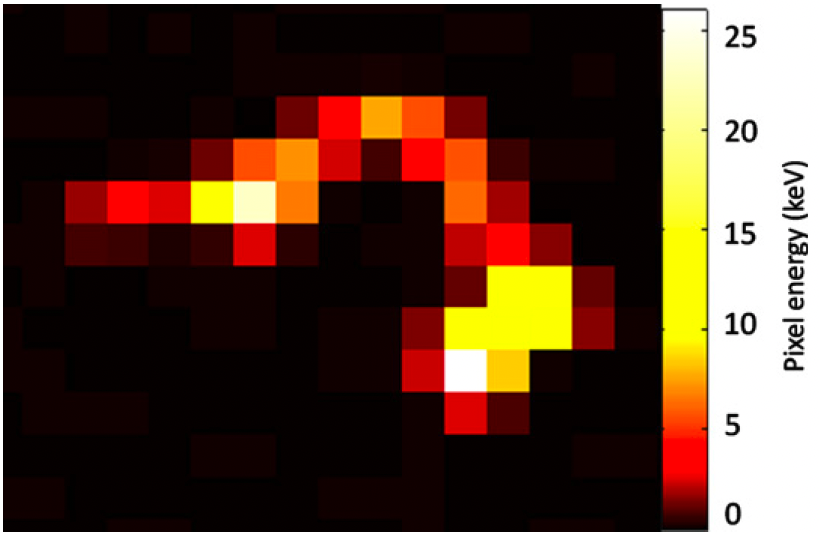}
    \end{minipage}
    \hfill
    \begin{minipage}{0.49\textwidth}
    \captionof{figure}{The measured recoil electron track from~\cite{2011NIMPA.652..595P} shows the curving nature of the path and the complication of track reconstruction (see Section~\ref{sec:performance_electrontracking}). 
    This high spatial resolution measurement, with each pixel being $10.5\times10.5$~$\mu$m$^2$, is the first demonstration of a reconstructed recoil electron track within a 650~$\mu$m-thick silicon charge-coupled device (CCD). The deposited energy from the recoil electron is 227~keV. Based on the Bragg peak, the larger energy deposit on the bottom right is likely the end of the track.}
    \label{fig:electron_track_example}
    \end{minipage}
\end{figure}

There are two main purposes for electron track recognition in event reconstruction: the initial Compton-scattering interaction location is needed for Compton sequencing (Section~\ref{sec:sequencing}), and the direction of motion of the electron can be used to kinematically constrain the incoming photon direction (Section~\ref{sec:electron_tracking} and \ref{sec:performance_electrontracking}). Unfortunately, the initial interaction location and electron trajectory are not always easy to discern from the measured energy deposit as ionization tracks left by recoil electrons are rarely straight and often short, as seen in Figure~\ref{fig:electron_track_example}. The paths of recoil electrons can make U-shaped tracks, and with poor spatial resolution or coarse sampling in a multi-layer Compton telescope, the discrete sampling of the energy deposits complicates the reconstruction. 
Higher energy electrons ($\gtrsim$1~MeV) will have long and initially straight tracks, but at lower energies, the track length decreases and the non-linearity increases.

There have been multiple approaches to the electron track reconstruction. The simplest version relies on the assumption that as the electron loses energy, it will tend to deposit more energy, as defined by the Bethe-Bloch equation and the Bragg peak~\cite{pdg}. Therefore, the track end point with the smaller measured energy is more likely to be the beginning of the electron's path.
The electron track reconstruction has been performed using different algorithms: a figure-of-merit approach \cite{zoglauerthesis}, Bayesian approach~\cite{zoglauerthesis}, momentum analysis~\cite{BLACK2007755}, and graph theory~\cite{yoneda2018development}. Different approaches are used based on the relative spatial resolution of the track.

Measuring the electron-scatter direction allows one to further constrain the kinematics of the Compton scattering interaction, reducing the photon direction from a back-projected circle to an arc. The precision of the electron recoil direction determines the length of the event arc.  Since the precision of the electron's direction is often much worse than that achieved through the measurements of the scattered photons, this additional kinematic information does not in general improve the angular resolution but can serve as a method to reduce the observational background. 

As described in Section~\ref{sec:electron_tracking} and shown in Section~\ref{sec:physics}, the kinematics of the recoil electron can be treated in the same way as the scattered photon. \citet{doi:10.1063/1.4898087} demonstrated that the source location can be determined entirely by the recoil electron without needing to measure the Compton-scattered photon, increasing the measurement efficiency. However, the angular resolution and sensitivity of this detection technique is limited.

After pair and high-energy charged particle events have been identified, and Compton recoil electron tracks have been reconstructed to determine the initial interaction position, the measured energy deposits can be tested for compatibility with Compton scattering. The processes of determining the most probable sequence of interactions for a single photon is called Compton sequencing. 

\subsection{Compton Sequencing}
\label{sec:sequencing}

Compton event reconstruction starts with a collection of hits within the detector volume, each with energy and position information. 
With no prior knowledge or assumptions about the first interaction location, all $N$!~possible combinations of the hit order, where $N$ is the number of interactions, must be analyzed. 
The first algorithm developed for kinematic event reconstruction was introduced by \citet{APRILE1993216}, and later formalized by \citet{boggs2000}. 
It is referred to as Compton Kinematic Discrimination, the chi-squared approach, or mean squared difference method, and as it remains one of the more popular approaches for events with 3$+$ interactions, it will be described here.

\begin{figure}[tb]
\centering
\begin{minipage}{0.49\textwidth}
\includegraphics[width=0.8\textwidth]{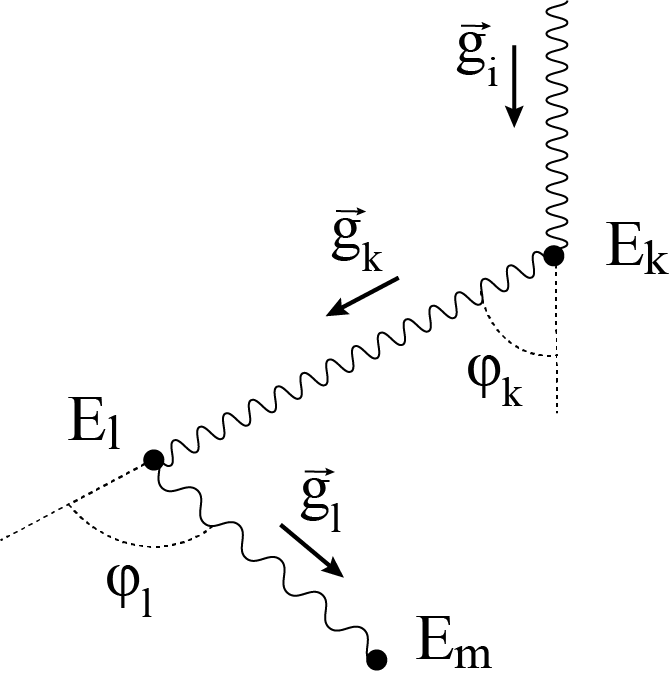}
\end{minipage}
\begin{minipage}{0.5\textwidth}
\caption{In Compton Kinematic Discrimination, the Compton scatter angle $\varphi_l$ of a central interaction $l$ is determined in two ways: kinematically with the energies $E$ and geometrically with the scatter directions $\vec{g}$. Minimizing the difference between these two measures for each Compton scatter leads to the most probable sequence of interactions. Modified from \citet{zoglauerthesis}.}
\label{fig:comptonkinematicreconstruction}
\end{minipage}
\end{figure}

For gamma-ray events which have at least 3 interactions (two Compton scatters and one photoabsorption; see Figure~\ref{fig:comptonkinematicreconstruction}), the Compton scatter angle of the central interaction(s), denoted by $l$, can be determined in two ways. The first is kinematically with the Compton equation
\begin{equation}
\cos \varphi^{kin}_l = 1 - \frac{m_ec^2}{E_{l_{+}}} + \frac{m_ec^2}{E_{l} + E_{l_{+}}},
\label{eq:phikin}
\end{equation}
where $E_{l_+}$ is the total energy of all interactions following $l$; $E_{l}$ is not included in $E_{l_+}$. Second, the Compton scatter angle can be determined geometrically considering the angles
\begin{equation}
\cos \varphi^{geo}_l = \frac{\vec{g_k} \cdot \vec{g_l}}{| \vec{g_{k}} | | \vec{g_l} | },
\label{eq:phigeo}
\end{equation}
where $\vec{g_k}$ is the incoming gamma-ray direction and $\vec{g_l}$ is the outgoing direction for interaction $l$. 

The two measures of $\cos \varphi_l$, from Equation~\ref{eq:phikin} and \ref{eq:phigeo}, should be identical for the correct order of interactions in an ideal instrument. 
In the classic Compton Kinematic Discrimination approach, a $\chi^{2}$ quality factor $Q$ is assigned to each permutation of $N$ interactions~\citep{zoglauerthesis}
\begin{equation}
\label{eq:recon_qfactor}
Q = \sum^{N-1}_{i=2} \frac{ ( \cos \varphi^{kin}_i  - \cos \varphi^{geo}_i )^2 }{(d \cos \varphi^{kin}_i)^2 +( d \cos \varphi^{geo}_i )^2},
\end{equation}
where $d \cos \varphi_i$ are the uncertainties determined through error propagation and $i$ is the interaction index. 
The sequence with the lowest quality factor is the best estimate of the correct kinematic ordering of the event. 

Another common method for Compton sequencing uses a deterministic approach based on the Klein-Nishina formula~\cite{10.1117/12.563905}. Again, all $N$! possible combinations of the interaction sequence must be checked, but the most probable sequence of energy deposits is based on the physics of the interaction.
Recent work was done to improve upon the deterministic event reconstruction techniques by developing a more efficient probabilistic method that can account for escape gamma-ray events~\cite{2021arXiv210701846Y}.
For events with three interactions or more, it has been shown that the gamma-ray energy can be estimated even if the event is not fully absorbed~\cite{kurfess2000, TASHENOV2010592}.

In real measurements, the reconstruction becomes more complicated due to the finite spectral and spatial resolution of the detectors, sub-threshold interactions, and energy deposits in passive material. Any missing energy will result in a mis-reconstructed event and the calculated event circle will not align with the initial photon direction (see Section~\ref{sec:psf_errors}).
The level of accuracy for event reconstruction is therefore a very important performance parameter for Compton telescopes. The standard Compton Kinematic Reconstruction algorithm was initially found to properly reconstruct $50-70\%$ of 3$+$ site events~\cite{2000SPIE.4141..168O, boggs2000}.
The standard approach has been modified for improved efficiency by a number of groups, for example, by assuming the first interaction is the largest energy deposit, the sequencing performance improves by 20\% for $\gtrsim$1~MeV gamma rays~\cite{SHY2020161443}. 
An evaluation of different Compton sequencing approaches can be found in~\cite{4774644, LEE20214080}. 
More recently, advancements in machine learning approaches have shown promise for a much higher reconstruction efficiency~\cite{zoglauer2007b, zoglauer2021, 2019ExA....47..129C}.

One of the main advantages of Compton sequencing is the capability to reject events which do not conform with the expected kinematics of Compton scattering.
This is effectively a method of background rejection.
Photons that do not deposit all of their energy in the active detector volume will not be properly reconstructed and are thus improperly imaged and contribute to the background during observations. 
By setting a maximum acceptable level for the quality of the event (e.g., given by Equation~\ref{eq:recon_qfactor} for the classic Compton Kinematic Discrimination), any improperly reconstructed Compton event can be rejected.
Likewise, pair production or $\beta$-decay events in the instrument can be rejected with this approach.
Only events that are statistically determined to be ``good'' Compton events are kept after Compton sequencing.
The challenges of high backgrounds in Compton telescopes are addressed further in Section~\ref{sec:background}.

\subsection{Two-Site Event Reconstruction}
\label{sec:twositerecon}

For two-site events, with one Compton scatter followed by a photoabsorption, only a small fraction of events can be sequenced unambiguously~\cite{boggs2000}. 
A simple comparison test, the Compton Edge Test~\cite{XuPhD}, can distinguish the order of some two-site events.
We know that the maximum energy deposited in Compton scatter is from backscattering given by Equation~\ref{eq:comptonedge}; this energy defines the Compton edge in Figure~\ref{fig:compton_spectrum}. 
It is therefore easy to compare the two measured energies with this relation, where $E_0$ is the sum of the two energies. If either of the measured energies is larger than the maximum allowable energy of the Compton edge for the given $E_0$, then that interaction is necessarily a photoabsorption and must be the second interaction in the sequence.

Another simple relation can be used if the initial photon energy is less than 256~keV~\cite{1323740, LEE20214080}. As will be derived below, if $E_0 \lesssim m_e c^2/2$ is satisfied, the energy deposit in the first interaction, or in the Compton scatter $E_S$, is  deduced to be smaller than the energy deposited through photoabsorption in the absorber $E_A$. With $E_0 = E_A + E_S$, this can be derived starting again with the maximum energy of the first interaction defined by the Compton edge in Equation~\ref{eq:comptonedge}:
\begin{eqnarray}
    \text{max}(E_S) &=& \left. E_e \right\rvert_{\varphi = 180^{\circ}} =  E_0\left( 1 - \frac{1}{1 + \frac{2E_0}{m_e c^2}} \right). \nonumber \\
\end{eqnarray}
If $2E_0 < m_e c^2$, then
\begin{equation}
    \text{max}(E_S) < E_0/2, \\
\end{equation}
and with $E_0 = E_A + E_S$ 
\begin{align}
    \text{max}(E_S) & < \frac{E_A + E_S}{2} \\
    \text{max}(E_S) & < E_A.
\label{eq:twosite_relative}
\end{align}
Therefore, when the initial photon has energy less than $m_e c^2/2$, $E_S$ must be less than $E_A$, and the smaller energy deposit must correspond to the Compton scatter interaction.
For all other two-site events there is no unique ordering; however, it has been shown through simulations that Equation~\ref{eq:twosite_relative} generally holds until $\sim$400~keV~\cite{boggs2000, 1323740}.
Conversely, if the incident photon energy is greater than $\sim$400~keV, the first interaction is likely to have a larger energy deposit.

For a more sophisticated approach, one can use a deterministic algorithm based on the Klein-Nishina cross-section to determine which hit order is more probable~\cite{10.1117/12.563905}. 
In doing so, one can also take into account the probability of
photoabsorption for a certain distance between interactions to better understand the most likely order.

A drawback of the two-site event reconstruction is the loss of background rejection; there is no redundant information available  to determine if the two measured interactions correspond to a fully absorbed photon. With the additional uncertainties with two-site sequencing, the sensitivity of a telescope may be improved by excluding two-site events for some observations. 


\section{Compton Telescope Performance Parameters}
\label{sec:performance}

The capabilities of a Compton telescope are driven by the precision of the energy and position measurements of each interaction within the detector volume, and the precision of the event reconstruction. 
The science question at hand drives the design of a Compton telescope. 
For example, gamma-ray burst polarization measurements can be done with simple scintillators to measure the azimuthal scatter direction without imposing strict requirements on the energy resolution or spatial resolution of the detectors. 
However, achieving a continuum sensitivity that is 1--2 orders of magnitude better than COMPTEL (see Figure~\ref{fig:MeVGap}) requires a detector with excellent spectral and spatial resolution and strong background rejection. In this section, the point spread function for Compton telescopes will be introduced, and the standard Compton telescope performance parameters, such as the angular resolution and sensitivity, will be discussed.

\subsection{Point Spread Function}
\label{sec:cds}
\begin{figure}[tb]
    \centering
    \begin{minipage}{0.53\textwidth}
    \includegraphics[width=\textwidth]{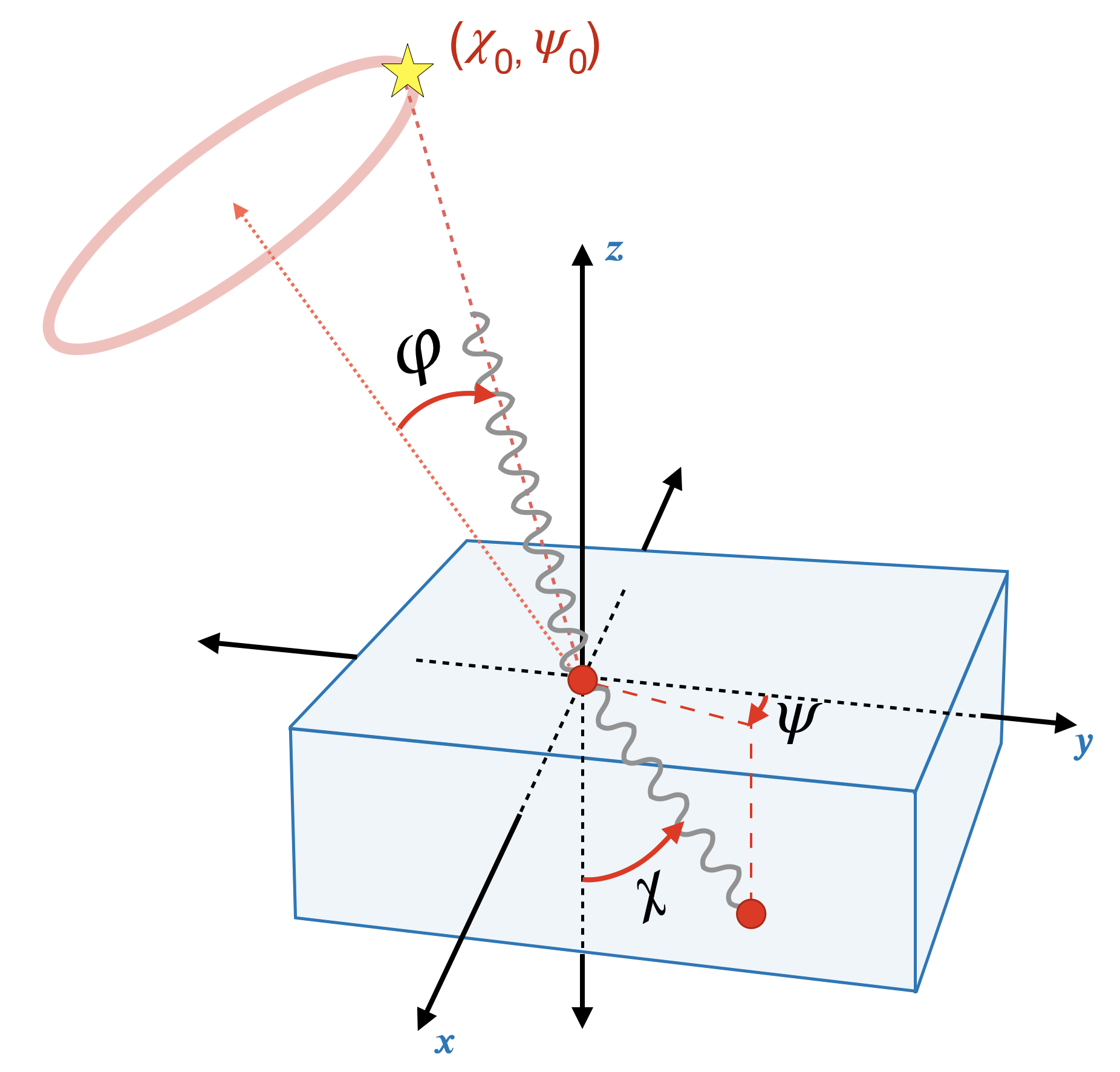}
    \subcaption{}\label{fig:CDS_detcoords}
    \end{minipage}
    \hfill
    \begin{minipage}{0.46\textwidth}
    \centering
    \includegraphics[width=0.9\textwidth]{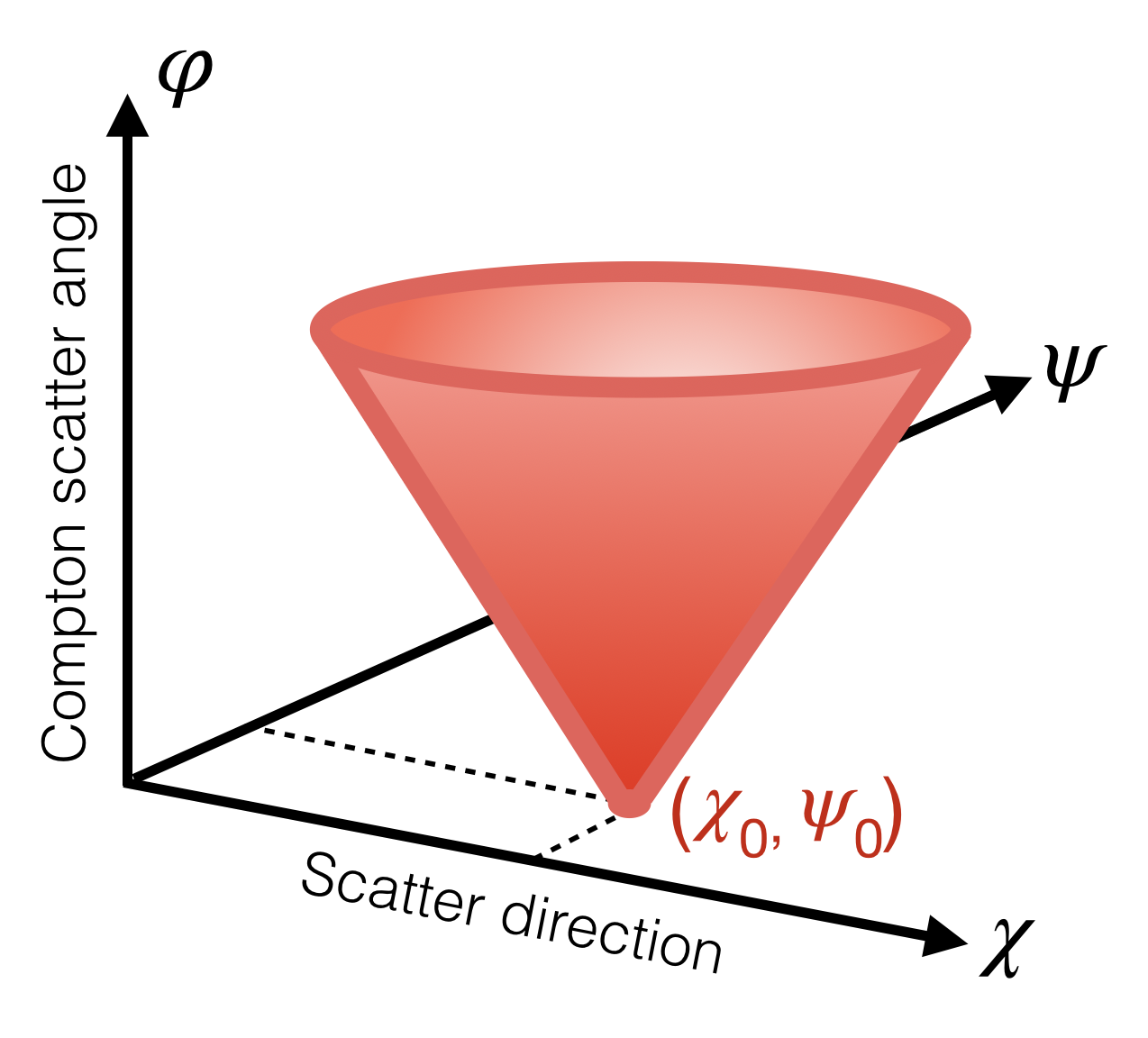}
    \subcaption{}\label{fig:CDS}
    \end{minipage}
    \caption{\textbf{(a)} Schematic diagram of the first two interactions of a Compton event showing the definition of the CDS angles ($\chi, \psi, \varphi$) for a gamma-ray source located at ($\chi_0, \psi_0$). The Compton scatter angle of the first interaction, $\varphi$, is calculated from the energy deposits. The polar and azimuthal angles of the Compton-scattered gamma ray direction ( ($\chi, \psi$) are measured relative to the detector coordinate system. These three parameters define the kinematics and geometry of a Compton event. Modified from \citet{kieransthesis}. 
    \textbf{(b)} The point spread function in the CDS is an easily discernible cone with apex at the source location ($\chi_0, \psi_0$) and a 90$^{\circ}$ opening angle.}
    \label{fig:comptondataspace}
\end{figure}

The raw data space for Compton telescopes consists of all possible measurements of the energy and position of Compton-scattering interactions within the detector volume. 
However, only the first two interactions are needed for imaging capabilities, where the key parameters are the Compton-scattered photon direction and the Compton scattering angle, as described in Section~\ref{sec:operating}.
Therefore, Compton telescope analysis is generally done in a reduced data space consisting of three-dimensions: the Compton scatter angle $\varphi$, and the polar and azimuthal angles of the Compton-scattered photon direction, $\chi$ and $\psi$, respectively.
These angles define the three orthogonal axes of the data space, as shown in Figure~\ref{fig:comptondataspace}. 
The total energy of the gamma ray is generally considered a fourth dimension of this data space, though it will not be explicitly written here. 
This data space, along with the fundamentals of modern Compton telescope analysis techniques, was pioneered for the COMPTEL mission~\cite{schonfelder1993, vonballmoos1989, 1992NASCP3137...95D} and is sometimes referred to as the COMPTEL Data Space, or simply the Compton Data Space (CDS).

The point spread function in the Compton Data Space is a cone with the apex at the location of the source (Figure~\ref{fig:CDS}). 
The opening angle of the cone is 90$^{\circ}$ since the Compton scatter angle is equal (within measurement error) to the deviation of the scattered gamma-ray direction from the known source location, as seen in Figure~\ref{fig:CDS_detcoords}.
Therefore, the CDS contains redundant information about the Compton scatter angle: the $\varphi$ dimension is calculated based on the measured energies with Equation~\ref{eq:Compton_scatter_angle}, 
and the scatter direction ($\chi, \psi$) is defined relative to the detector coordinates, but is related to the geometric measurement of the Compton scatter angle. 
In contrast to the projected event circles in image space, each Compton event is a point in the CDS at ($\chi, \psi, \varphi$), and the accumulation of properly reconstructed Compton events from point-source emission fills the CDS along the walls of the three-dimensional cone.
The CDS cone is connected to the familiar Compton event circle in image space: the probability of any region of the sky contributing to one bin in the CDS ($\chi, \psi, \varphi$) is represented by the back-projected event circle.

The response density along the cone follows the Klein-Nishina cross-section.
For example, the smaller Compton scatter angles near the cone apex are likely to be populated by higher energy events.
Furthermore, the photon polarization information is encoded in the CDS cone. The modulation in the Compton azimuthal scatter angle $\eta$, defined in Equation~\ref{eq:klein-nishina-pol} and shown in Figure~\ref{fig:pol_basics}, can be defined in terms of the scatter direction ($\chi, \psi$) in detector coordinates ($\eta \equiv \psi$ for on-axis sources).
The population of the CDS cone is also impacted by the detector geometry where some scattering directions can be suppressed.

\begin{figure}[tb]
    \centering
    \begin{minipage}{0.53\textwidth}
    \includegraphics[width=\textwidth]{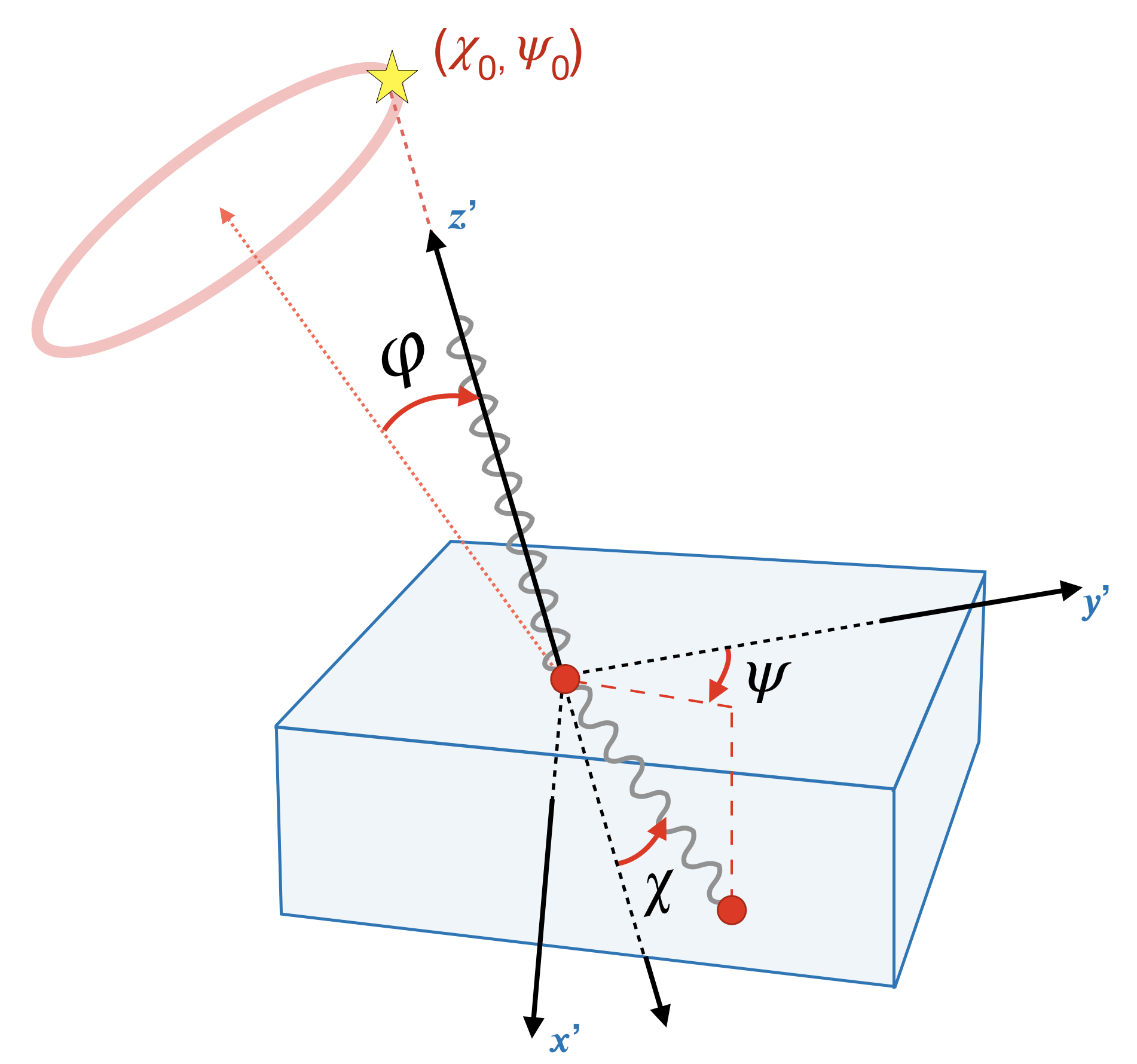}
    \subcaption{}\label{fig:CDS_detcoords_onaxis}
    \end{minipage}
    \hfill
    \begin{minipage}{0.46\textwidth}
    \centering
    \includegraphics[width=0.9\textwidth]{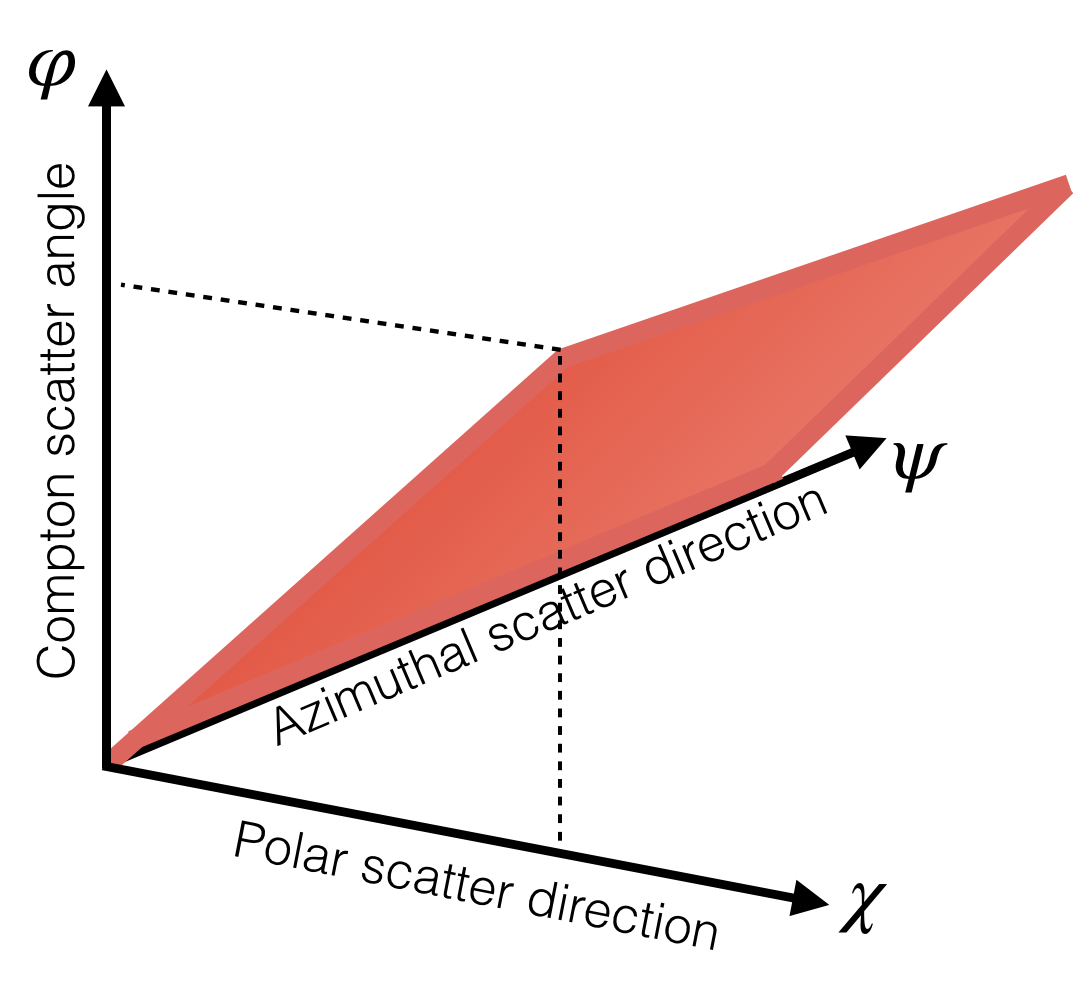}
    \subcaption{}\label{fig:CDS_OnAxis}
    \end{minipage}
    \caption{\textbf{(a)} It can be illustrative to rotate the CDS coordinates so that the measurement of the scattering direction is made relative to the incident photon direction from a known source location ($\chi_0, \psi_0$). Then the scattering angles ($\chi, \psi$) in the CDS equate to the geometrically measured polar Compton scattering angle ($\varphi^{geo}$) and the azimuthal Compton scattering angle ($\eta$), as defined in Figure~\ref{fig:polarized_angles}.
    \textbf{(b)} For on-axis sources, the point spread function in the CDS translates into a two-dimensional plane at 45$^{\circ}$. Modulation from a polarized source will result in a sinusoidal distribution along the $\psi$ dimension.}
    \label{fig:comptondataspace_onaxis}
\end{figure}

\begin{figure}[tb]
    \centering
    \begin{minipage}{0.5\textwidth}
    \centering
    \includegraphics[width=\textwidth]{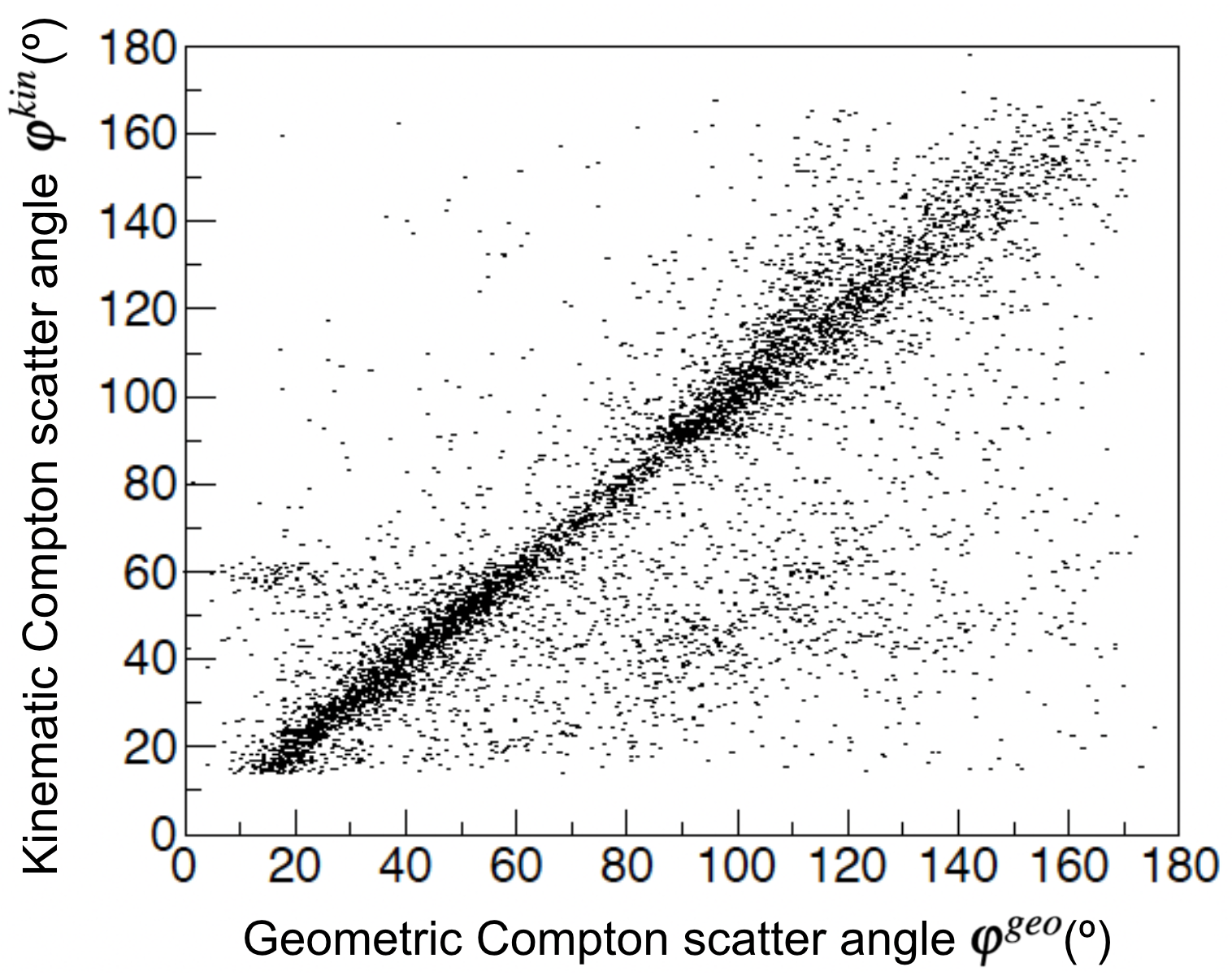}
    \subcaption{}\label{fig:projectedcds_sims}
    \end{minipage}
    \hfill
    \begin{minipage}{0.49\textwidth}
    \centering
    \includegraphics[width=0.8\textwidth]{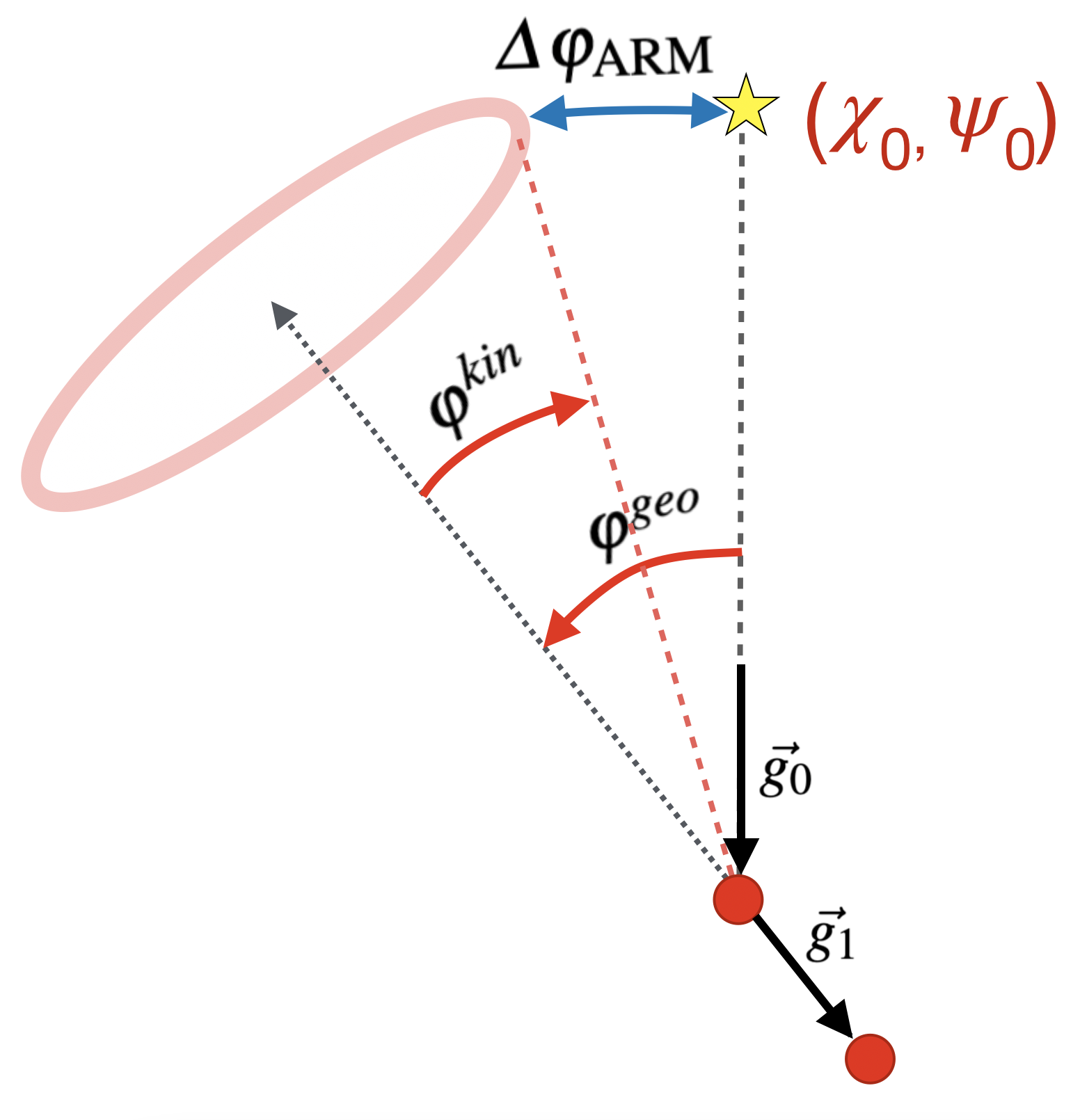}
    \subcaption{}\label{fig:armangledef}
    \end{minipage}
    \caption{When the source position is known, projection of the CDS into two dimensions allows for a direct relation between the kinematically measured Compton scatter angle ($\varphi^{kin}$) and the geometrically measured angle ($\varphi^{geo}$). Events that are properly reconstructed lie along the $\varphi^{kin} = \varphi^{geo}$ line, where the spread is defined by uncertainties in the energy and positions measurements. (a) A simulation of a 511~keV point source measured in a COSI-like instrument shows this 2D projection of the CDS point spread function. All of the events located off of the  $\varphi^{kin} = \varphi^{geo}$ line are improperly reconstructed, most often due to missing energy. Modified from \citet{kieransthesis}. (b) The geometrically measured Compton scattering angle is given by the angle between the initial photon direction $\vec{g_0}$ and the scattered gamma-ray direction $\vec{g_1}$. The difference between the kinematic and geometric Compton scatter angle gives the Angular Resolution Measure (ARM), $\Delta \varphi_{ARM}$ for each event.}
    \label{fig:projectedcds}
\end{figure}

To better understand the CDS, it can be illustrative to rotate into a reference frame where the scattering direction ($\chi, \psi$) is defined relative to the initial photon direction of a known source position ($\chi_0, \psi_0$). 
This transformation, as visualized in Figure~\ref{fig:comptondataspace_onaxis}, means that the polar scattering angle $\chi$ becomes a geometric measure of the Compton scattering angle $\varphi^{geo}$, and the azimuthal scattering direction $\psi$ becomes equivalent to azimuthal scatter angle $\eta$. 
In the CDS, the distribution of events fills a two-dimensional plane at 45$^{\circ}$, as shown in Figure~\ref{fig:CDS_OnAxis}. The polarization response and geometric asymmetries will now show as a sinusoidal modulation along the $\psi$ ($= \eta$) dimension.
For unpolarized sources at a known location, the CDS can be further simplified by projecting into two dimensions (with some loss of discriminating power); by integrating over the azimuthal scattering angle $\eta$, a simplified 2D CDS can be defined by $\varphi^{kin}$ and $\varphi^{geo}$~\cite{schonfelder1993}.
A simulated point spread function in this projected 2D Compton Data Space is shown in Figure~\ref{fig:projectedcds}, where the deviations from  $\varphi^{kin} = \varphi^{geo}$ are from uncertainties in the measured positions and energies close to the line, and the width of the CDS cone corresponds to the angular resolution of the instrument.
The events that are far from the $\varphi^{kin} = \varphi^{geo}$ line are improperly reconstructed: an incompletely absorbed recoil electron results in a measure of $\varphi^{kin} < \varphi^{geo}$, while incompletely absorbed gamma rays result in $\varphi^{kin} > \varphi^{geo}$.

The 3D CDS point spread function (Figure~\ref{fig:comptondataspace}) forms the basis for sensitive Compton telescope data analysis. 
Compared to the image space with overlapping Compton event circles causing source confusion, the CDS provides strong discriminating power and allows for a better separation of background and source emission regions~\cite{2021arXiv210213158Z}.



\subsubsection{Angular Resolution Measure}
\label{sec:arm}

\begin{figure}[tb]
    \centering
    \begin{minipage}{0.39\textwidth}
    \centering
    \includegraphics[width=\textwidth]{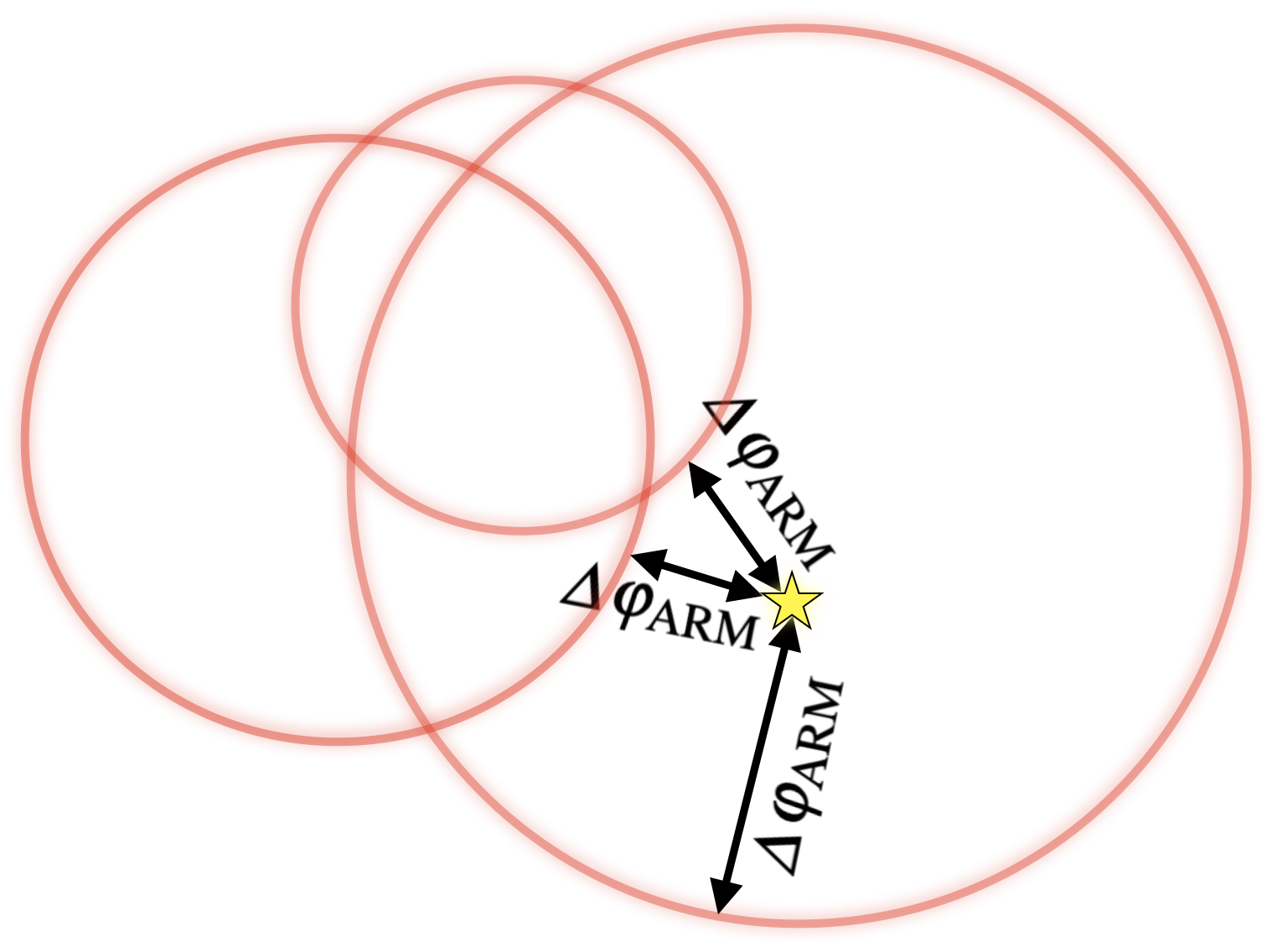}
    \subcaption{}\label{fig:imagespace_arm}
    \end{minipage}
    \hfill
    \begin{minipage}{0.6\textwidth}
    \centering
    \includegraphics[width=\textwidth]{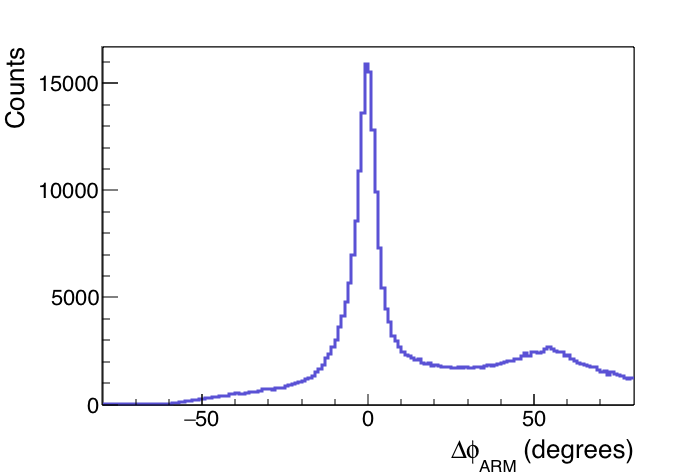}
    \subcaption{}\label{fig:armhistogram}
    \end{minipage}
    \caption{\textbf{(a)} The smallest angular distance between the known source location and each Compton event circle defines the ARM, $\Delta \varphi_{ARM}$. 
    \textbf{(b)} A histogram of the $\Delta\varphi_{\text{ARM}}$ values gives a one-dimensional projection of the point spread function, and the FWHM of the ARM distribution defines the angular resolution of a Compton telescope. As an example, the histogram shown here is from a measurement of a far-field laboratory 511~keV source with the COSI balloon detector~\cite{2017arXiv170105558K} and has a 6$^{\circ}$ angular resolution.}
    \label{fig:arm}
\end{figure}

The angular resolution of a Compton telescope given by the width of the CDS cone walls and is generally dominated by the uncertainty in the measured energy and position of interactions. 
The most common definition of the angular resolution is the full width at half maximum (FWHM) of the Angular Resolution Measure (ARM) distribution, which is a one-dimensional projection of the CDS cone.
The ARM for each event is defined as the minimum distance between the geometrically calculated Compton scatter angle and the kinematically calculated Compton scatter angle:
\begin{equation}
    \Delta\varphi_{\text{ARM}} = \varphi^{geo} - \varphi^{kin}.
\end{equation}
The kinematic Compton scatter angle is determined from the measured energy deposits using the Compton equation:
\begin{equation}
    \varphi^{kin} = \arccos \left[ 1 - m_e c^2 \left( \frac{1}{E _{scat}} - \frac{1}{E _0} \right) \right]
\end{equation}
with $E_0$ being the total deposited energy, as detailed in Section~\ref{sec:physics}. The geometrically determined Compton scatter angle is found by taking the dot product of the initial photon direction $\vec{g_0}$ and the measured scattered gamma-ray direction $\vec{g_1}$:
\begin{equation}
    \varphi^{geo} = \arccos \left[  \frac{\vec{g_0} \cdot \vec{g_1}}{| \vec{g_{0}} | | \vec{g_1} | }  \right],
\end{equation}
as defined in Figure~\ref{fig:armangledef}. 
Note that this is similar to the redundant information used in  Compton Kinematic Discrimination (Equations~\ref{eq:phikin} and \ref{eq:phigeo}); however, here we must know the initial source location to define $\vec{g_0}$.
This definition of $\Delta\varphi_{\text{ARM}}$ is equivalent to the shortest angular distance between the known source location and the Compton event circle in image space. This is depicted for three event circles in image space in Figure~\ref{fig:imagespace_arm}. 

The FWHM of the ARM distribution is the standard definition for the angular resolution.
A histogram of the measured $\Delta\varphi_{\text{ARM}}$ values from a COSI-like detector is shown in Figure~\ref{fig:armhistogram}. For an ideal detector, one expects a sharp feature at $\Delta\varphi_{\text{ARM}} = 0$ indicating that the event circle correctly overlaps the known source location. Measurement errors in the position and energy of interactions within the detector broaden this response. Events that are improperly reconstructed, often due to incomplete absorption, end up in the wings of this ARM distribution $\sim50^{\circ}$. Sophisticated event reconstruction techniques, as discussed in Section~\ref{sec:eventrecon}, and a selection on the quality of the Compton event can reduce the prominence of these off-ARM-peak events~\cite{zoglauerthesis,2021arXiv210213158Z}.

\subsubsection{Scatter Plane Distribution}
\label{sec:performance_electrontracking}

Compton telescopes that have electron tracking allow for a more defined point spread function in image space, as shown in Figure~\ref{fig:electron_tracking_images}. By measuring the recoil electron trajectory from the first Compton interaction in a telescope, the standard event circle can be reduced to an arc.

\begin{figure}[t]
    \centering
    \begin{minipage}{0.5\textwidth}
    \includegraphics[width=0.9\textwidth]{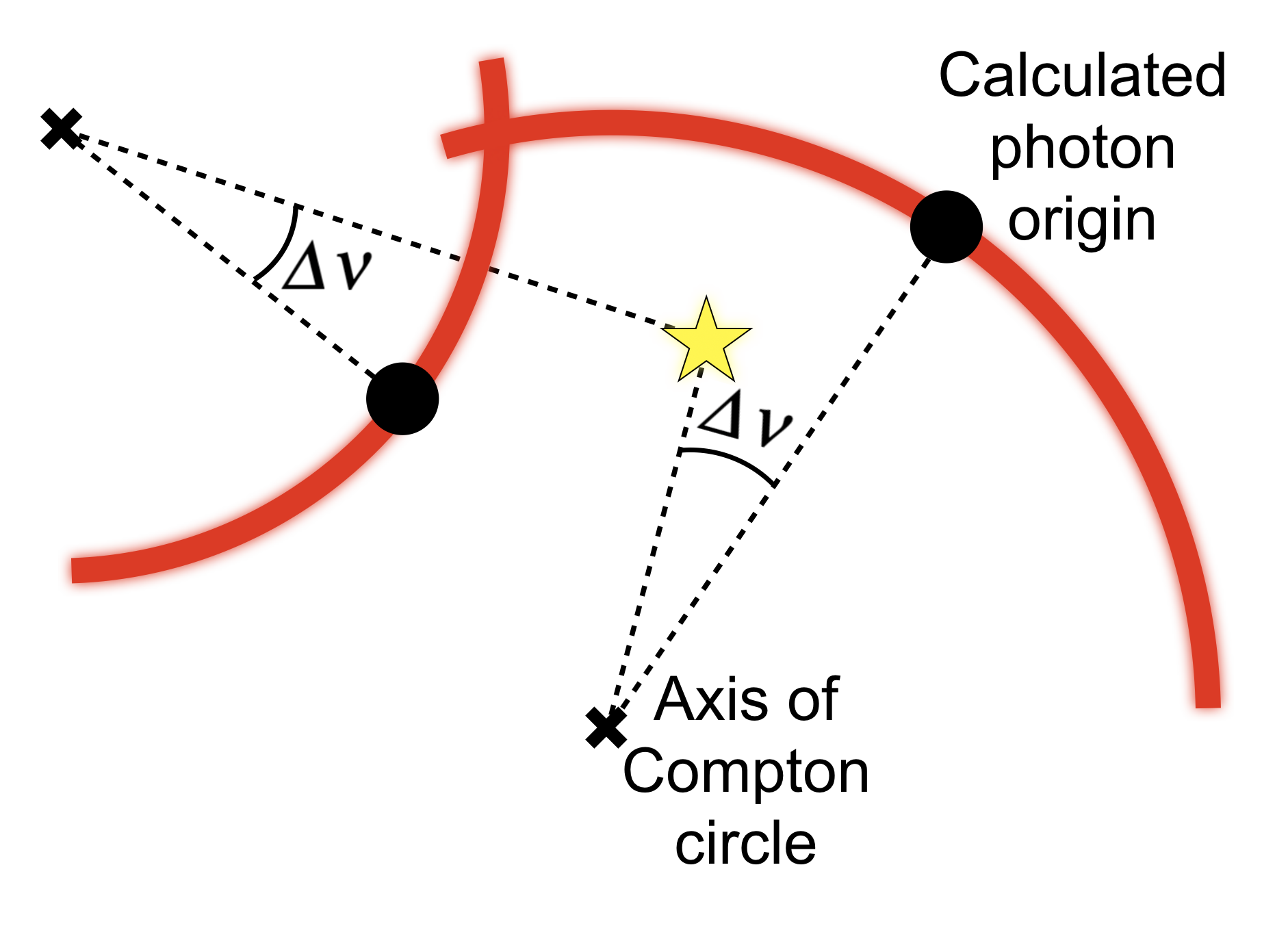}
    \end{minipage}
    \hfill
    \begin{minipage}{0.49\textwidth}
    \captionof{figure}{The SPD $\Delta \nu_{SPD}$ is defined as the angular distance between the known source location (yellow star) and the measured origin from the scattered electron and gamma-ray kinematics. Adapted from~\citet{zoglauerthesis}. }
    \label{fig:spd_definition}
    \end{minipage}
\end{figure}

The measured direction of the recoil electron is less precise than the direction of the scattered gamma ray (longer lever arm) and, therefore, the ARM is still used to describe the angular resolution of an electron-tracking Compton telescope. It is convenient to define the electron's kinematics relative to the center of the photon cone defined by $\vec{g_1}$, and this is commonly done in terms of the Scatter Plane Deviation (SPD). The SPD describes the angle between the scattered plane defined by the known source location and the true scattered gamma-ray direction ($\vec{g_0}$ and $\vec{g_1}$ in Figure~\ref{fig:armangledef}), and the measured scatter plane defined by the scattered gamma-ray direction and the recoil electron direction ($\vec{e}$):
\begin{equation}
    \Delta \nu_{SPD} = \arccos \left[ (\vec{g_1} \times \vec{g_0} ) \cdot  ( \vec{g_1} \times \vec{e})    \right].
\end{equation}
This is equivalent to the angular distance between the known source location and the calculated photon origin for each event circle, as depicted in Figure~\ref{fig:spd_definition}. The SPD is a measure of the length of the Compton event arc.

The Compton data space for a tracking telescope includes two more dimensions to describe the recoil electron direction. While a 5+ dimensional data space can be used for the most accurate Compton telescope analysis with enough computation resources, it is currently more common for the SPD to be used as an event selection in list-mode imaging approaches (see Section~\ref{sec:imaging}). By only using events with a small SPD, the photon direction can be more precisely constrained  and the image-space representation of the source is much more defined. In the presence of background, a selection on the SPD for each gamma-ray event significantly reduces the number of background photons that overlap with the source region, increasing the sensitivity of the telescope.

Electron tracking capabilities have been achieved in both gaseous and solid-state detectors. 
The low density in gaseous TPCs allows for the recoil electron to subtend a few centimeters in the detector volume, and thus can result in a fairly accurate measure of the initial scattering direction.
The SMILE-II instrument demonstrated a FWHM of measured SPDs $\sim75^{\circ}$ for events from a $^{137}$Cs source that transferred more than 80~keV to the recoil electron~\cite{Tanimori_2015}.
However, low-density gaseous TPCs require a large detector size and are limited in terms of the energy resolution, which in turn limits the angular resolution.
Semiconductor detectors have advantages with high density and high spectral resolution; however, the path length for recoil electrons is generally much shorter.
The recoil electron direction can be measured in multi-layer silicon Compton telescopes \cite{oneil2003, Kanbach_etal_2005}, but the effects of Moli\`{e}re scattering are a limitation; \citet{Kanbach_etal_2005} achieved a SPD FWHM of $\sim80^{\circ}$ for a $^{88}$Y source.
To more precisely measure the recoil electron direction in silicon, the detector requires a spatial resolution around 10 $\mu$m. Track reconstruction within a single solid-state detector layer was first demonstrated for Charged Coupled Devices (CCDs) with $10.5 \times 10.5$~$\mu$m$^{2}$ pixel size \cite{vetter2011, 2011NIMPA.652..595P}; see Figure~\ref{fig:electron_track_example}. 
More recently, a hybrid detector with complementary metal–oxide–semiconductor (CMOS) 20~$\mu$m pitch readout on one side of a 500~$\mu$m-thick silicon wafer, and strip electrodes on the other side, was developed to achieve high spatial resolution tracking information, improved energy resolution, and $\mu$sec timing information~\cite{yoneda2018development, YabuPhD}.




\subsubsection{Uncertainties in the Angular Resolution}
\label{sec:psf_errors}

The angular uncertainty for each event can be determined from the uncertainties in the measured energies, which contribute to the Compton scatter angle calculation, and the uncertainties in the measured positions of the first two interactions, which determine the axis of the event circle~\cite{vonballmoos1989}. 
While there is a fundamental limit to the angular resolution of Compton telescopes (see Section~\ref{sec:doppler}), in most modern designs, the dominant factor is usually either the position resolution or energy resolution of the detectors. 


\begin{figure}[tb]
    \centering
    \includegraphics[width=0.7\textwidth]{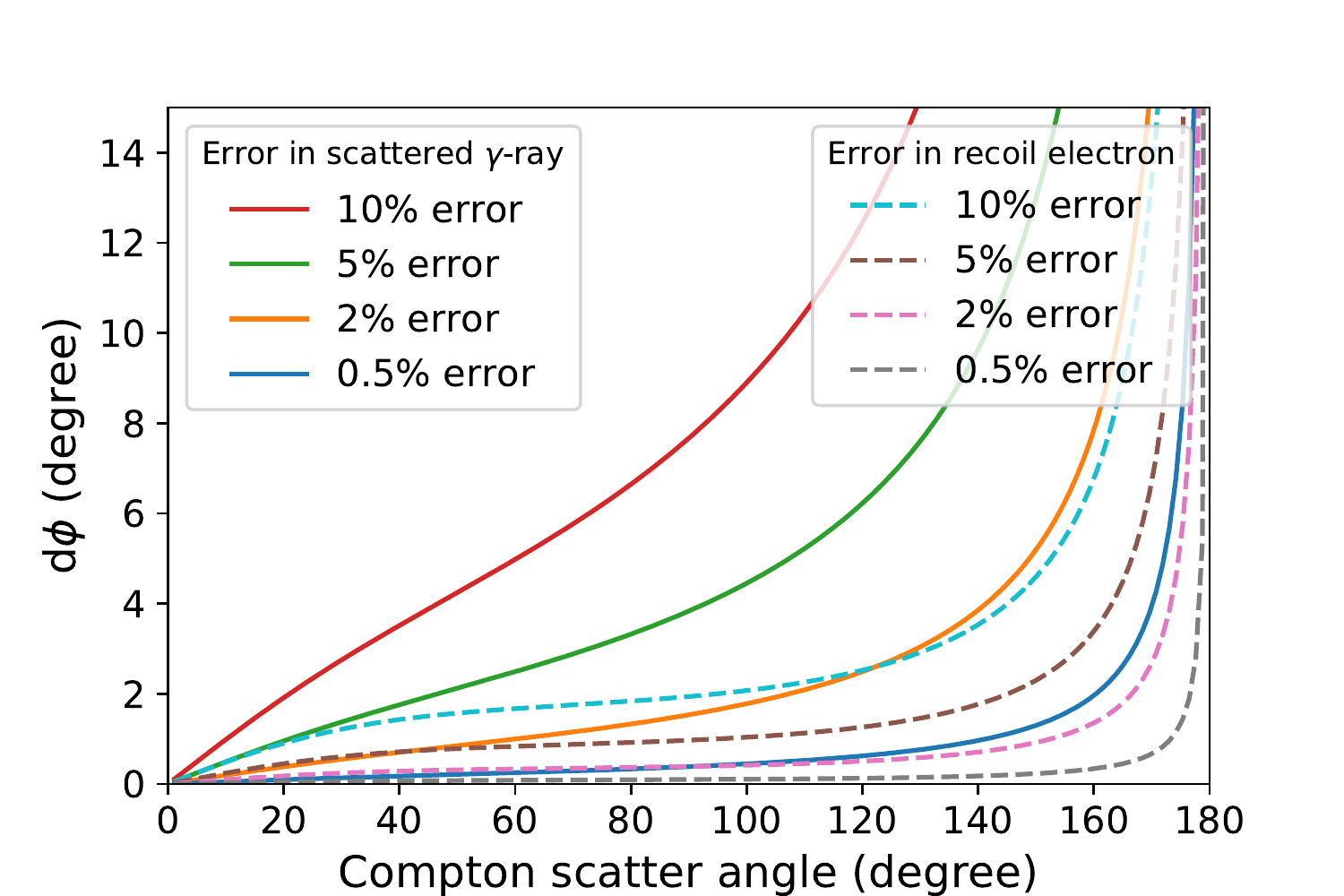}
    \caption{The error in the angular resolution is affected by both the uncertainty in the measured energy of the scattered gamma ray and the recoil electron. These are plotted separately assuming a fraction error: the solid lines show the error in the angular resolution from only the error from the scattered gamma-ray energy of $\Delta E/E_{scat} = 0.5, 2, 5,$ and 10\% of $E_g$, given by Equation~\ref{eq:armerr}. The dashed lines show the contribution from the uncertainty in the recoil electron energy with the same fractional error. The total photon energy was assumed to be 2~MeV. Adapted from \citet{zoglauerthesis}.}
    \label{fig:dphierr}
\end{figure}

It can be seen from Equation~\ref{eq:Compton_scatter_angle} that the measured energy of each interaction plays a role in calculating the Compton scattering angle for each event, and thus is a defining factor for the ARM.
The precision with which the energy can be measured, i.e. the energy resolution, depends on the detector properties and readout electronics.
A propagation of energy measurement errors for Equation~\ref{eq:Compton_scatter_angle} gives the following error on the Compton scattering angle~\cite{zoglauerthesis,1323740}:
\begin{equation}
d\varphi^{kin} = \frac{E_0}{\sin \varphi} \sqrt{ \left( \frac{1}{E_{scat}^2} - \frac{1}{(E_e + E_{scat})^2} \right)^2dE_{scat}^2 + \frac{1}{(E_e + E_{scat})^4} dE_e^2 }
\label{eq:armerr}
\end{equation}
where $E_0$ is the initial photon energy, $dE_{scat}$ is the uncertainty in the Compton-scattered photon energy, and $dE_e$ is the uncertainty in the recoil electron energy, using the same notation in Section~\ref{sec:physics}.
For events with more than two interactions, $dE_{scat}$ will include the measured energy error from each additional interaction added in quadrature.
This relation is shown in Figure~\ref{fig:dphierr} where the initial photon energy is assumed to be 2~MeV, and the two different contributions, $dE_{scat}$ and $dE_{e}$, are plotted separately for set errors of $\Delta E/ E$ = 0.5, 2, 5 and 10\%. 
The angular resolution is significantly impacted by a poor energy measurement of the scattered gamma ray, especially at higher Compton scatter angles.
It is clear that to achieve an angular resolution $\sim1^{\circ}$, the energy resolution of the detector must be $\lesssim1-2$\%, and by selecting only events which have a smaller Compton scatter angle, the angular resolution can be improved.

The uncertainty in the position measurements for the first two interactions translates to an error in the axis of the Compton circle. 
When the distance between these two interactions is small, this effect becomes more pronounced. For example, for interactions that are 0.5~cm apart, a spatial resolution of 1~mm results in a significant error on the scattering direction compared to events that are separated by $>1$~cm.
The size of this effect can be simply estimated as $d \varphi^{geo} \lesssim \text{tan}(\frac{d x}{D})$, where $d x$ is the spatial resolution and $D$ is the distance between interactions. For a 1~mm spatial resolution, the potential angular deviation is as much as 11$^{\circ}$ for a separation of 0.5~cm and only 6$^{\circ}$ at separation of 1~cm.
This is only an approximation, and likely an upper limit since there is a higher probability of interacting near the center of a spatial resolution element than near the corner when the scattering direction is off-axis~\cite{1323740}.
For a detailed calculation of the angular uncertainty from the detector position resolution considering different resolutions in each dimension, see \cite{XuPhD, TakedaPhD, 10.1117/12.563905}.
In general, the uncertainty in the angular resolution from the error in the position measurements is more pronounced at lower energies since these events will have a smaller distance between the first and second interactions.

\begin{figure}[tb]
    \centering
    \includegraphics[width=0.7\textwidth]{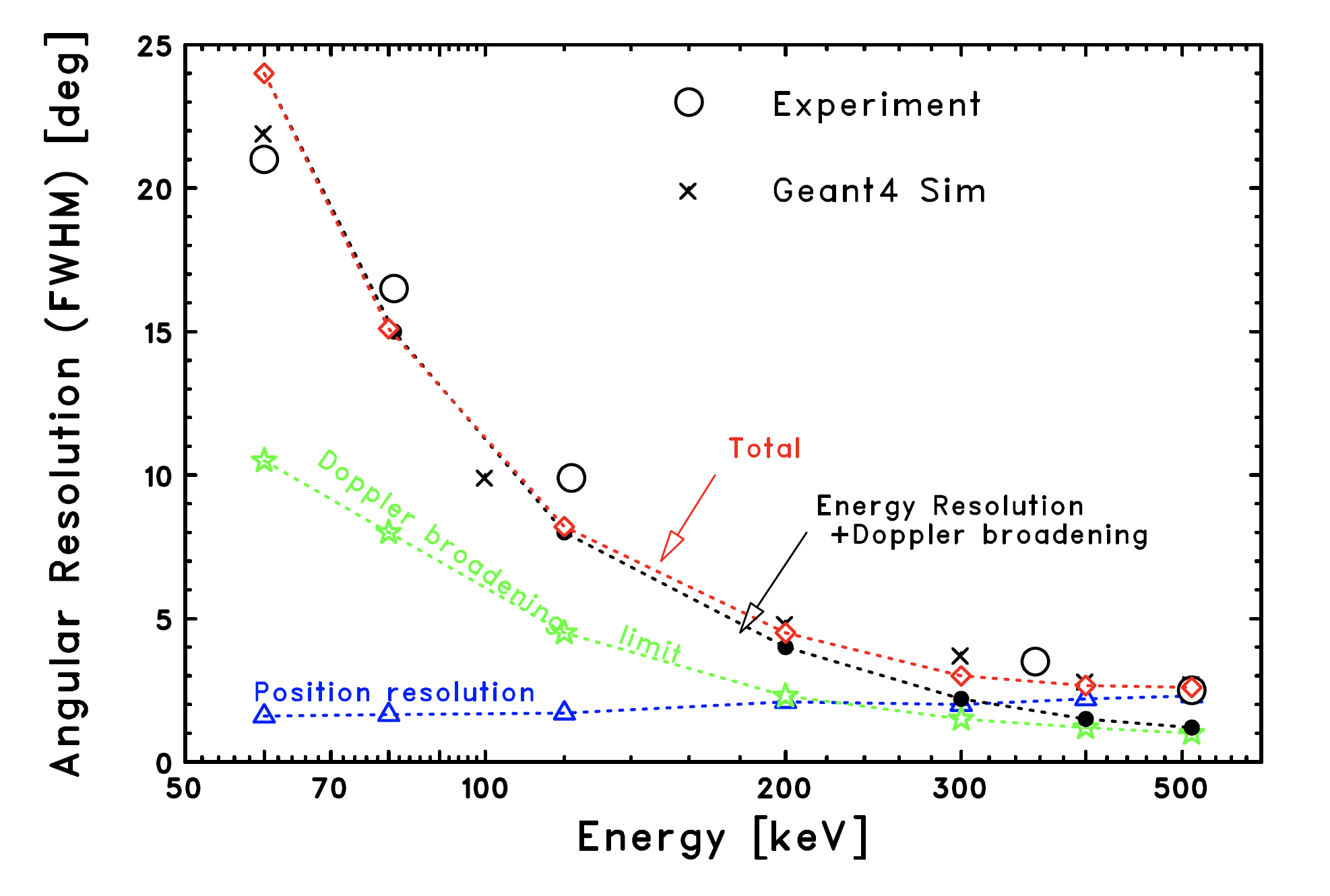}
    \caption{The simulated angular resolution for a silicon and cadmium telluride instrument is shown broken out into the main sources of error. The uncertainty from the position resolution ($\sim$2$^{\circ}$) is relatively independent of energy and starts dominating above 300~keV. The energy resolution and Doppler broadening is highly energy-dependent and dominates the ARM below 300~keV. Figure from \citet{TakedaPhD}.}
    \label{fig:angulareserrors}
\end{figure}

The total angular uncertainty is given by a sum of the squares of the two independent components:
\begin{equation}
d\varphi_{total} = \sqrt{ (d\varphi^{kin})^2 + (d \varphi^{geo})^2  }.
\end{equation}
As an example of how the different contributions affect the angular resolution of a telescope, Figure~\ref{fig:angulareserrors} shows the measured and simulated FWHM of the ARM distribution for a balloon-borne silicon and cadmium telluride Compton telescope~\cite{10.1117/12.733840}.
The simulated response is broken down into the three main contributing factors: the energy resolution, position resolution, and Doppler broadening (Section~\ref{sec:doppler}). The angular resolution is dominated by uncertainties in the energy $<300$~keV, and the position uncertainty, which is relatively independent of energy, starts to dominate at higher energies.

The effect of the position and energy error on the angular resolution can be reduced through event selections. If events with larger Compton scatter angle are rejected, then the contribution of the error measurement on the scattered gamma-ray energy will be reduced.
Similarly, if events with a small separation distance between the first and second interaction are rejected, then events with a more precise measurement of the scattering direction will remain.
Both examples can result in improved angular resolution at the cost of using fewer, higher-quality events in the analysis. 
There is a constant trade-off between the angular resolution and the efficiency of the telescope and different observations will call for different optimizations.


\subsubsection{Doppler broadening as a Lower Limit to the Angular Resolution}
\label{sec:doppler}

The Klein-Nishina cross section (Equation~\ref{eq:klein-nishina}) was derived with the assumption that the target electron is a free particle at rest. 
However, in a Compton telescope the electrons are bound in an atom with finite momentum.
Therefore, there is an additional uncertainty in the interaction energy  which fundamentally limits the angular resolution that can be achieved with a Compton telescope~\cite{zoglauer2003}.
This effect is referred to as Doppler broadening~\cite{PhysRev.33.643}.

The Klein-Nishina formula can be modified to include the effects of the bound electron's momentum.
According to the modified formula~\cite{ribberfors1975, 10.1117/12.187266}, the magnitude of Doppler broadening depends on the initial photon direction, Compton scattering angle, and the atomic number~\cite{zoglauer2003, TakedaPhD}.
The angular uncertainty is generally small for low-Z material, such as plastic scintillator or silicon, but can be a dominant factor for high-Z material, such as germanium, CdTe, or CZT.

\begin{figure}[tb]
    \centering
    \includegraphics[width=0.9\linewidth]{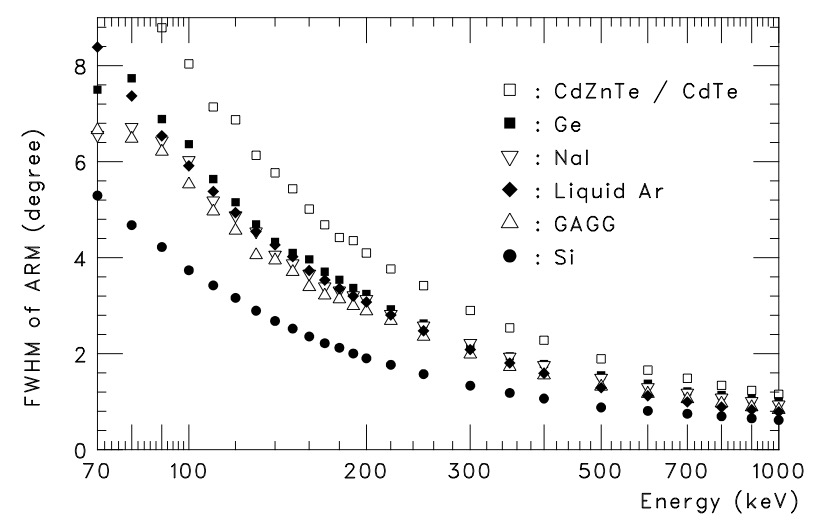}
    \caption
    {Doppler broadening is a fundamental limit to the angular resolution of a Compton telescope, and the magnitude of the effect is highly dependant on the material atomic number Z and the initial photon energy. The angular resolution, as measured by the FWHM of the ARM distribution, is simulated in Geant4 for common detector materials. Figure from \citet{GoroYabu2021}.
    }
    \label{fig:doppler2}
\end{figure}

\citet{GoroYabu2021} performed a study of the Doppler-broadening effects in silicon and other common detector materials. 
Figure~\ref{fig:doppler2} shows the angular resolution simulated with the Geant4~\cite{geant4} toolkit as a function of energy for CdTe, CZT, germanium (Ge), NaI, liquid argon (Ar), gadolinium aluminum gallium garnet crystal (GAGG), and silicon. 
Silicon, with its low atomic number, has the lowest achievable angular resolution among the simulated materials, with the Doppler-limited angular resolution almost a factor of two lower than CZT. There is also a strong dependence on the shell structure of the atoms: the worst uncertainty is obtained when the electron orbitals are completely filled and the best is obtained for alkaline or alkaline earth metals~\cite{zoglauerthesis}. Figure~\ref{fig:angulareserrors} shows the Doppler broadening contribution to the total angular resolution of a silicon and cadmium telluride instrument, and the relation to other uncertainty factors from the measured energy and position resolution of the detectors.

\subsection{Sensitivity}
\label{sec:sens}

The most impactful performance characteristic of a Compton telescope is its sensitivity (see Figure~\ref{fig:MeVGap}). It is equivalent to the lowest  source flux detectable and is derived from the standard way of calculating detection significance. 
The sensitivity of a telescope is given by~\cite{1996A&AS..120C.673J, 1985ICRC....3..477S}:
\begin{equation}
    F_{min}(E) = \frac{n\sqrt{N_S + N_B}}{A_{eff}T_{obs}} = \frac{n^2 + n\sqrt{n^2 + 4 N_B}}{2A_{eff}T_{obs}}
    \label{eq:sensitivity}
\end{equation}
where $n$ is the detection significance (typically chosen to be 3 or 5$\sigma$), $N_S$ is the number of source photons detected, $N_B$ is the number of background photons, $A_{eff}$ is the effective area, and $T_{obs}$ is the observation time. 

At first glance, the main contributions to the sensitivity are the number of detected background counts, the effective area, and the time. 
However, $N_B$ is defined by the size of the source resolution element and depends on the angular resolution measure, the energy bin width, and the SPD (for tracking detectors), which all in turn depend on the detector energy resolution and/or position resolution.
Only the source and  background photons whose event circles overlap with the source position are included. The standard assumption is to use a selection on the ARM distance from the source position of 1$\times$FWHM of the ARM distribution.
The background counts are calculated by integrating the expected background spectrum in the measured energy bin $\Delta$E. 


The narrow-line sensitivity can be calculated for gamma rays with known energy, such as nuclear lines. Equation~\ref{eq:sensitivity} can be used; however, only the number of background photons within the energy resolution of the instrument are included, where $\Delta E$ is usually taken to be 1.5--2$\times$FWHM. The narrow-line sensitivity, therefore, can be enhanced by having an excellent energy resolution as fewer background photons will lie beneath the photopeak.

To properly calculate the sensitivity of a Compton telescope, detailed instrument simulations with an accurate detector response are necessary. 
The background count rate in a certain resolution element can only be determined through simulations since a significant fraction of the background contribution comes from activation of the instrument material (Section~\ref{sec:background}). 
And an accurate measure of the effective area is only possible through detailed instrument simulations, where event selections can significantly affect the outcome.
Furthermore, calculations performed in the Compton data space instead of the image space allow for a much lower sensitivity since the source and background regions are much easier to separate.  
For examples of sensitivity calculations for Compton imagers, see \cite{2001A&A...376.1126B, zoglauerthesis, 10.1117/12.312882}.





\subsection{Imaging Capabilities}
\label{sec:imaging}

\begin{figure}[tb]
    \centering
    \begin{subfigure}[b]{0.7\textwidth}
    \centering
    \includegraphics[width=\textwidth]{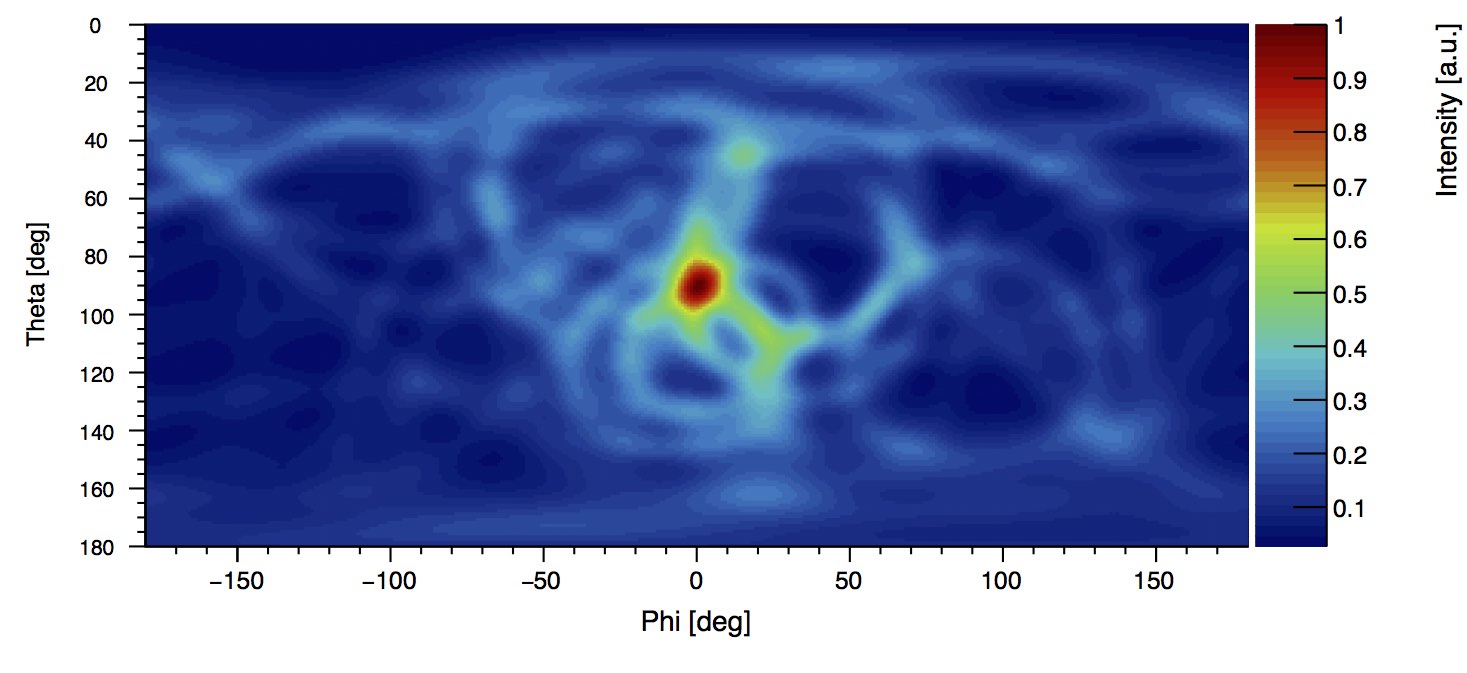}
   \caption{Back-projection of a point source simulation with 200 events.}
    \label{fig:listmodeimaginga}
    \end{subfigure}
    \begin{subfigure}[b]{0.70\textwidth}
    \centering
    \includegraphics[width=\textwidth]{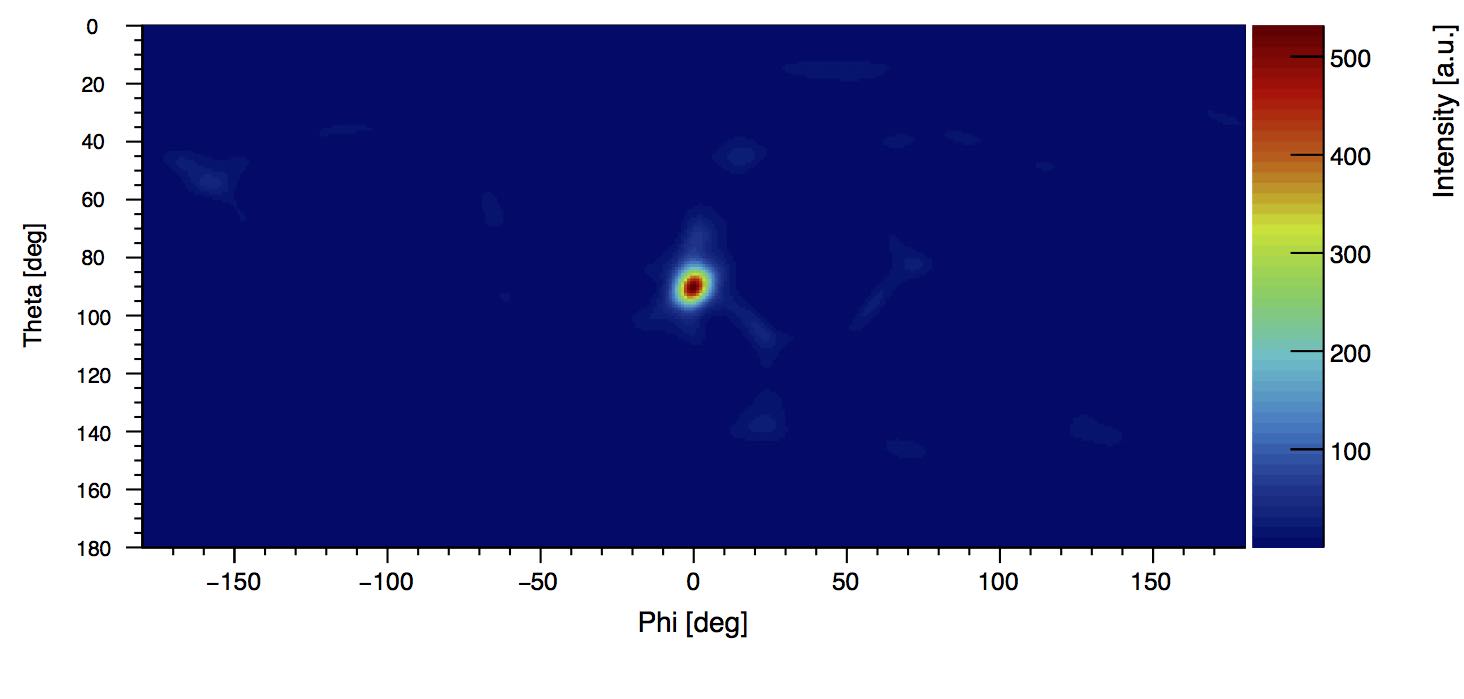}
    \caption{Image obtained after 5 iterations of the LM-ML-EM algorithm.}
    \label{fig:listmodeimagingb}
    \end{subfigure}
    \caption{\textbf{(a)} The back-projection of a 500 keV point source simulation with a Compton telescope detecting 200 events. The event circles overlap at the location of the source in the center of the image, shown with the red hot spot. \textbf{(b)} After 5 iterations of the LM-ML-EM imaging algorithm on the image in (a), the original source is reconstructed. From \citet{kieransthesis}.}
    \label{fig:listmodeimaging}
\end{figure}

The goal of Compton telescope imaging is to recover an accurate measurement of the spatial distribution of source emission. 
The simplest imaging approach for Compton telescopes is the back-projection of event circles onto the sky (see Figure~\ref{fig:compton_basics} and \ref{fig:electron_tracking_images}). 
One can discover the position of gamma-ray sources by detecting regions in which multiple event circles overlap one another; the density of the event circles in each bin can be used to identify sources.
Figure~\ref{fig:listmodeimaginga} shows that the back-projection of 200 simulated gamma-ray events clearly reveals the source in the center of the field-of-view, but there are many additional features in the image due to the extended nature of Compton event circles. Considering the high-background environments in space observations, more sophisticated techniques are needed to reveal the true source distribution.

The measured imaging response of a Compton telescope to a source with intensity distribution $I(l,b)$ can be written~\cite{vonballmoos1989, 2021arXiv210213158Z}:
\begin{equation}
    D(\chi, \psi, \varphi) = R(\chi, \psi, \varphi; l, b) \times I(l,b) + B(\chi, \psi, \varphi)
    \label{eq:imaging}
\end{equation}
where $D$ is the measured data in the Compton Data Space (see Section~\ref{sec:cds}), $R$ is the instrumental imaging response, and $B$ is the measured background in the data space. The response $R$ defines the probability that a photon emitted from the Galactic image-space bin $(l, b)$ is detected in a given data space bin $(\chi, \psi, \varphi)$.
The task of image reconstruction is to invert the measurement process to determine $I(l,b)$. 
As the problem is not directly invertible, iterative approaches must be used.

There are two general classes of iterative imaging reconstruction approaches used for Compton telescopes~\cite{Frandes2016}: binned mode and list mode. 
In binned mode imaging, the imaging response $R$ is determined through laboratory measurements and supplemented by simulations with enough statistics to fill all bins of the data space.
While Equation~\ref{eq:imaging} shows a five-dimensional response (three dimensions for the data space, and two for the image space), each additional parameter used to describe the event (energy, time, recoil electron direction, etc.) adds another dimension to this data space.
List-mode methods~\cite{Barrett:97} use a list of events with all of their measured parameters to define the data space, and the image response, i.e. the back-projected circles or arcs, is calculated on an event-by-event basis. The size of the data space is proportional to the number of events.

One of the main advantages of list-mode imaging is its capacity to save all of the information from photon interactions with full precision. Event parameters that are not traditionally in the image data space, like the distance between the first two interactions or the electron recoil direction, can be saved and used to further select the quality of events.
A common list-mode algorithm used in Compton telescopes is the List Mode Maximum-Likelihood Expectation-Maximization (LM-ML-EM)~\cite{wilderman1998, zoglauer2011}. 
Figure~\ref{fig:listmodeimagingb} shows the results of five iterations of the LM-ML-EM algorithm for the 200 simulated events shown in~\ref{fig:listmodeimaginga}; the expected point source is recovered with very little variation in the field of view outside of the source location.
Other list-mode imaging algorithms include Filtered Back-Projection~\cite{10.1117/12.563905, doi:10.1063/1.4898087} and stochastic origin ensembles~\cite{Andreyev2011, 2020EPJWC.22506006D}, and many variations of these approaches are used.

Unfortunately, there are significant limitations to list-mode imaging, especially for astrophysical observations. 
Without taking into account the full response of the telescope, asymmetries in the data space not modeled in list-mode (for example the peak at large $\Delta\varphi_{ARM}$ in Figure~\ref{fig:armhistogram}) can cause systematics in the imaging.  
Furthermore, list-mode imaging does not allow for the determination of the absolute flux.
So while list-mode imaging is often used for Compton imaging analysis~\cite{zoglauer2011}, especially for terrestrial applications~\cite{10.1117/12.563905}, astrophysical observations strongly benefit from a binned-mode imaging technique.

In a binned image space, the imaging response describes the probability that an event emitted from a Galactic location ($l,b$) in image space is detected in a given bin ($\chi, \psi, \varphi$) in the Compton Data Space. 
Each bin in the response matrix $R$ must be filled with sufficient statistics, and therefore, computationally intensive simulations are required. 
The simulated response matrix must however be calibrated and benchmarked with high-statistics measurements of laboratory sources to understand systematics.
Additional dimensions can be included, such as the gamma-ray energy, time, and recoil electron scatter direction, to better discern the source emission, but the number of bins can quickly scale beyond the capabilities of state-of-the-art personal computers. 

\begin{figure}[tb]
    \centering
    \begin{minipage}{0.49\textwidth}
    \centering
    \includegraphics[width=\textwidth]{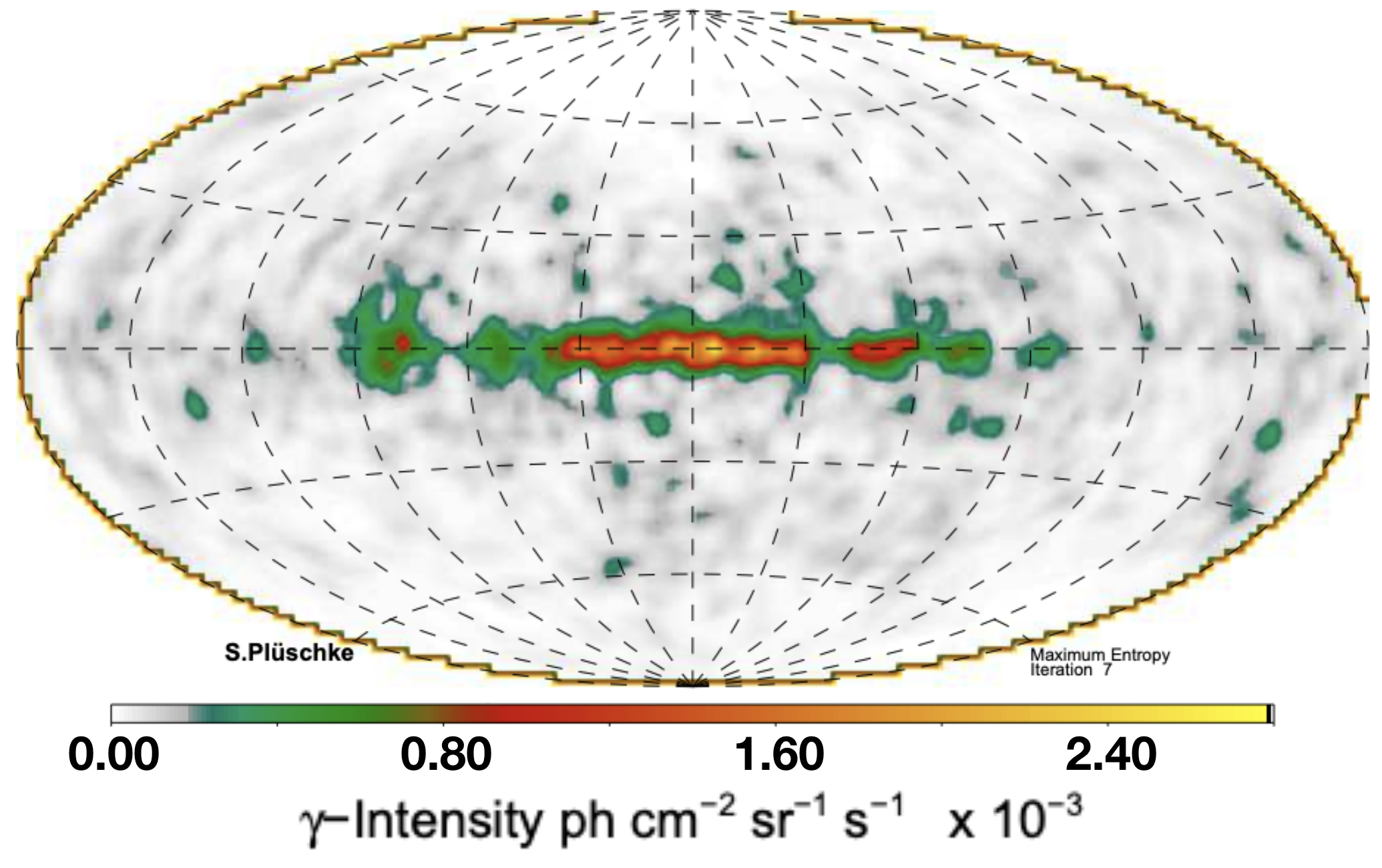}
    \subcaption{MEM image reconstruction}\label{fig:al26mem}
    \end{minipage}
    \begin{minipage}{0.49\textwidth}
    \centering
    \includegraphics[width=\textwidth]{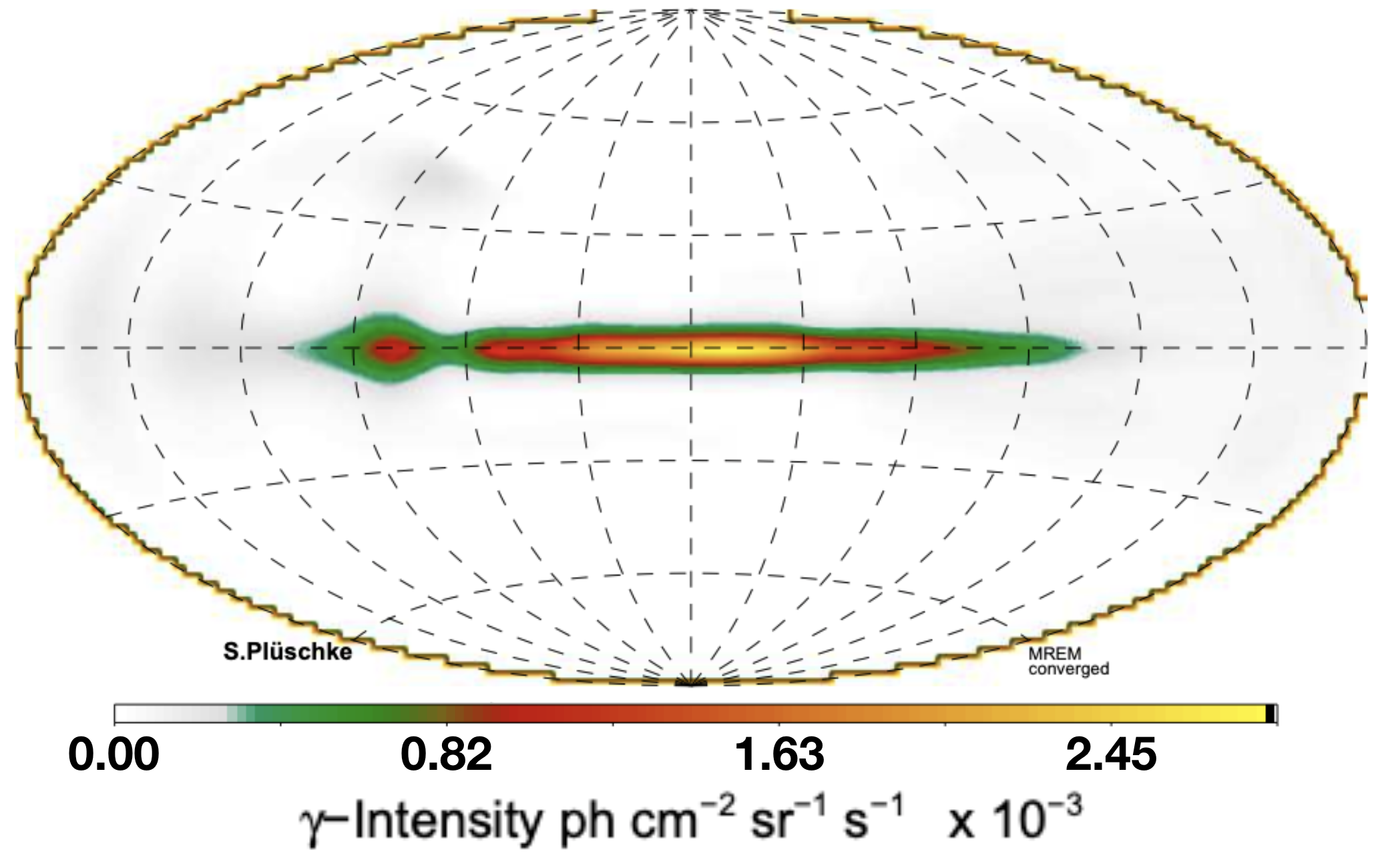}
    \subcaption{MREM image reconstruction}\label{fig:al26mrem}
    \end{minipage}
    \caption{The $^{26}$Al emission measured by COMPTEL over 9 years shows emission along the Galactic disk with two different imaging techniques. \textbf{(a)} The Maximum Entropy Method (MEM) creates a lumpy, structured image enhancing point-source and noise features. \textbf{(b)} The Multi-resolution Regularized Expectation Maximization (MREM) algorithm was developed to create the smoothest distribution consistent with the observed data. Both images have the same underlying data. From \citet{2001ESASP.459...55P}.}
    \label{fig:comptelimaging}
\end{figure}

A number of different binned-mode algorithms have been developed for Compton imaging~\cite{2014NIMPA.760...46I}.
COMPTEL predominantly used the Maximum Entropy Method (MEM)~\cite{strong1990} and the Richardson-Lucy approach~\cite{richardson1972, lucy1974} for many of their imaging analyses (e.g. see \cite{knodlseder1996, strong1995, 1995A&A...298..445D}). 
Additionally, the Multi-resolution Regularized Expectation Maximization (MREM) approach was developed to suppress noise structures in the imaging~\cite{knodlseder1999}. 
Each algorithm has different advantages: the Richardson-Lucy approach is better suited for point sources, the MREM method is optimized for very smooth images, and MEM is a middle ground~\cite{2021arXiv210213158Z}.
Figure~\ref{fig:comptelimaging} shows a comparison for the $^{26}$Al emissions as measured by COMPTEL using the MEM and MREM approaches. The underlying data for both of these images are the same, and while there are similarities in the large-scale emission, they show different levels of small scale emission features due to the optimizations of the iterative methods.

The broad range of Compton imaging techniques all have slightly different advantages. List-mode approaches give a straightforward way to find bright point sources, without needing a predefined imaging response, and are sufficient for terrestrial applications and quick checks in astronomical observations. The most accurate and sensitive imaging is achieved through binned-mode algorithms which can account for any measurements systematics in the imaging response. Different binned-mode techniques are better suited for enhancing point source emission features, while others are optimized for smooth extended emission. Additionally, model fitting and forward-folding approaches are also used for Compton telescope imaging analysis.
Depending on the observation, the best practice is to compare multiple imaging methods to achieve a clear understanding of the underlying emission.

\subsection{Polarization Capabilities}
\label{sec:polarization_caps}

As introduced in Section~\ref{sec:polarimeter_basics}, Compton telescopes are inherently sensitive to polarization. 
The response of a polarized source is defined by a sinusoidal modulation in the azimuthal Compton scattering angle $\eta$ from Equation~\ref{eq:klein-nishina-pol}. 

\begin{figure}[tb]
    \centering
    \includegraphics[width=0.7\textwidth]{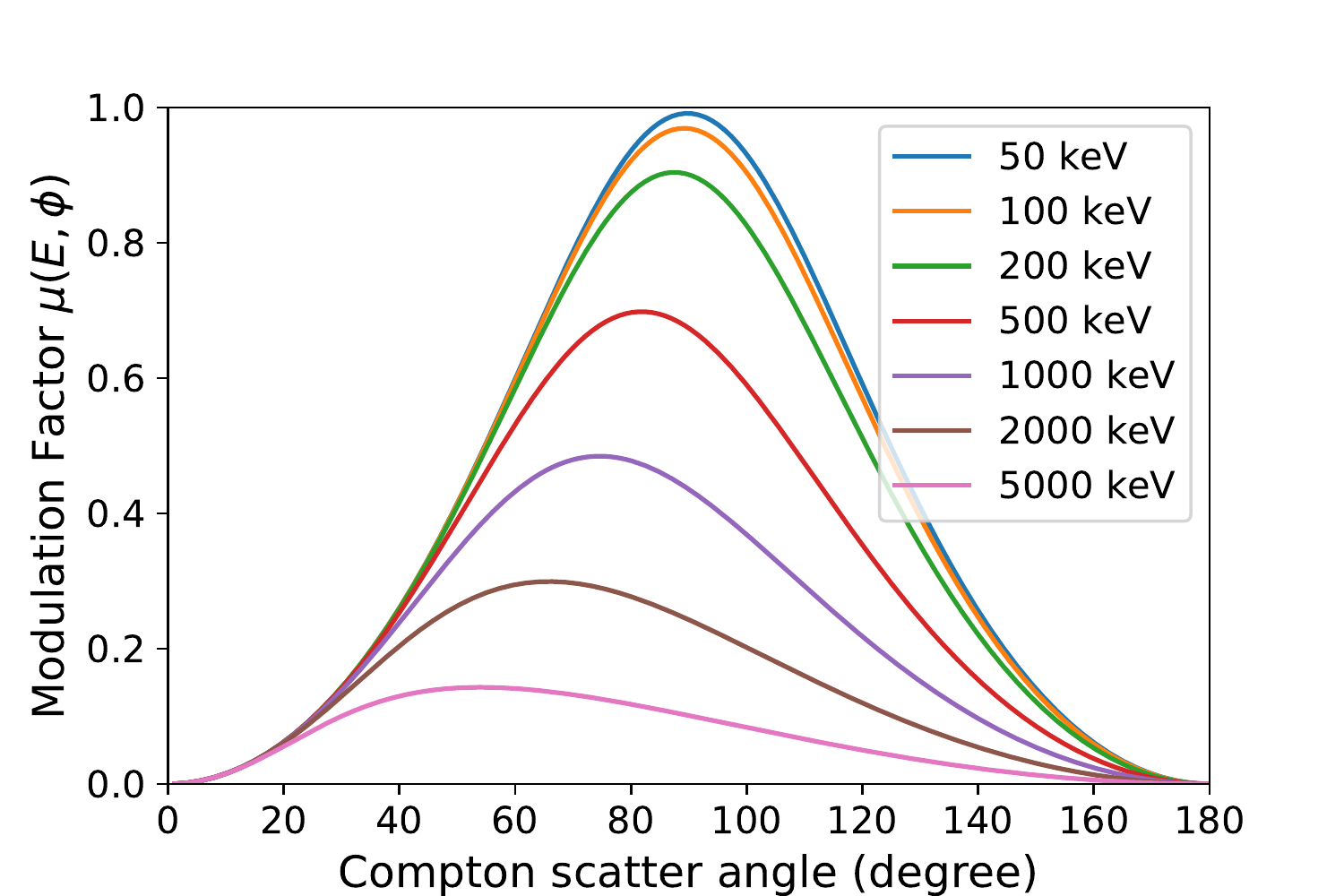}
    \caption{The modulation factor describes the amplitude of the polarization response. The modulation factor, and therefore the polarization response, is maximized at Compton scattering angles $\varphi = 90^{\circ}$ and for lower incident photon energies. Adapted from \citet{lei1997}.}
    \label{fig:modulationfactor}
\end{figure}

The standard method for polarization analysis is to fit the histogram of azimuthal scattering angles $\eta$ with a cosine function to determine the polarzation fraction and polarization angle of the incident photons~\cite{lei1997, Tatischeff2019}. 
The modulation amplitude of the azimuthal scattering angle distribution (ASAD), depicted in Figure~\ref{fig:ASAD}, is directly proportional to the polarization fraction of the incident beam.
The level of polarization is quantified by the modulation factor:
\begin{equation}
    \mu = \frac{A}{B} = \frac{C_{max} - C_{min}}{C_{max} + C_{min}}
\end{equation}
 where $A$ is the amplitude of the sinusoidal response, and $B$ is the offset, and $C_{max}$ and $C_{min}$ are the maximum and minimum values in count space as indicated in Figure~\ref{fig:ASAD}. Since the maximum modulation response $C_{max}$ occurs when $\eta = 90^{\circ}$, and the minimum at $\eta = 0^{\circ}$, we can evaluate Equation~\ref{eq:klein-nishina-pol} to find the modulation factor as a function of the energy and Compton scattering angle: 
\begin{equation}
    \mu(E, \varphi) = \frac{\frac{d\sigma}{d\Omega} (\eta = 90^{\circ}) - \frac{d\sigma}{d\Omega} (\eta = 0^{\circ})}{\frac{d\sigma}{d\Omega} (\eta = 90^{\circ}) + \frac{d\sigma}{d\Omega}(\eta = 0^{\circ})} 
    = \frac{\sin^2\varphi}{\frac{E_{scat}}{E_0} + \frac{E_0}{E_{scat}} - \sin^2\varphi}.
\end{equation}
This relation is shown in Figure~\ref{fig:modulationfactor}.
The modulation from a polarized source is largest at lower energies and for Compton scattering angles $\varphi \sim 90^{\circ}$. 
Geometric effects in the detector and asymmetries in the instrument response can give rise to false modulation signals in the ASAD; therefore, it is common to normalize the measured ASAD with a simulated ASAD from unpolarized emission.

In the standard analysis, the degree of linear polarization $\Pi$ is derived by normalizing $\mu$ in such a way that
\begin{equation}
    \Pi = \frac{\mu}{\mu_{100}},
\end{equation}
where $\mu_{100}$ is the modulation factor for 100\% linearly polarized photons.

The polarization sensitivity of a telescope is generally reported in terms of the minimum detectable polarization (MDP). The MDP defines the polarization for which there is only a small chance that the measured modulation is due to statistical fluctuations:
\begin{equation}
    \text{MDP} = \frac{4.29}{\mu_{100}S}\sqrt{\frac{S+B}{T_{obs}}}
\end{equation}
The 4.29 factor corresponds to a 99\% confidence level~\cite{2010SPIE.7732E..0EW}, $\mu_{100}$ is the modulation factor from a 100\% polarized source determined by simulations, $S$ and $B$ are the source and background count rates, respectively, and $T_{obs}$ is the observation time.

The general steps for the standard Compton polarization analysis method are as follows:
\begin{enumerate}
\setlength{\itemindent}{1em} 
    \item Simulate 100\% polarized source to find $\mu_{100}$.
    \item Determine event selections that give the lowest MDP.
    \item Subtract background ASAD from the measured ASAD.
    \item Simulate unpolarized source and divide background-subtracted ASAD by the unpolarized ASAD to correct for geometric systematics.
    \item Fit background-subtracted and geometry-corrected ASAD with
    \begin{equation}
        C(\eta) = A\cos(2(\eta - \eta_0)) + B,
    \end{equation}
    where $A$, $B$, and $\eta_0$ are free parameters. The modulation factor is given by $\mu = \frac{A}{B}$.
    \item Scale the modulation factor by $\mu_{100}$ to find the polarization fraction $\Pi = \frac{\mu}{\mu_{100}}$.
\end{enumerate}
The standard method, and other more sophisticated techniques, rely heavily on simulations to find the response to an unpolarized and 100\% polarized source. Therefore, instrument simulations need to be benchmarked with real measurements and pre-flight polarization calibrations are essential~\cite{2005NIMPA.540..158M, 2009NIMPA.600..424B, 8069846}. As systematic errors almost always result in an overestimation of polarization, and off-axis emissions tend to produce false modulation signals, ASAD measurements  of unpolarized sources are essential to understand  systematics.

As the number of gamma-ray polarization measurements is rapidly increasing~\cite{Chattopadhyay_2019, 2020A&A...644A.124K}, the analysis techniques are quickly becoming more sensitive and sophisticated.
For telescopes with three-dimensional position resolution that measure the Compton polar angle in addition to the azimuthal scattering angle, \citet{KRAWCZYNSKI2011784} show that a significant improvement in the sensitivity is achieved when both the polar and azimuthal scattering angles are used with a Maximum Likelihood analysis approach. This has recently been demonstrated for real measured data with the COSI balloon instrument~\cite{lowell2017a, lowell2017b}.
Analysis for the dedicated GRB polarimeter POLAR has leveraged the  existing Multi-Mission Maximum Likelihood (3ML) framework to simultaneously model the spectral and polarization data~\cite{2019A&A...627A.105B, 2020A&A...644A.124K}. 
\citet{2015NIMPA.799..155B} show similar improvements in the sensitivity by using a moments method to compute a simple estimator
of the polarization fraction.

\subsection{Limitations and Challenges}

While the science in the MeV range is enticing and continues to drive the scientific community, observations in this range have been slow to progress compared to other wavelengths (see Figure~\ref{fig:MeVGap}).
This is due to a number of reasons that have been addressed throughout this Chapter. 
The low Compton scattering cross section (see Figure~\ref{fig:mass_attenuation_coefficients}) requires large detector volumes to fully contain the high-energy photon events, which can be difficult to instrument with sufficient position resolution while maintaining a good energy response.
The incident photon direction can only be constrained to a large circle on the sky, or at best an event arc for instruments with tracking capabilities (Section~\ref{sec:operating}).
Therefore, while Compton telescopes provide powerful single-photon detection, they are not true direct imagers.
With each photon depositing its energy in multiple discrete interactions throughout the detector volume, sophisticated algorithms are required for event sequencing and reconstruction (see Section~\ref{sec:eventrecon}) to determine the initial photon parameters.
The high-level analysis for Compton telescopes requires the use of large data spaces with a heavy reliance on simulations to understand the instrument response (see Section~\ref{sec:imaging}). 
And finally, even an ideal Compton telescope, with infinite position and energy resolution, as well as perfect event reconstruction and response determination, has only a modest angular resolution since there is a fundamental limit based on the unknowable momentum of bound electrons (see Section~\ref{sec:doppler}).
A further complication for Compton telescopes is the high background radiation that is prevalent in the MeV range.

\subsubsection{Background Radiation}
\label{sec:background}

Background radiation is one of the dominant factors that determines a Compton telescope's sensitivity (see Section~\ref{sec:sens}). 
The large volume detectors that are required to fully absorb Compton events, and the large FOV, result in  large  background signals~\cite{vonBallmoos2005}.

\begin{figure}[b]
    \centering
    \includegraphics[width=0.9\textwidth]{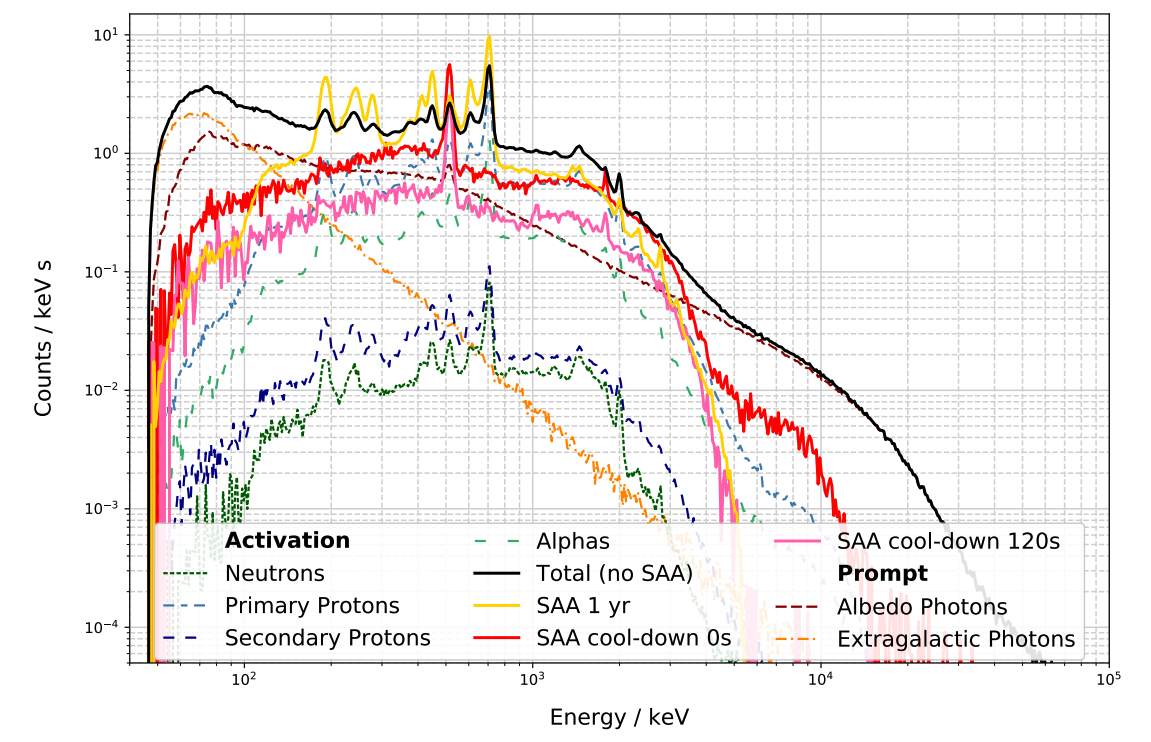}
    \caption{Simulated spectra of the reconstructed background Compton events show the separate background contributions of the eASTROGAM mission concept. An orbit of $15^{\circ}$ is used, but lower inclinations $\lesssim 5^{\circ}$ have been shown to reduce the SAA induced background.  Figure from~\citet{2019ExA....47..273C}.}
    \label{fig:cumanibkg}
\end{figure}

The nuclear activation lines that are produced in interesting cosmic sources can also plague the instrument as background lines. 
After high-energy cosmic-ray collisions occur within the instrument mass, which is inevitable at balloon-altitudes and in satellite orbits, the material around the detector will become activated and the radiation from these instrumental nuclear decays contribute a source of significant background. 
Furthermore, cosmic-ray interactions with particles in the atmosphere will produce a broadband spectrum of gamma rays that is orders of magnitude more intense than Galactic sources. 
This atmospheric emission, referred to as albedo radiation, is strongest near the
horizon.

\citet{2019ExA....47..273C} present an overview of the background components from 10~keV to 100~GeV in low-Earth orbit. Simulations of the response of a proposed gamma-ray mission (eASTROGAM \cite{2017ExA....44...25D}) are performed to understand the contributions of each background component, including activation from charged cosmic rays and the passage through the South Atlantic Anomaly (SAA). Figure~\ref{fig:cumanibkg} shows the reconstructed  background Compton events for an orbit inclination of 15$^{\circ}$.



A number of lessons about the reduction of background radiation have been learned from previous MeV telescope designs \cite{schonfelder2003}: anti-coincidence detectors (ACDs) surrounding the Compton telescope can significantly reduce the background by vetoing signals caused by cosmic-rays; 
low-activation materials near the detector element will reduce instrumental background; 
and discrimination of neutrons and cosmic rays, either in electronics (pulse-shape identifiers), through vetoing, or in the software analysis chain can lower the dead-time and false identification of gamma-ray signals. 
Electron tracking capabilities and selections on the quality of the events also provide a way to reduce the background. 
Additionally, detailed simulations for a proper benchmarking of the instrument prior to launch are essential to understand the source and background response of the telescope.


\section{Notable Compton Telescope Designs}
\label{sec:instruments}

The development of Compton telescopes began in the 1960s and 1970s and there have been significant improvements in design and technologies since then. In this section, the focus will be on notable, mature instruments with space or balloon-flight experience, or substantial laboratory measurements and calibration campaigns.
This is not meant to be an exhaustive list of all instrument development efforts, but a representative sample.
Reference~\cite{2017SSRv..212..429C, Tatischeff2019, vonBallmoos2005} for an overview of past instruments, \cite{2022arXiv220307360E} for a summary of some future missions, and other chapters in this handbook for further details on past and present Compton telescope designs. 

The variety of Compton telescopes is extensive, with different detector materials, geometries, spatial and spectral resolution capabilities, and general design optimizations. We have attempted to organize the many development efforts into four main categories: semiconductor-based Compton imagers (Section~\ref{sec:semiconductorinstruments}), gaseous/liquid detectors (Section~\ref{sec:gasdetectors}), dedicated polarimeters (Section~\ref{sec:polarimeterdetectors}), and instruments capable of detecting both Compton scattering and pair creation  (Section~\ref{sec:comptonandpairdetectors}). We also include a brief description of instrument applications in other fields, such as medical imaging and nuclear defense, in Section~\ref{sec:applicationsinotherfields}.

\subsection{Semiconductor-based Compton Imagers}
\label{sec:semiconductorinstruments}

\begin{figure}[tb]
    \centering
    \begin{minipage}{0.34\textwidth}
    \centering
    \includegraphics[width=0.95\textwidth]{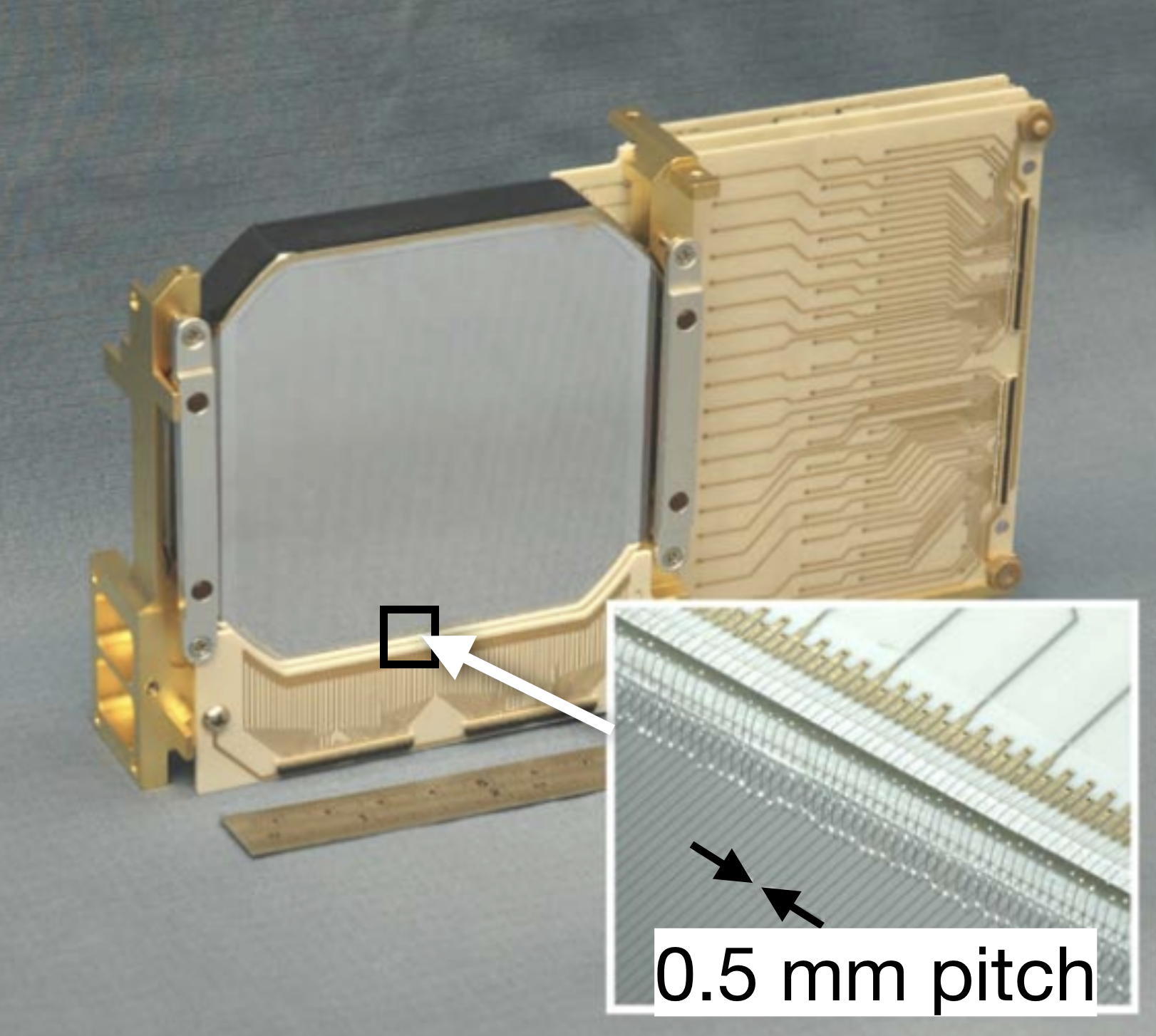}
    \subcaption{Double-sided strip Ge detector.}\label{fig:gesemiconductor}    
    \end{minipage}
    \hfill
    \begin{minipage}{0.31\textwidth}
    \centering
    \includegraphics[width=\textwidth]{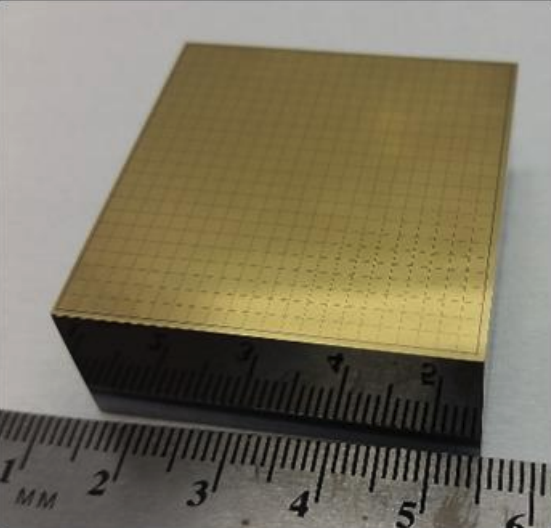}
    \subcaption{Pixelated CZT detector.}\label{fig:cztsemiconductor}    
    \end{minipage}
    \hfill
    \begin{minipage}{0.32\textwidth}
    \centering
    \includegraphics[width=0.95\textwidth]{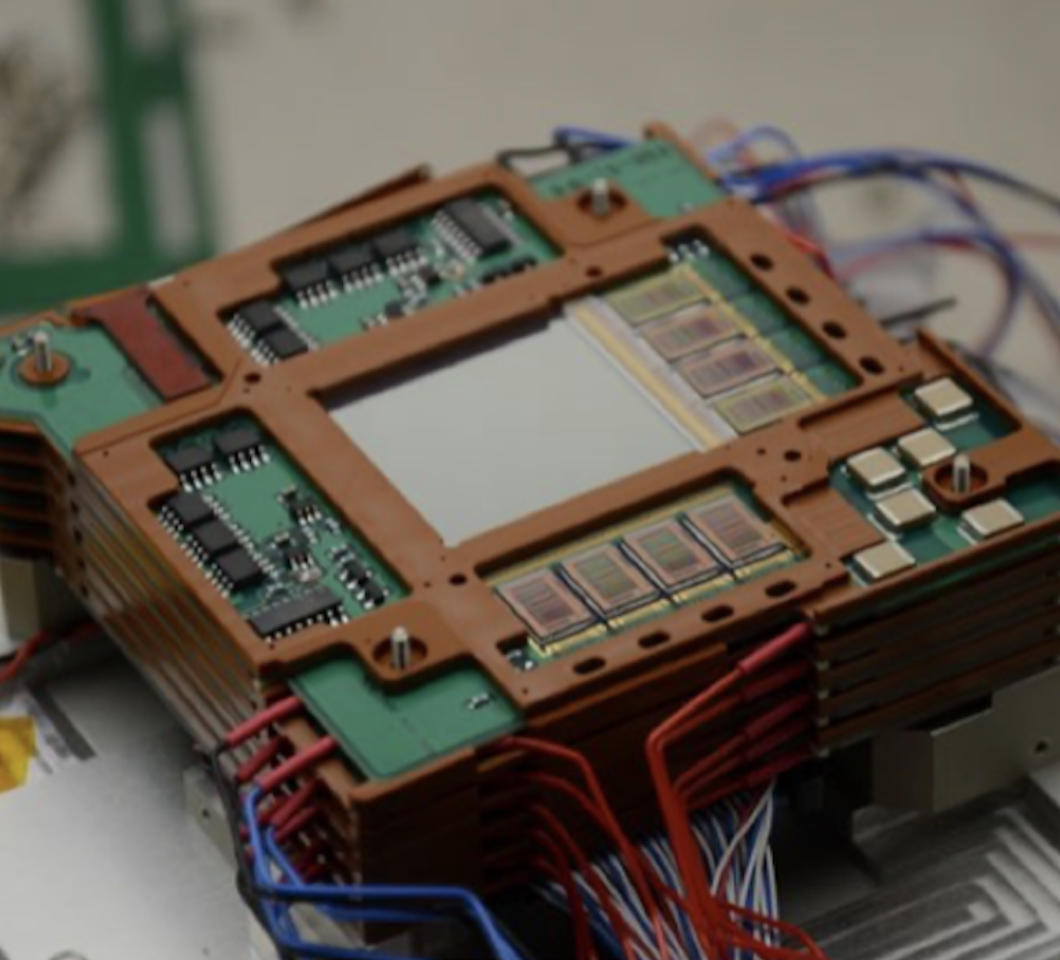}
    \subcaption{Multi-layer double-sided strip Si and CdTe detectors.}\label{fig:sisemiconductor}    
    \end{minipage}
    \caption{
    \textbf{(a)} Large volume double-sided strip germanium detectors ($8 \times 8 \times 1.5$~cm$^3$) have been developed with 0.5~mm strip pitch for the Gamma-Ray Imager/Polarimeter for Solar flares (GRIPS) instrument~\cite{10.1117/12.926450}. Modified from~\cite{https://doi.org/10.48550/arxiv.2006.05471}.
    \textbf{(b)} Large volume CZT detectors are an emerging technology. The ($4\times4\times1.5$~cm$^3$) monolithic detector with 1.8~mm pitch pixelated electrodes has demonstrated excellent position and energy resolution. Figure from~\cite{10.1117/12.2321244}.
    \textbf{(c)} As silicon detectors cannot be made thicker than $\sim$1~mm, multi-layer designs are more prevalent. An instrument which combines layers of $32\times32$~mm double-sided strip silicon and double-sided strip CdTe with positional resolution of 250~$\mu$m has been used in environmental monitoring~\cite{Takahashi_2012,Takeda_2015} and  medical imaging~\cite{GoroYabu2021,YabuPhD}.}
    \label{fig:semiconductor_examples}
\end{figure}


Many of the modern Compton imaging telescopes employ semiconductor detectors. The desired energy resolution is inherent to these detectors, and the two or three-dimensional good position resolution within each detector volume is achieved by segmented electrodes.  
The most common semiconductor materials used are germanium, CdTe, CZT, and silicon, and three examples of Compton telescope detector designs with these materials are shown in Figure~\ref{fig:semiconductor_examples}.
Recent work into artificial diamond detectors has shown excellent energy, position, and most notably timing resolution, that can potentially enable a compact and efficient Compton telescope using time-of-flight~\cite{9059643}.
One challenge with semiconductor detectors is the difficulty to make large volumes, and the detection efficiency is limited compared to Compton telescopes using scintillators.

There are two general designs for semiconductor Compton telescopes: multi-layer silicon detectors with 2D position resolution, and thick, large-volume Ge or CZT detectors with internal 3D position sensitivity. We highlight examples of each concept here by focusing on the Soft Gamma-ray Detector on board Hitomi, which employed a multi-layer design, and the newly selected Compton Spectrometer and Imager, which utilizes thick germanium detectors.



\subsubsection{Soft Gamma-ray Detector on Hitomi}
\label{sec:collimatedinstruments}


The Soft Gamma-ray Detector (SGD) on board the Hitomi satellite was designed to cover an energy range of $60-600$~keV with a sensitivity 10 times better than that of the {\em Suzaku} Hard X-ray Detector~\cite{takahashi2018hitomi,Hitomi_SGD}. 
The SGD combined a multi-layer Compton telescope with an active well-type shield to reduce the FOV and significantly increase the signal-to-background ratio~\cite{ICHINOHE20165}. 
Hitomi was launched in February 2016, and sadly the SGD only had a few days of nominal observations before the spacecraft ceased to function in March 2016.

\begin{figure}[tb]
    \centering
    \includegraphics[width=\textwidth]{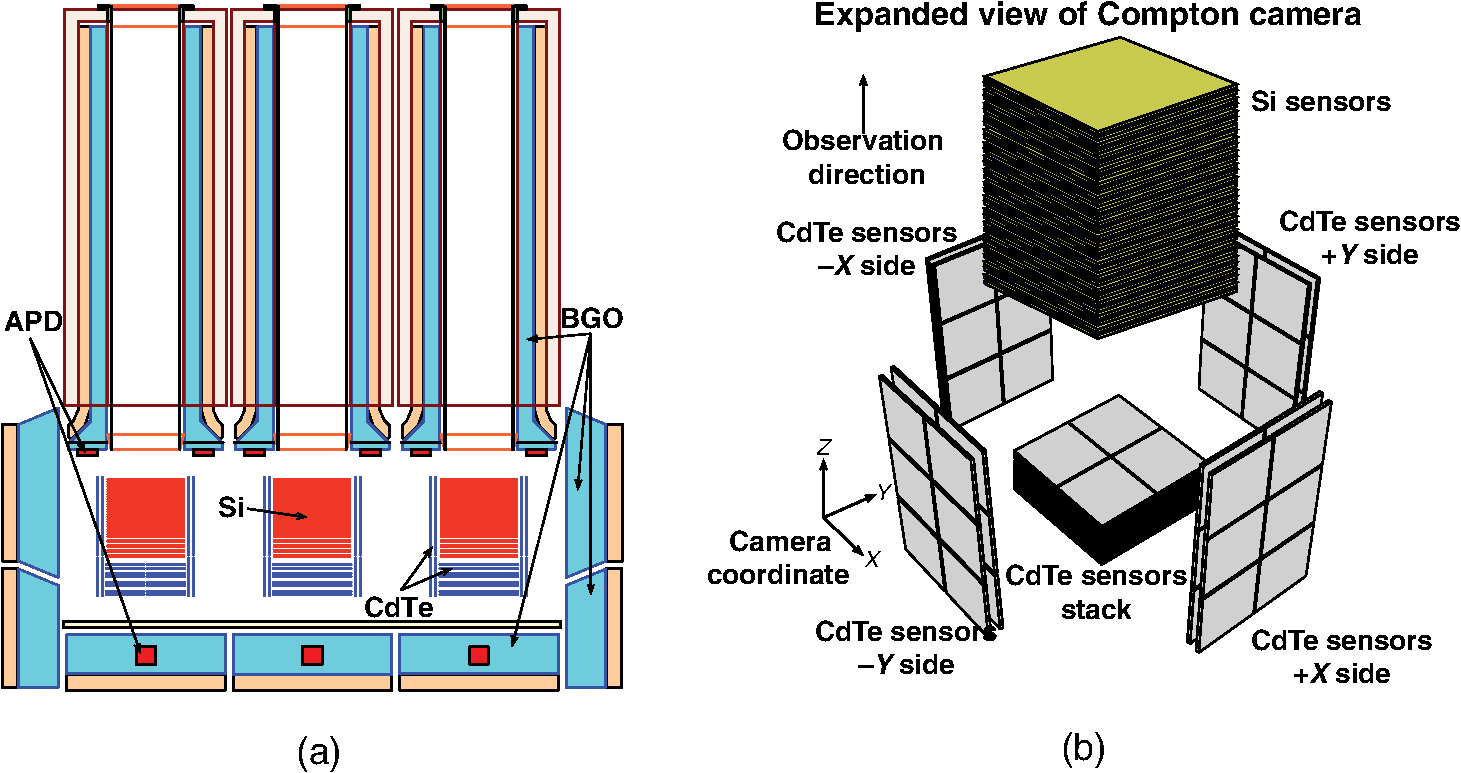}
    \caption{\textbf{(a)} The Hitomi SGD combined a multi-layer Compton telescope with an active BGO well-type shield to achieve excellent background rejection from 60-600~keV. Three identical Compton telescope modules composed the SGD. 
    \textbf{(b)} The Compton camera was composed of 32 layers of silicon pad detectors and 8 layers of CdTe pad detectors, both with a pixel pitch of 3.2~mm. Additional CdTe sensors surrounded the silicon detectors to increase the detection efficiency for events with large Compton scattering angles. Figure from~\citet{Hitomi_SGD}.}
    \label{fig:Hitomi_SGD}
\end{figure}

The Compton telescope for the SGD was made of 32 layers of 0.625~mm thick silicon pad detectors and 8 layers of 0.75~mm thick CdTe pad detectors, as shown in Figure~\ref{fig:Hitomi_SGD}. 
The sides of the silicon detectors were also surrounded by two layers of CdTe detectors to measure the Compton-scattered photons with large scattering angles. 
Each silicon detector had a $16\times16$ array of $3.2 \times 3.2$~mm$^2$ pads (called pads instead of pixels due to their size), and the CdTe sensors, which are one quarter of the size of the silicon, had an $8\times8$ array of $3.2 \times 3.2$~mm$^2$ pads.
The design combined a low-Z material (Si) optimized for Compton scattering and a high-Z material (CdTe) to act as an efficient absorber to obtain high detection efficiency. 
Photons which Compton scattered in the silicon detector and were fully absorbed in the CdTe sensors were used for Compton imaging, but in principle each layer could act as a scatterer or absorber.
An additional advantage of the silicon sensors was the limited effect of Doppler broadening (see Section~\ref{sec:doppler}).

The multi-layer silicon and CdTe detector was mounted at the bottom of a well-type active shield, as shown in Figure~\ref{fig:Hitomi_SGD}.   
The major advantage the narrow FOV was enhanced background rejection.
If the reconstructed event circle for a specific event did not align with the FOV defined by the active-well shield, then that event was considered background and rejected. 
Most of the background could be rejected by requiring this condition.
The opening angle provided by the bismuth germanate (BGO) shield was $\sim$10$^{\circ}$ at 500~keV. 
An additional fine phosphor bronze collimator installed in the opening of the BGO well restricted the FOV of the telescope to 33$^\prime$ for photons below 100 keV,  to minimize the flux from the cosmic X-ray background in the FOV. 

\begin{figure}[tb]
    \begin{minipage}{0.45\textwidth}
    \centering
    \includegraphics[width=\textwidth]{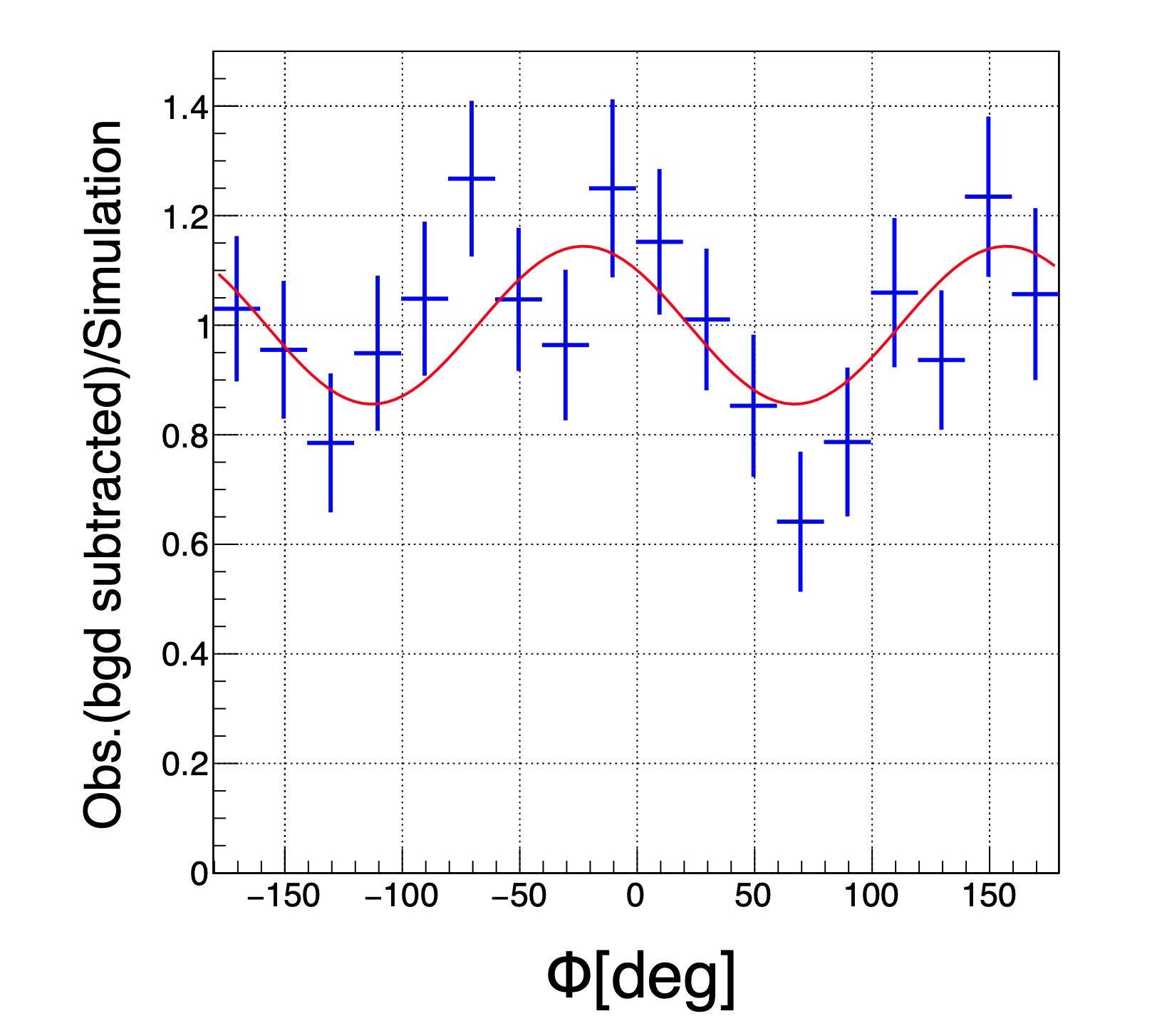}
    \end{minipage}
\hfill
    \begin{minipage}[b!]{0.55\textwidth}
    \captionof{figure}{The SGD observed the Crab nebula for a total of 5~ks during the testing phase of Hitomi. With this observation, the polarization capabilities and functionality of the SGD were demonstrated. The background-subtracted ASAD from the Crab observations shown here corresponds to a polarization fraction of $22\pm11$\% with a polarization angle of $111\pm 13^\circ$ in the energy range of 60--160 keV. Figure from \citet{hitomi2018detection}.}
    \label{fig:sgd_crab_polarization}
\end{minipage}
\end{figure}

The SGD was also designed to function as a Compton polarimeter~\cite{KATSUTA201651}. Fortunately, it observed the Crab nebula for 5~ks during the initial Hitomi test phase~\cite{hitomi2018detection}.
From this short detection, the polarization fraction of the phase-integrated Crab emission (sum of pulsar and nebula emissions) was found to be $22\pm11$\% with the polarization angle being $111\pm 13^\circ$ in the energy range of 60--160 keV (errors correspond to the 1$\sigma$ deviation). 
Figure~\ref{fig:sgd_crab_polarization} shows the measured background-subtracted ASAD for the SGD Crab measurement, where the analysis procedure was similar to that described in Section~\ref{sec:polarization_caps}.
The polarization angle measured by SGD  is one sigma from the projected spin axis of the pulsar, $124.0^\circ \pm 0.1^\circ$.
While the observation time was much less than planned, the results show the capability of polarization measurement with a multi-layered semiconductor Compton telescope and the functionality of the SGD in flight.


\subsubsection{The Compton Spectrometer and Imager}
\label{sec:thicksemiconductorinstruments}

The Compton Spectrometer and Imager (COSI) is a Compton telescope that was recently selected to be NASA’s next Small Explorer mission with a launch in 2026~\cite{Tomsick2021}. 
COSI is sensitive in the 0.2-5 MeV range and is designed to study sources of line emissions, such as Galactic positrons and nucleosynthesis, and perform polarization studies of gamma-ray bursts and compact objects.
The heart of COSI consists of an array of 16 high-purity double-sided strip germanium detectors which provide excellent energy resolution and $\sim$1.5~mm$^{3}$ position resolution.
The Compton imaging capabilities give COSI a narrow line sensitivity that is better than INTEGRAL/SPI, despite its smaller size.
The COSI instrument was advanced through technology development as a balloon-borne instrument~\cite{2017arXiv170105558K, Beechert_2022}, and measurements from balloon flights provide a proof of principle of the telescope capabilities~\cite{2021arXiv210213158Z, lowell2017b, Siegert_2020}.

\begin{figure}[t]
    \centering
    \includegraphics[width=\textwidth]{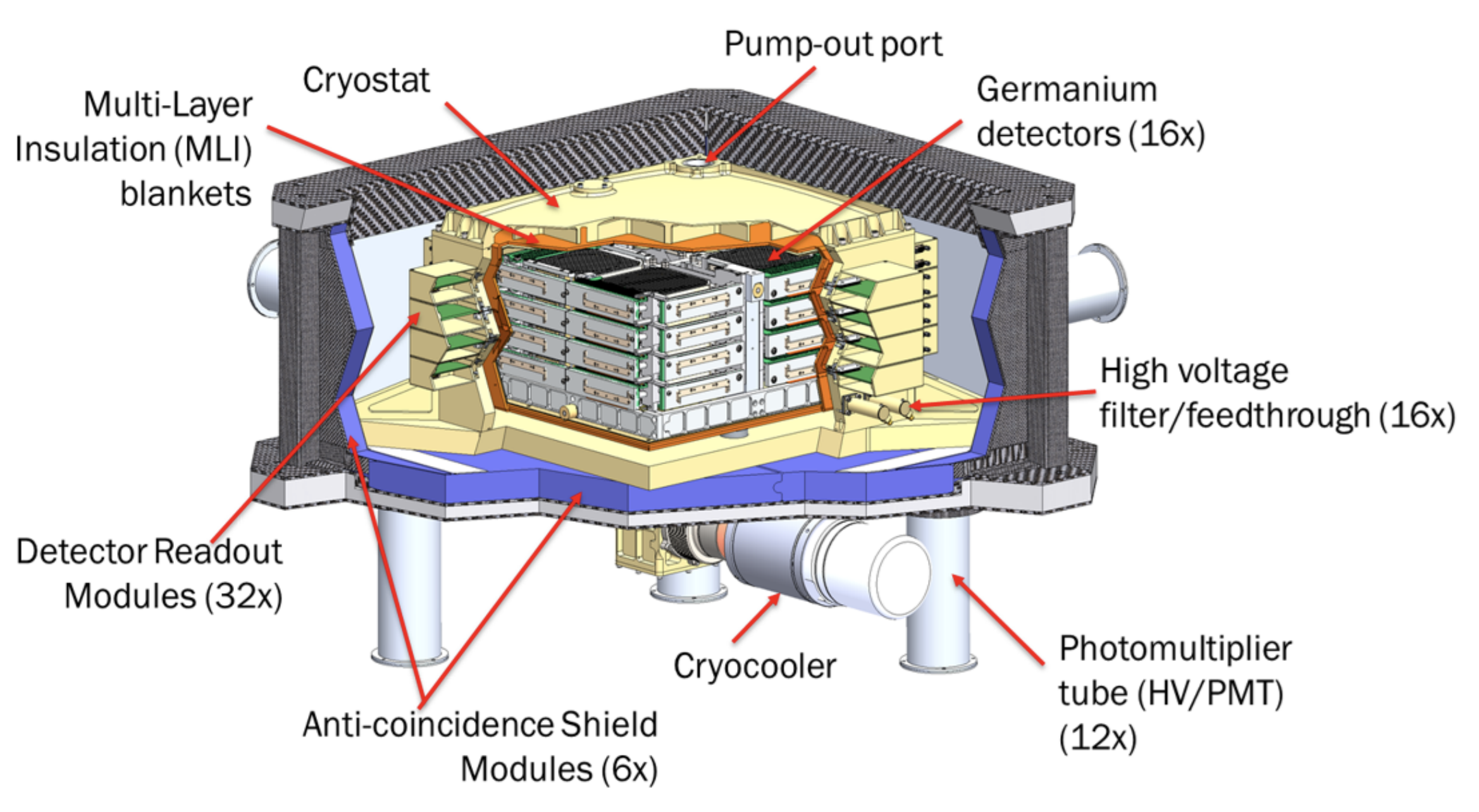}
    \caption{COSI uses 16 double-sided strip germanium detectors to measure Compton interaction with high spectral and spatial resolution. Each detector has 1.16~mm strip pitch and a $<1.5$~mm$^3$ internal position resolution. The detector array is housed in an aluminum cryostat and cooled to liquid nitrogen temperatures with a mechanical cryocooler. The cryostat is surrounded on five sides with an anti-coincidence shield to effectively constrain the FOV and reduce the background. Figure from \citet{Tomsick2021}.}
    \label{fig:COSI}
\end{figure}

The COSI instrument is shown in Figure~\ref{fig:COSI}. The 16 germanium detectors are each $8\times8\times1.5$~cm$^{3}$ with a strip pitch of 1.16~mm (see Figure~\ref{fig:gesemiconductor}). 
Each of the COSI germanium detectors can act as a Compton imager alone, but combined in an array increases the active detector volume and detection efficiency.
While the germanium gives COSI its excellent energy resolution, it must be cooled to liquid nitrogen temperatures. 
The COSI detector array is housed in an evacuated cryostat, with readout electronics mounted  externally, and detectors are cooled by a mechanical cryocooler. 

The COSI cryostat is surrounded on five sides by a BGO anti-coincidence shield read out with photomultiplier tubes (PMTs). The BGO shield effectively constrains the instantaneous FOV to 25\% of the sky, and reduces the background.
Interactions detected in the BGO shields are used to veto photons from outside the FOV, and reject events that may not be fully absorbed in the germanium array. 
With the all-sky FOV and high detection efficiency, the BGO detectors act as a sensitive transient monitor.

\subsection{Gaseous and Liquid Time Projection Chambers}
\label{sec:gasdetectors}

The large sensitive detector volumes needed for Compton imaging are easily achieved with gaseous and liquid time projection chambers (TPCs), and these designs have been pursued since the early 1990's alongside semiconductor concepts.
Gaseous and liquid TPCs require significantly fewer electronics channels, and therefore lower power consumption, compared to the multi-layer or segmented design of semiconductor-based instruments. 
Liquid designs provide high detection efficiency, while the tracks of Compton recoil electrons in  low density gaseous TPCs are long and easy to reconstruct. Therefore, the additional background rejection capabilities of electron tracking is an additional design driver.
The most common designs use either liquid or gaseous argon or xenon~\cite{tanimori2004mev, Aramaki2020,9059643, Hunter_2014}.



When a gamma ray interacts with liquid argon or xenon via Compton scattering, it produces an ionization cloud and simultaneously generates ultraviolet scintillation light. Liquid argon and liquid xenon TPCs (LArTPC and LXeTPC) generally measure the scintillation light with PMTs or silicon photomultipliers (SiPMs) and the ionization tracks of electrons with two dimensional electrodes, such as wires or segmented electrode pads~\cite{Aramaki2020}. 
These detectors have three-dimensional spatial sensitivity due to the combination of the horizontal electrodes and the drift time between the light and the electron signals. The first fully realized TPC Compton imaging instrument was LXeGRIT.



\subsubsection{Liquid Xenon Gamma-Ray Imaging Telescope}

\begin{figure}[t]
    \centering
    \includegraphics[width=0.8\textwidth]{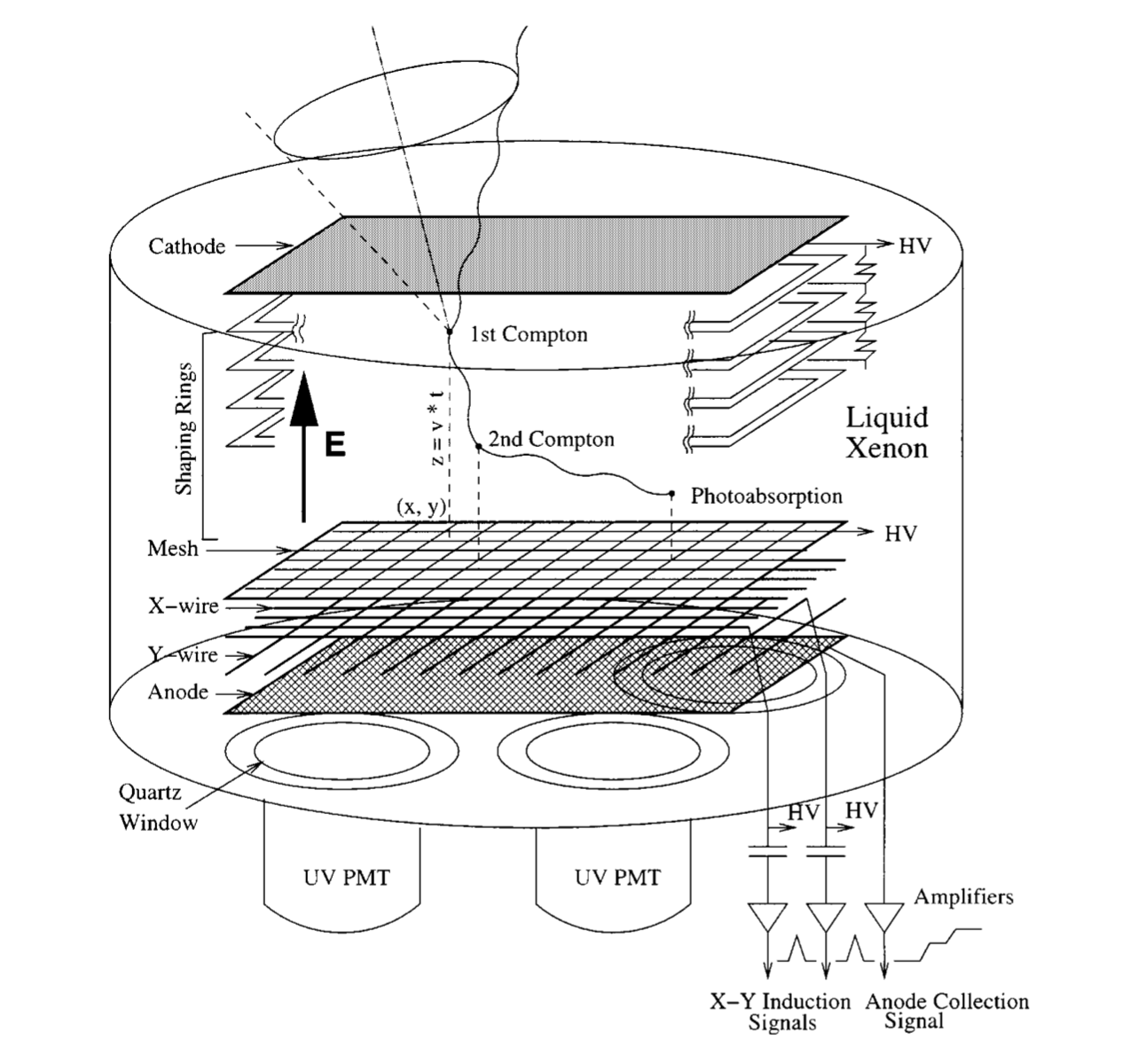}
    \caption{The LXeGRIT time projection chamber used UV scintillation light to provide the event trigger and timing, and the drift of the ionization charge cloud past orthogonal sensing wires gave a measure of the X-Y position of the interaction. The drift time gave the interaction depth. Figure from \citet{2001NIMPA.461..256A}.}
    \label{fig:LXeGRIT}
\end{figure}

The Liquid Xenon Gamma-Ray Imaging Telescope (LXeGRIT) was developed at Columbia University and the University of New Hampshire~\cite{10.1117/12.187266} in the 1990's and early 2000's. The instrument was based on a high-purity liquid xenon time projection chamber, as shown in Figure \ref{fig:LXeGRIT}.

The detector measured ionization and scintillation from Compton-scattering interactions. The fast UV scintillation light, measured by PMTs at the bottom of the chamber, gave the initial event trigger and the time of the interaction. 
The position of the interaction was read out on X-Y directional sets of wires through which the ionization charge drifts; the timing difference provided the Z-distance of the interaction sites. 
The demonstrated position resolution was $\sim$1~mm, and the energy resolution was 14\% FWHM at 511~keV~\cite{CurioniPhD}.
LXeGRIT pioneered the use of event pattern recognition to reject background events, and the initial concept of the Compton Kinematic Discrimination sequencing approach was developed for the LXeGRIT instrument~\cite{APRILE1993216, 2000SPIE.4141..168O}.
LXeGRIT was tested at balloon altitudes in 1999 and 2000~\cite{Curioni_2007}, and while LXeGRIT is no long in active development, the instrument demonstration has enabled more modern and capable designs using liquid TPCs~\cite{Aramaki2020}.

\subsection{Dedicated Polarimeters} 
\label{sec:polarimeterdetectors}


There is a sub-class of Compton telescopes that are optimized for detection efficiency and polarization sensitivity, particularly for transient events, as described in Section~\ref{sec:polarimeter_basics}. 
In general, these instruments use scintillator detectors which have poor position resolution and/or energy resolution and cannot perform full event reconstruction for imaging.
However, to be an effective Compton polarimeter, only the azimuthal scattering angle needs to be measured, and scintillator detectors provide excellent detection efficiency in large volumes with sufficient position resolution.

The first gamma-ray polarization measurements in fact came from instruments not designed for Compton scattering~\cite{2003Natur.423..415C, 2005A&A...439..245W}. By measuring coincident interactions in separate detector volumes, an azimuthal scattering angle distribution can be determined; however, without careful, dedicated polarization calibrations of the instrument before launch, systematics are difficult to address. 

The Cadmium Zinc Telluride Imager (CZTI) on Astrosat~\cite{2017JApA...38...31B}, India's first satellite, is designed as a hard X-ray imaging instrument with a coded-mask, but the mask becomes transparent above 100~keV. While not a dedicated polarimeter, CZTI was calibrated prior to launch and has been used to constrain the polarization of GRBs~\cite{2019ApJ...884..123C} and the Crab pulsar~\cite{2018NatAs...2...50V} by measuring the azimuthal scattering direction between neighboring pixels~\cite{2015A&A...578A..73V}. 

More recently, there has been a significant efforts devoted to the development of next-generation gamma-ray polarimeters~\cite{leap2021} (e.g. see Figure~\ref{fig:leap}).
The first prolific dedicated Compton polarimeter was the POLAR instrument, which has reported polarization measurements for a similar number of GRBs as CZTI. 

\subsubsection{POLAR}
POLAR was a dedicated Compton polarimeter launched in 2016 to the Chinese space laboratory Tiangong-2, and during its 6 month operation it detected 55 GRBs~\cite{2020A&A...644A.124K}.
POLAR was comprised of 25 modules (in a 5$\times$5 layout), each housing 64 thin and elongated plastic scintillator bars (8$\times$8 array) with dimensions $5.8\times5.8\times176$~mm$^3$. Scintillators inside the module were then read out with a 64 channel multi-anode photo-multiplier (MAPMT). 

\begin{figure}[b]
    \centering
    \begin{minipage}{0.56\textwidth}
    \centering
    \includegraphics[width=\textwidth]{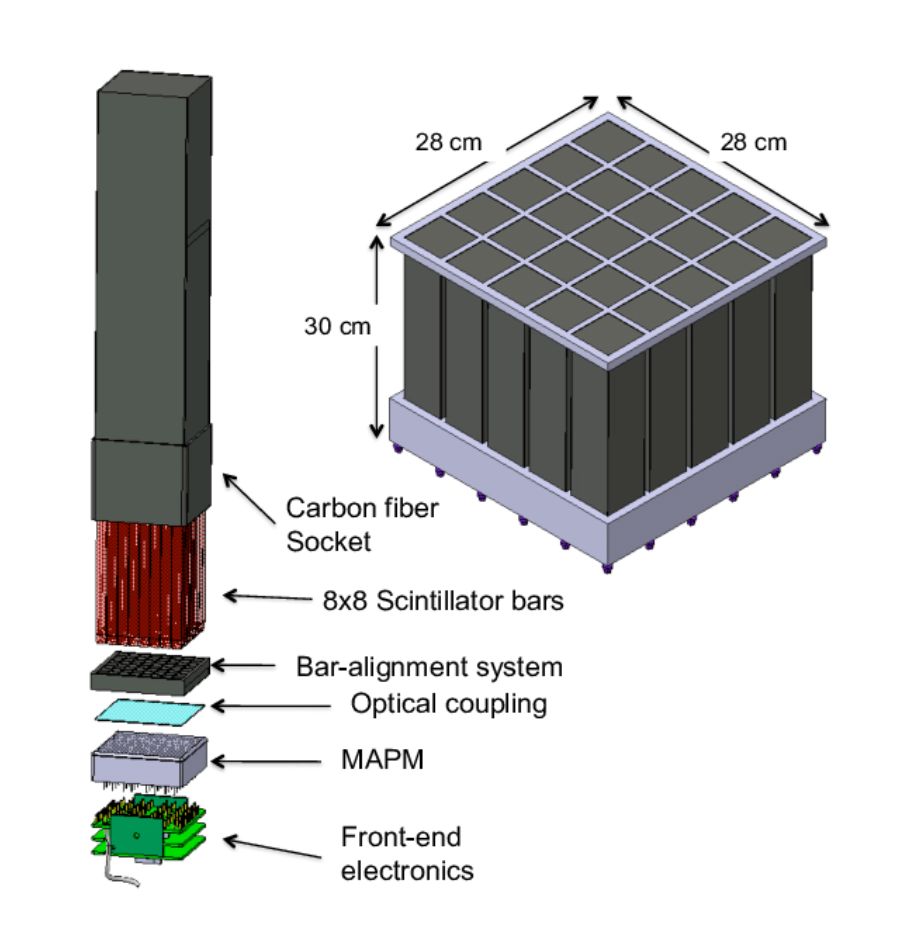}
    \subcaption{}\label{fig:POLAR_instrument}  
    \end{minipage}
    \begin{minipage}{0.43\textwidth}
    \includegraphics[width=\textwidth]{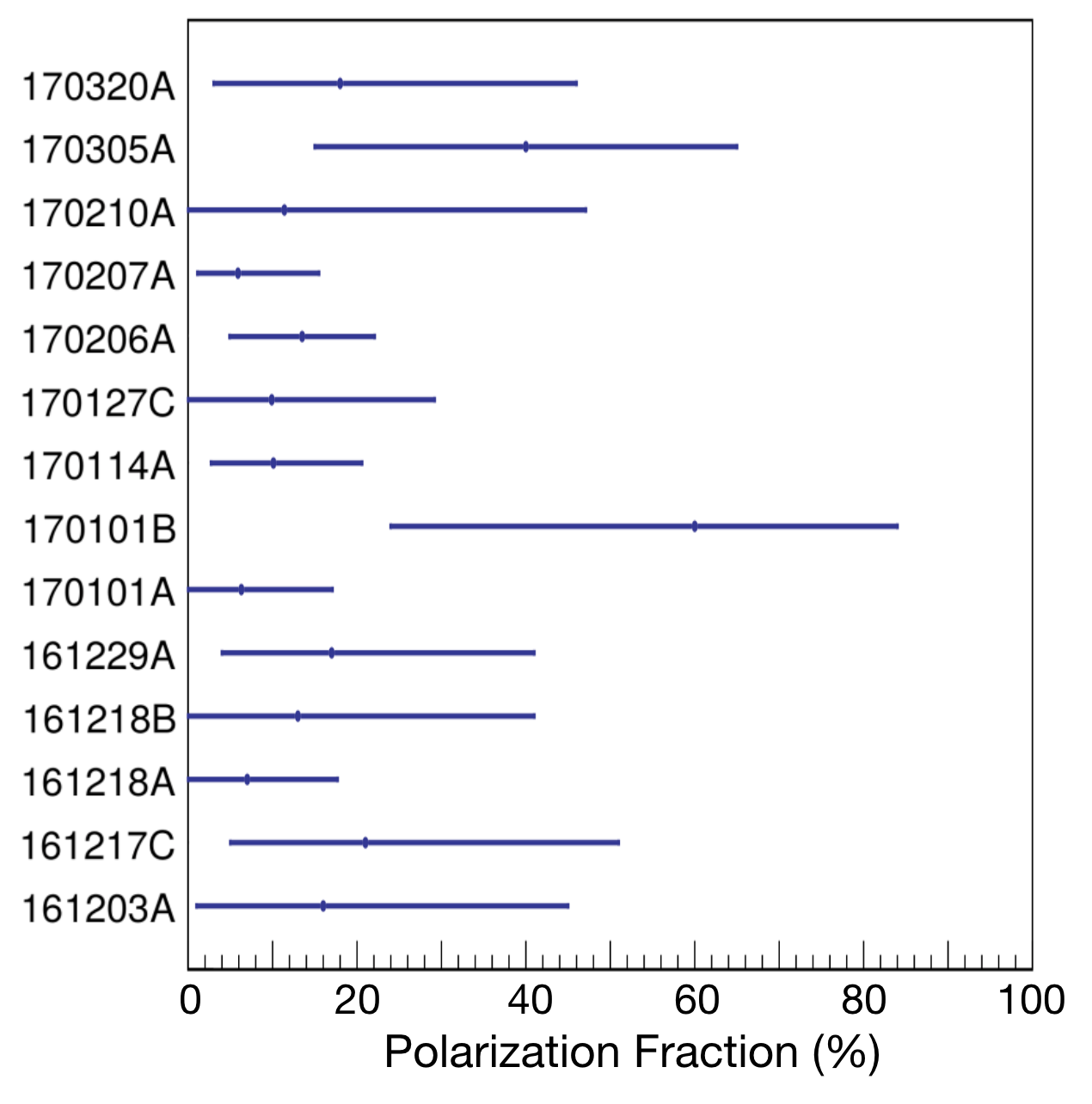}
    \subcaption{}\label{fig:POLAR_results}  
    \end{minipage}
    \caption{\textbf{(a)} The POLAR instrument consisted of 25 modules, each with an array of $8\times8$ plastic scintillating bars. With each scintillator being $5.8\times5.8\times176$~mm$^3$, the total instrument volume was segmented into 2034 bars individually read out through a multi-anode photo-multiplier (MAPM) to determine the azimuthal Compton scattering angle.  Figure from \citet{Suarezpolar}.
    \textbf{(b)} POLAR operated on the Chinese space laboratory Tiangong 2 for 6 months and detected 55 GRBs, where 14 bursts  had enough statistics to constrain the polarization fraction, as shown here. Modified from \citet{2020A&A...644A.124K}.
    }
    \label{fig:POLAR}
\end{figure}

A schematic of the POLAR instrument and one of its modules is shown in Figure \ref{fig:POLAR_instrument}. The instrument had a total effective area of $400$~cm$^2$ at 350~keV. Polarization information was obtained from photons which scatter between two scintillators within 100~ns. The second interaction can be either Compton scattering or photoabsorption~\cite{8069846}.
POLAR has released the largest catalog of GRB polarization measurements to date, demonstrating the polarization sensitivity of the telescope~\cite{2019A&A...627A.105B}. Looking forward, an enhanced instrument, POLAR-2, with an order of magnitude increase in effective area, is manifested for launch on board the China Space Station in 2024~\cite{2020SPIE11444E..2VH}.


\subsection{Compton and Pair Telescopes}
\label{sec:comptonandpairdetectors}



\begin{figure}[b]
    \centering
    \includegraphics[width=8cm]{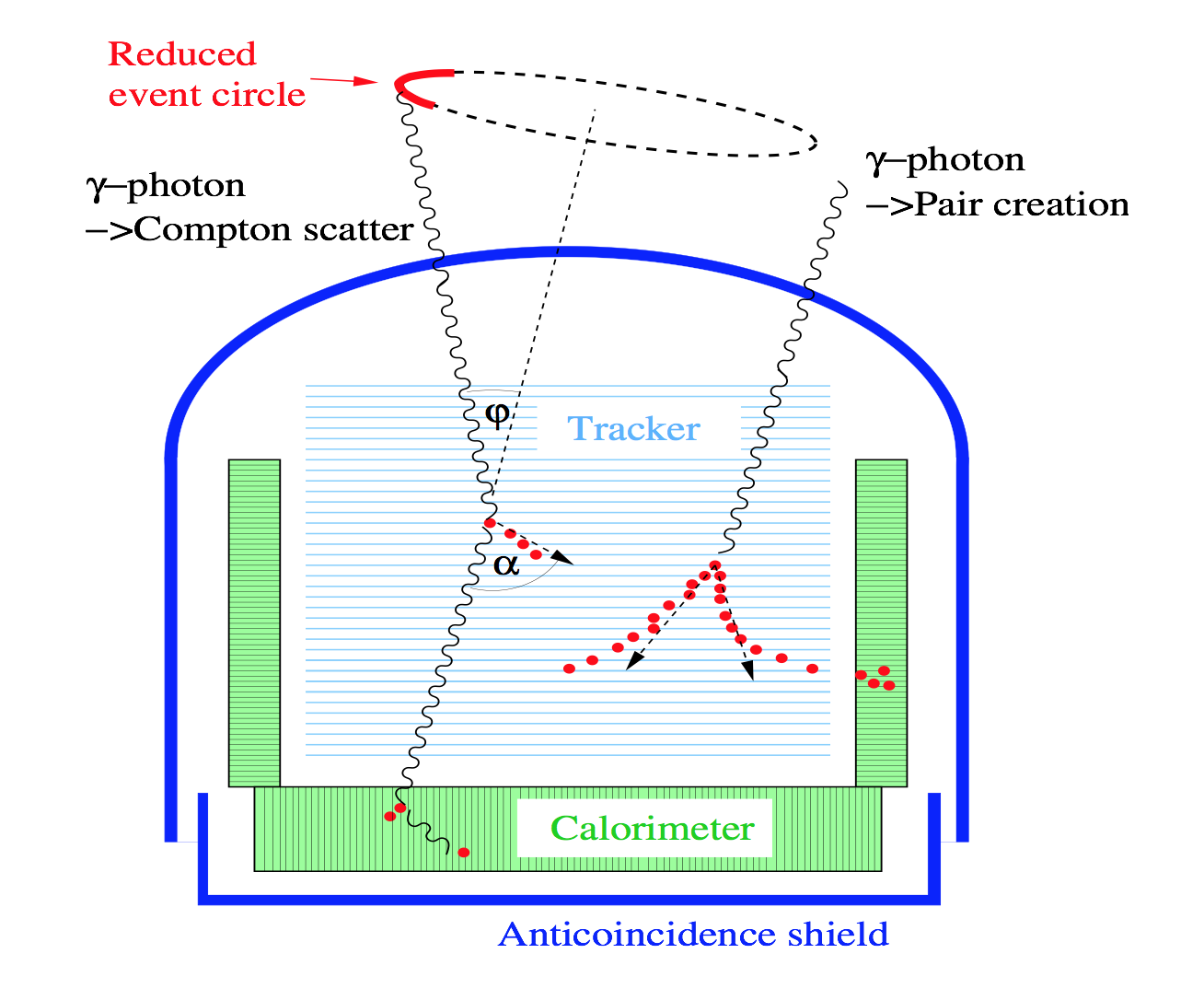}
    \caption{Schematic of a combined Compton and pair-creation telescope. A multi-layer tracker and thick calorimeter together detect and characterize Compton and pair events over a large energy range from $\sim$0.1 to 100 MeV. Figure from \citet{Kanbach_etal_2005}.}
    \label{fig:MEGA_principle}
\end{figure}

Detectors that have electron tracking capabilities can also be used to track the ionization from electron and positron pair-creation products.
This allows a single instrument to be sensitive to events across 3-4 orders of magnitude in energy from $\sim$100~keV in the Compton regime, to $>100$~MeV in the pair regime, depending on the size of the instrument.
Typically, combined Compton and pair-creation telescopes are designed with two separate detectors: a tracker, in which the initial Compton scatter or pair conversion takes place, and a large, high-Z calorimeter, which absorbs and measures the energy of the Compton-scattered photons and pair-products components. 
This concept is shown in Figure~\ref{fig:MEGA_principle}.
To contain the events, and provide enough volume for efficient tracking, these instruments are generally the largest scale of Compton telescope, and combined with the broad energy range, these instruments aim to bring the general capabilities of the Large Area Telescope on Fermi~\cite{Atwood_2009} into the MeV range.

The first Compton telescopes of this design were pursued in the latter 1990's and early 2000's.
Using a similar design TIGRE~\cite{10.1117/12.187266, oneil2003} and MEGA~\cite{2002NewAR..46..611B, Kanbach_etal_2005} demonstrated the electron tracking and pair capabilities of a multi-layer Compton telescope.
While MEGA was never flown, the significant laboratory and beam calibrations~\cite{Andritschke_etal_2004, zoglauerthesis}, and accompanying software~\cite{zoglauer2006}, have built a solid foundation for future missions. 
Similar designs with modern advancements are still being explored~\cite{DeAngelis_etal_2018, DeAngelis_etal_2021, McEnery_etal_2019, Fleischhack_etal_2021}.

\subsubsection{Medium Energy Gamma-ray Astronomy Telescope}

\begin{figure}[tb]
    \centering
    \begin{minipage}{0.5\textwidth}
    \centering
    \includegraphics[width=\textwidth]{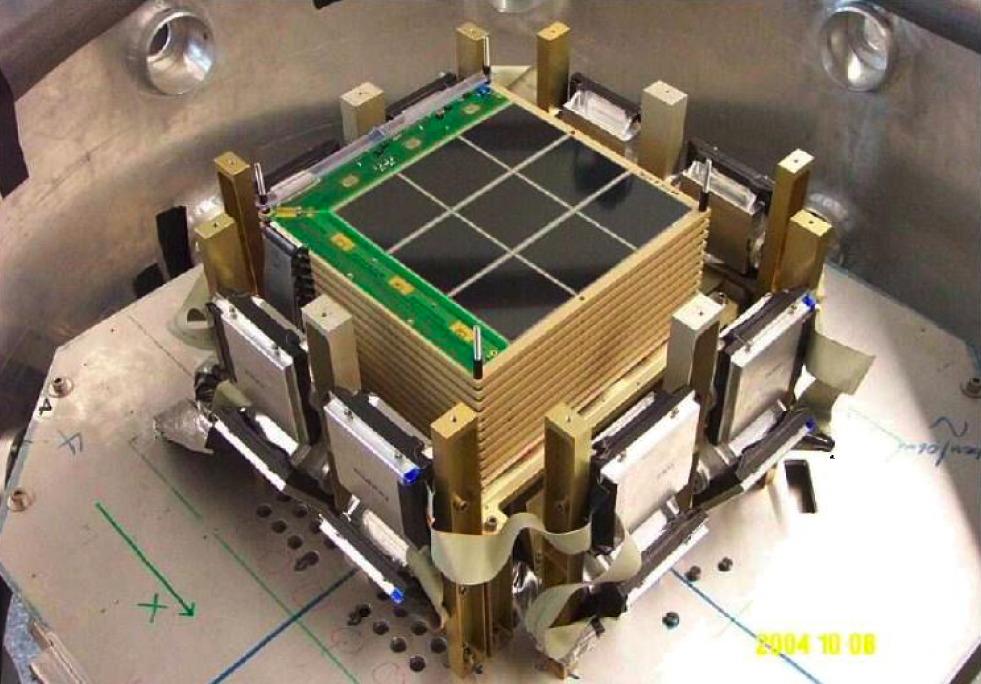}
    \end{minipage}
    \hfill
    \begin{minipage}{0.48\textwidth}
    \caption{The MEGA prototype consisted of a tracker with 11 layers of $3 \times 3$ arrays of  double-sided silicon strip detectors surrounded on the sides and the bottom with CsI calorimeter modules. The MEGA prototype was extensively tested and calibrated, but never flown. Image from \citet{2002NewAR..46..611B}.}
    \label{fig:megaprototype}
    \end{minipage}
\end{figure}

The Medium Energy Gamma-ray Astronomy (MEGA) Telescope prototype was built at the Max Planck Institute for Extraterrestrial Physics (MPE) in Garching, Germany.
The prototype detector, described in detail in \cite{Kanbach_etal_2005}, is shown in Figure~\ref{fig:megaprototype}. 
The instrument was built with 11 layers of double-sided silicon strip detectors (DSSD), arranged in $3 \times 3$ arrays of 500 $\mu$m thick silicon wafers, each $6 \times 6$~cm$^2$ in size with 470 $\mu$m strip pitch. 
The pixelated calorimeter consisted of 20 modules each with an array of $10 \times 12$ CsI(Tl) bars~\cite{SCHOPPER2000394}: the bottom calorimeter was 8~cm thick, and the side calorimeters were either 4 or 2 cm. The CsI bars were read out with silicon PIN-diodes. 

Beyond the significant technology maturation with MEGA which serves as a proof of concept for larger modern missions~\cite{Andritschke_etal_2004}, the lasting legacy of the MEGA prototype work is in the Medium Energy Gamma-ray Astronomy library (MEGAlib) toolkit~\cite{zoglauer2006, zoglauer2011, zoglauer2013}. MEGAlib was developed for MEGA and has become the state-of-the-art simulation and analysis software for Compton telescope development around the world~\cite{zoglauer2006}.

\subsection{Applications in Other Fields}
\label{sec:applicationsinotherfields}

While astrophysical telescopes detect gamma rays from distant celestial objects, accelerator experiments, medical imaging, and nuclear non-proliferation and environmental monitoring, require imaging targets in the near field. 
Compton imagers, referred to often as Compton cameras for near field and terrestrial applications, provide a wide FOV, high detection efficiency, and the potential for real-time imaging.
Compton cameras were first pursued in these other fields at the same time as early Compton telescopes were being developed for astrophysics.

One prevalent application for Compton cameras is environmental monitoring and nuclear non-proliferation.
A recent example of the power of Compton cameras was the environmental monitoring of radioactive contaminants after the March 2011 Fukushima nuclear power plant accident. 
Dust containing radioactive materials dispersed throughout the Fukushima Prefecture, and gamma rays emitted through the decay of unstable nuclei gave a way to find contaminated regions. 
A silicon cadmium telluride (Si/CdTe) Compton camera referred to as the ``Ultra-Wide-Angle Compton Camera'' was used to image the distribution of radioactive substances in the Fukushima area~\cite{Takahashi_2012}, where the radiation level ranged from 1 to $\sim$30 $\mu$Sv/h. 
The camera consisted of 2 layers of double-sided strip silicon detectors (DSSDs) and 3 layers of CdTe double-sided strip detectors (CdTe-DSDs), each with a detector area of $3.2 \times 3.2$~cm$^2$ and a strip pitch of 250~$\mu$m. 
With an energy resolution of 2.2\% FWHM at 662 keV, and a wide FOV corresponding to 2$\pi$ steradian (180$^{\circ} \times 180^{\circ}$), the Ultra-Wide-Angle Compton Camera was sensitive to nuclear line emissions across a large area. 
The camera had an angular resolution of 3.8$^{\circ}$ FWHM in the energy range from 500 to 800~keV, which gave the camera the ability to localize radioactive hot spots to areas $\sim$100~cm$^2$ at a distance of 10~m~\cite{Takeda_2015}.
Other Compton cameras also contributed to radiation monitoring in Fukushima~\cite{VETTER2018159, 2017NatSR...741972T, doi:10.1080/00223131.2019.1581111}

Similar to the instrumentation for astrophysical Compton telescopes, the designs of telescopes for environmental monitoring are diverse (e.g.~\cite{2018EPJWC.17006003I, 2018JaJAP..57b6401W}). 
Recent techniques have been developed to increase the size of high-quality CZT crystals, resulting in 3D-position sensitive detectors with a thickness over 1~cm and a volume of several cm$^{3}$~\cite{SHY2020161443, LEE20214080}.
By using the ratio of the charge collected on the anode and cathode electrodes, the depth of the interaction can be calculated in pixel coordinates. 
Even in the case of multiple interactions in a single detector volume, it is possible to obtain three-dimensional position and energy information at each interaction site. 
CZT is a promising candidate for Compton cameras for radiation monitoring due to its high detection efficiency and excellent energy resolution without the complication of cooling~\cite{10.1117/12.563905}.

Compton cameras are also being used in nuclear medicine to image radioactive tracers used for positron emission tomography (PET) and single photon emission computed tomography (SPECT) scans~\cite{10.1117/12.187266,KABUKI20071031,GoroYabu2021,10.1117/12.2321244,timpixSPECT,5410004, MOCHIZUKI201943}.
SPECT relies on a pin-hole or parallel-hole collimator and becomes less efficient for energies above a few hundred keV because those collimators become transparent.
Compton cameras are expected to advance medical imaging technology of radioactive tracers due to their increased efficiency and relatively precise imaging capabilities.

\begin{figure}[tb]
    \centering
    \begin{minipage}{0.4\textwidth}
    \centering
    \includegraphics[width=0.9\textwidth]{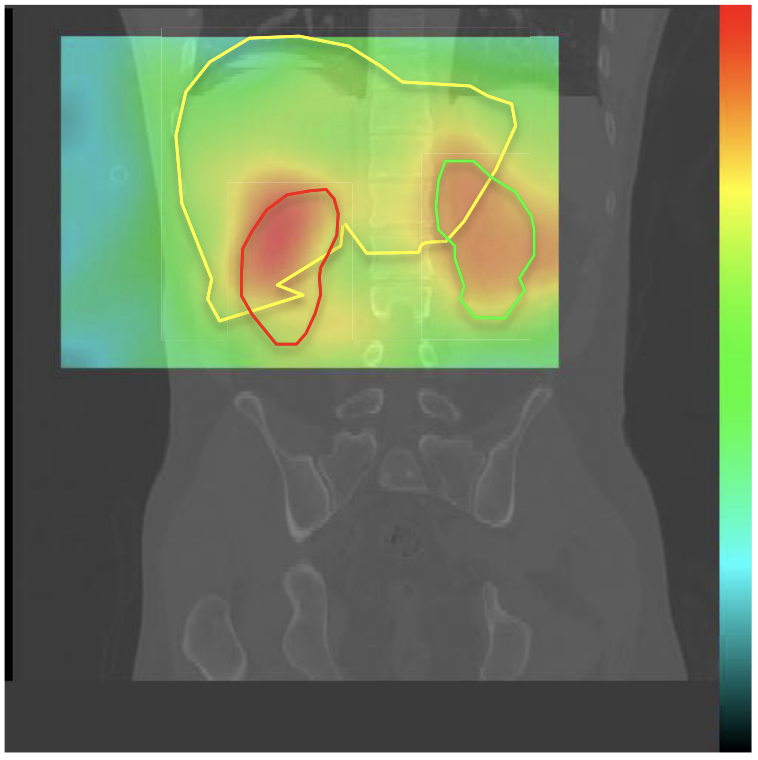}
    \end{minipage}
    \hfill
    \begin{minipage}{0.59\textwidth}
    \centering
    \caption{The Compton image of a human torso after the administration of $^{99m}$Tc-DMSA is shown in color contours. Overlaid on a computerized tomography (CT) image, these concentrations are consistent, as expected, with the left and right kidneys outlined in red and green. Figure from~\citet{Nakano2019}.}
    \label{fig:medicalimaging}
    \end{minipage}
\end{figure}

Figure~\ref{fig:medicalimaging} shows the Compton image of a patient after administering Technetium-99m dimercaptosuccinic acid ($^{99m}$Tc-DMSA), 
commonly used in the field of nuclear medicine, measured with a silicon cadmium telluride Compton camera. The concentration of the emission is consistent with the left and right kidneys, as expected~\cite{Nakano2019}. 
The range of technology developments is seen through various types of Compton cameras emerging for small animal {\it in vivo} imaging~\cite{Takeda2012,Motomura2013,Suzuki2013,kishimoto2017first}. 
In addition to high sensitivity, which minimize the radiation dose in the body, a practical application in medicine requires accurate estimation of the distribution of radionuclides in the body which is achievable through Compton imaging.

\section{Conclusions}


The MeV range is ripe with scientific potential, but remains relatively unexplored. Due to the low interaction cross-section, unavoidably high background radiation, and the inherent difficulties with Compton imaging, progress has been slower than the neighboring energy bands.

COMPTEL pioneered Compton imaging in the 1990's, and set a baseline performance for the next-generation telescope to exceed. The Hitomi/SGD was the second Compton imaging telescope to be launched. Sensitive in the range of 60 to 600 keV, SGD would have provided crucial observations in the soft gamma-ray energy range, but unfortunately, issues with the spacecraft cut the mission short in 2016 and only allowed for a few days of nominal science observations.
The dedicated polarimeter POLAR mission was also cut short, but was able to produce a catalog of GRB polarization measurements from its 6 months of observing in 2016--2017.
Looking forward, Compton telescopes again have a chance to make significant progress in the low MeV range with the launch of POLAR-2 in 2024 and COSI in 2026. 
POLAR-2 is expected to provide strong constraints on the measured polarization for many more GRBs.
COSI will perform a sensitive all-sky survey of the 0.2--5~MeV sky, with a particular focus on sources of gamma-ray line emissions.
Despite all of this progress, the COMPTEL's sensitivity has yet to be exceeded by an order of magnitude and there have been no further observations from $\sim$5 to 30~MeV after CGRO's termination in 2000.

The high observational backgrounds are one of the biggest challenges in Compton imaging. Activation from interactions of charged particles in the detector and spacecraft presents a challenging, and so far unavoidable, background component. 
The effects of activation can be reduced by minimizing the amount of passive material around the active detector volume, and avoiding high-Z materials in the instrument design.
Furthermore, a low-inclination ($<5^{\circ}$) low earth orbit avoids the SAA and minimizes the incident charge particle flux.
A measure of the recoil electron direction can reduce the impact of background radiation as the direction of each incoming photon can be more precisely constrained.
And sophisticated event reconstruction and background identification methods can be employed to reduce the background in post-processing.
Accurate modeling of the particle injection to the spacecraft is still a limitation, however, and coincident background radiation monitoring is necessary.

Doppler broadening remains a fundamental limit to the angular resolution of a Compton telescope. Without introducing an additional imaging technique, for example a coded mask, a telescope relying only on Compton imaging will not achieve an imaging resolution better than $\sim$1$^{\circ}$. 
Combined with the high backgrounds, this sets a hard limit on the achievable sensitivity of future missions.

Many current Compton telescope designs focus on the increase in effective area to increase sensitivity; however, a more impactful approach is potentially through the development of new concepts for low background observations. Techniques that go beyond active shielding, minimal passive material, and event reconstruction may be needed to make significant gains.


Despite progress in Compton imaging over the past few decades, observations in the MeV range continue to be a challenge. The MeV Gap cannot be fully filled with current instrument concepts and technology. For continued progress and improvements, technology developments are key. Borrowing insight from a leader in another field, Dr. Sydney Brenner, who won the 2002 Nobel Prize in Physiology or Medicine, said~\cite{brennerquote}:
\begin{quote}
    Progress in science depends on new techniques, new discoveries and new ideas, probably in that order. Innovation comes only from an assault on the unknown. I think one of the things about creativity is not to be afraid of saying the wrong thing.
\end{quote}
The large number of diverse instrument development efforts pursued around the world gives hope for the future of Compton telescopes to excel beyond the current challenges.

\section{Cross-References}
For further reading, please see the related entries from the handbook:
\begin{itemize}
\item ``Astrosat,'' Section III: X-ray Missions, Kulinder Pal Singh
\item ``Telescope concepts in gamma-ray astronomy,'' Section IV: Optics and Detectors for Gamma-ray Astrophysics,  Deirdre Horan,  Gottfried Kanbach \& Thomas Siegert
\item ``Orbits and background of gamma-ray space instruments,'' Section IV: Optics and Detectors for Gamma-ray Astrophysics, Vincent Tatischeff, Pietro Ubertini, Tsunefumi Mizuno \& Lorenzo Natalucci
\item ``Time projection chambers for gamma-ray astronomy,'' Section IV: Optics and Detectors for Gamma-ray Astrophysics, Denis Bernard, Stanley Hunter \& Toru Tanimori 
\item ``Design of gamma-ray polarimeters, Section IV: Optics and Detectors for Gamma-ray Astrophysics, Denis Bernard, Tanmoy Chattopadhyay, Fabian Kislat \& Nicolas Produit
\item ``Gamma-ray detector and mission design simulations,'' Section IV: Optics and Detectors for Gamma-ray Astrophysics, Eric Charles, Henrike Fleischhack \& Clio Sleator
\item ``The COMPTEL instrument on the CGRO mission,'' Section V: Space-based Gamma-ray Observatories, James Ryan and Volker Sch\"{o}nfelder
\item ``General introduction and history,'' Section XIX: Polarimetry, Enrico Costa
\item ``Maximum likelihood analysis of the data from COSI,'' Section XIX: Polarimetry, John A. Tomsick, Clio C. Sleator, Andreas C. Zoglauer, Hadar Lazar \& Alexander W. Lowell
\item ``Analysis of the data from POLAR,'' Section XIX: Polarimetry, Jianchao Sun \& Merlin Kole
\end{itemize}

\begin{acknowledgement}
The authors thank Hiroki Yoneda and Yutaka Tsuzuki for their assistance with the preparation of the manuscript. 
\end{acknowledgement}

\bibliographystyle{abbrvunsrtnat}
\bibliography{Citations}

\end{document}